\documentclass[useAMS,usenatbib,letterpaper]{mn2e}
\usepackage{times}
\usepackage{amssymb}
\usepackage{amsmath}
\usepackage{lscape,graphicx}

\def\M{{\rm M}}
\def\R{{\rm R}}

\title[Galactic conformity]{Galactic conformity and central / satellite quenching, from the satellite profiles of M$^{\ast}$ galaxies at $0.4<z<1.9$ in the UKIDSS UDS}
\author[W. G. Hartley et al.]{W.~G. Hartley$^{1,2}$\thanks{E-mail:
    william.hartley@phys.ethz.ch}
, C.~J. Conselice$^{1}$, A. Mortlock$^{1,3}$, S. Foucaud$^{4}$, C. Simpson$^{5}$\\
  $^{1}$School of Physics and Astronomy, University of Nottingham, University Park, Nottingham NG7 2RD, UK \\
  $^{2}$ETH Z\"urich, Institut f\"ur Astronomie, Wolfgang-Pauli-Str. 27, 8093, Z\"urich, Schweiz\\
  $^{3}$Institute for Astronomy, University of Edinburgh, Royal Observatory, Edinburgh, EH9 3HJ, UK\\
  $^{4}$Center for Astronomy and Astrophysics, Department of Physics \& Astronomy, Shanghai Jiao Tong University, 800 Dongchuan Road, Shanghai, 200240, China\\
  $^{5}$Astrophysics Research Institute, Liverpool John Moores University, ic2 Building, Liverpool Science Park, 146 Brownlow Hill, Liverpool, L3 5RF, UK  \\
}

\begin{document}

\date{}

\pagerange{\pageref{firstpage}--\pageref{lastpage}} \pubyear{2014}

\maketitle

\label{firstpage}

\begin{abstract}

\noindent We explore the redshift evolution of a curious correlation between the star-formation properties of central galaxies and their satellites (`galactic conformity') at intermediate to high redshift ($0.4<z<1.9$). Using an extremely deep near-infrared survey, we study the distribution and properties of satellite galaxies with stellar masses, ${\rm log} (\M_*/\M_{\odot})>9.7$, around central galaxies at the characteristic Schechter function mass, $\M \sim \M^{\ast}$. We fit the radial profiles of satellite number densities with simple power laws, finding slopes in the range $-1.1$ to $-1.4$ for mass-selected satellites, and $-1.3$ to $-1.6$ for passive satellites. We confirm the tendency for passive satellites to be preferentially located around passive central galaxies at $3\sigma$ significance and show that it exists to at least $z\sim2$. Meanwhile, the quenched fraction of satellites around star-forming galaxies is consistent with field galaxies of equal stellar masses. We find no convincing evidence for a redshift-dependent evolution of these trends. One simple interpretation of these results is that only passive central galaxies occupy an environment that is capable of independently shutting off star-formation in satellite galaxies. By examining the satellites of higher stellar mass star-forming galaxies (${\rm log} (\M_*/\M_{\odot}) > 11$), we conclude that the origin of galactic conformity is unlikely to be exclusively due to the host dark-matter halo mass. A halo-mass-independent correlation could be established by either formation bias or a more physical connection between central and satellite star-formation histories. For the latter, we argue that a star-formation (or AGN) related outburst event from the central galaxy could establish a hot halo environment which is then capable of quenching both central and satellite galaxies.
\end{abstract}

\begin{keywords}
Infrared: galaxies -- Galaxies: High Redshift -- Galaxies: Evolution -- Galaxies: Formation.
\end{keywords}

\section{Introduction}
\label{intro}

Over the past decade or so, large data sets of both the near and far Universe have transformed the study of galaxy evolution. The statistical power that these data sets now afford has enabled the identification of subtle patterns in the photometry, emission line ratios, environments, star-formation properties and morphologies of the galaxy population \citep[e.g.][]{Bell04,Baldry06, Weinmann06, Mannucci10, Lani13, Lin14}. These patterns have, in turn, lead to phenomenological descriptions of how the galaxy population emerges and evolves through time \citep{Peng10, Peng12, Dave12, Behroozi13, Conselice13, Driver13, Lilly13}. Naturally, purely phenomenological models do not provide an understanding of the detailed physical processes involved, but they can indicate which avenues of investigation are likely to be of greatest interest.

A potentially instructive pattern within the Sloan Digital Sky Survey (SDSS, \citealt{York00}) data was discovered by \cite{Weinmann06}. They identified a correlation between both the colours and specific star-formation rates (sSFR) of galaxies that are at the centre of their respective groups, and those of the satellites that surround them. This correlation is in the sense that red central galaxies have a greater fraction of red satellites and vice-versa. Trends of this sort in groups had been noted in a limited sense previously \citep{Wirth83,Ramella87,Osmond04}, but the advent of large data sets, such as the SDSS, was necessary to show that it is a systematic effect. \cite{Weinmann06} termed this behaviour `galactic conformity', and similar behaviour has since been confirmed by clustering analysis \citep{Ross09}, at separations beyond the group radius \citep{Kauffmann13}, by spectroscopic galaxy pairs \citep{Phillips14} and even appears to extend to the spatial satellite distribution within halos \citep{Wang08}. An equivalent correlation has also been discovered in the {\em morphologies} of central and satellite galaxies \citep{Ann08}. Furthermore, \cite{Papastergis13} find an anti-correlation in the spatial distributions of HI-selected galaxies and red galaxies, indicating that neighbours of red galaxies are deficient in star-forming gas.

The physical processes which cause galaxies to cease forming stars remains one of the most controversial and important topics in extragalactic astrophysics. It has been clear for many years that red (quenched) populations are typically either of high stellar mass or located in regions of elevated galaxy overdensity \citep{Butcher78b, Faber79, Dressler80, Baldry06}. More recently, it has become possible to follow the redshift evolution of the environmental and stellar-mass related passive fractions. The quenched fraction of galaxies at high stellar masses is seen to be only weakly dependent on redshift (e.g. \citealt{Hartley13,Knobel13,Straatman13}). The colour-density correlation, meanwhile, has been observed in blank field surveys to at least $z\sim1.5$ \citep[e.g.][]{Chuter11, Quadri12}, and in proto-clusters to perhaps as high as $z\sim3.3$ \citep{Lemaux14}. Reproducing the passive galaxy fraction as a function of stellar mass and environment continues to pose challenges to our models of galaxy formation, even at low redshift. In particular, the majority of theoretical models over-produce the number of low-mass passive satellite galaxies \citep[e.g.][]{Guo11}, and are yet to fully reproduce galactic conformity. Perhaps the redshift dependence of these quantities will provide the necessary clue to which aspect of physics is currently being modelled incorrectly.

Some facets of the observational results have, however, been successfully encapsulated by the phenomenological formalism laid out by \cite{Peng10}. By identifying two independent channels of quenching (mass and environment quenching), together with a small number of observationally motivated assumptions, Peng et al. were able to reproduce the stellar mass functions and relative fractions of red and blue (or passive and star-forming) galaxies over a range of environments at $0<z<1$. An important aspect of their model is that the efficiency of the mass quenching channel is proportional to the stellar mass of the galaxy. Due to the shape of the star-forming galaxy mass function, the model implies that galaxies with stellar masses at the characteristic Schechter function mass, $\M \sim \M^{\ast}$, dominate the recently mass-quenched population. If we are to search for the cause of the correlation between passive central and satellite galaxies, it would therefore seem prudent to study galaxies of $\M^{\ast}$ mass and their satellites.

This approach was taken by \cite{Phillips14}, using carefully controlled samples of $\M \sim \M^{\ast}$ passive and star-forming galaxies from the SDSS, each with a single satellite. The properties of the satellite galaxies were then compared with similar mass but isolated galaxies. They found that around passive centrals, the fraction of satellites that are also passive is elevated relative to the field. The satellites of star-forming galaxies, meanwhile, were found to be consistent with having been drawn from the field, i.e. the quenching efficiency due to their environment is zero. This result is similar in essence to that of \cite{Weinmann06}, but is a much stronger statement. These observational results tell us that the star-formation state (star forming or passive) of central and satellite galaxies are more tightly correlated than was previously thought. Given this insight, it is tempting to ascribe a common cause or origin for the quenching of central and satellite galaxies. In truth, however, there are a number of reasons why we might expect the observed correlation, a few of which we discuss in Section \ref{disc}.

Thus far, galactic conformity has only been studied in detail at low redshift. The predictions of how this correlation may persist or diminish at higher redshift is dependent on the origin of the conformity. We might therefore be able to constrain which physical processes are involved if we study the relative star-forming properties of $\M^{\ast}$ galaxies and their satellites from $z=0$ out to $z\sim2-3$ (the earliest epoch at which substantial populations of passive galaxies are observed).
In this paper we investigate the radial distributions of passive and star-forming satellite galaxies around $\M \sim \M^{\ast}$ galaxies in two redshift intervals covering $0.4<z<1.9$. We do not attempt to link the objects in these intervals in any kind of evolutionary sequence, but can nevertheless begin understand which processes could be at play in causing galactic conformity. 

The structure of this paper is as follows: in Section \ref{data} we introduce the datasets and derived quantities used for this work together with our sample definitions. We describe our method for computing satellite surface-number-density profiles in Section \ref{rad_measurement}. In Section \ref{prof_mass} we show our measurements for $\M\sim \M^{\ast}$ central galaxies for two stellar mass intervals of satellite galaxies. We further consider the passive satellite population separately. We separate central galaxies into passive and star-forming sub-samples in Section \ref{prof_passsf} and identify the existence of galactic conformity up to $z\sim2$. In Section \ref{passfrac} we investigate the radially-dependent passive fraction of satellites around our central samples. We discuss these results in Section \ref{disc} and propose a simple picture for the origin of conformity. Finally, we conclude in Section \ref{conc}. Throughout this work, where appropriate, we adopt a $\Lambda$CDM cosmology with $\Omega_\M = 0.3$, $\Omega_{\Lambda} = 0.7$, h$ = {\rm H}_0/100 ~{\rm kms}^{-1}{\rm Mpc}^{-1} = 0.7 $. All magnitudes are given in the AB system \citep{Oke83} unless otherwise stated.

\section[]{Data sets, derived quantities and sample definitions}
\label{data}

The dataset we use in this work is based on the UKIRT Infrared Deep Sky Survey (UKIDSS, \citealt{Lawrence07}) Ultra Deep Survey (UDS, Almaini et al., in prep.), Data Release 8. The UDS DR8 allows us to construct a mass complete galaxy sample with stellar mass, ${\rm log}(\M_*/\M_{\odot}) > 9.7$ to $z\sim2$ \citep{Hartley13,Mortlock13}.

\subsection{UKIDSS UDS data}

The data set from which we draw our sample has been described extensively elsewhere \citep{Simpson12,Hartley13}. Here we provide just a brief overview. In our selection band, the K-band, the UDS DR8 data reach $24.6$ ($5\sigma$, AB), estimated from the RMS of flux between apertures of $2\arcsec$ diameter in source-free regions of the image. Covering an area of $0.77$ square degrees, the UDS is the deepest degree-scale near-infrared survey to date, with J and H-band depths, ${\rm J}=24.9$ and ${\rm H}=24.2$, in addition to the K-band data. 

The UDS field overlaps the Subaru-XMM Deep Survey (SXDS, \citealt{Furusawa08, Ueda08}) from which we use the ultra-deep optical data. The coincident area with the near-infrared data, after masking bright stars and bad regions, is $\sim0.62$ square degrees. There are additional data of comparable depth at $3.6{\rm \mu m}$ and $4.5{\rm \mu m}$ from the UDS Spitzer Legacy Program (SpUDS) and in the u$^{\prime}$-band (Foucaud et al., in prep.). Further details of these data and the description of the source extraction can be found in \cite{Hartley13}, while details of astrometric alignment and multi-wavelength catalogue construction are described in \cite{Simpson12}. Finally, X-ray \citep{Ueda08}, radio \citep{Simpson06} and $24{\rm \mu m}$ data (also from SpUDS) were used to remove AGN and refine our sample selection. This catalogue is henceforth referred to simply as the UDS catalogue.

\subsection{Photometric redshifts}
\label{photz}

\begin{figure}
\includegraphics[angle=0, width=240pt]{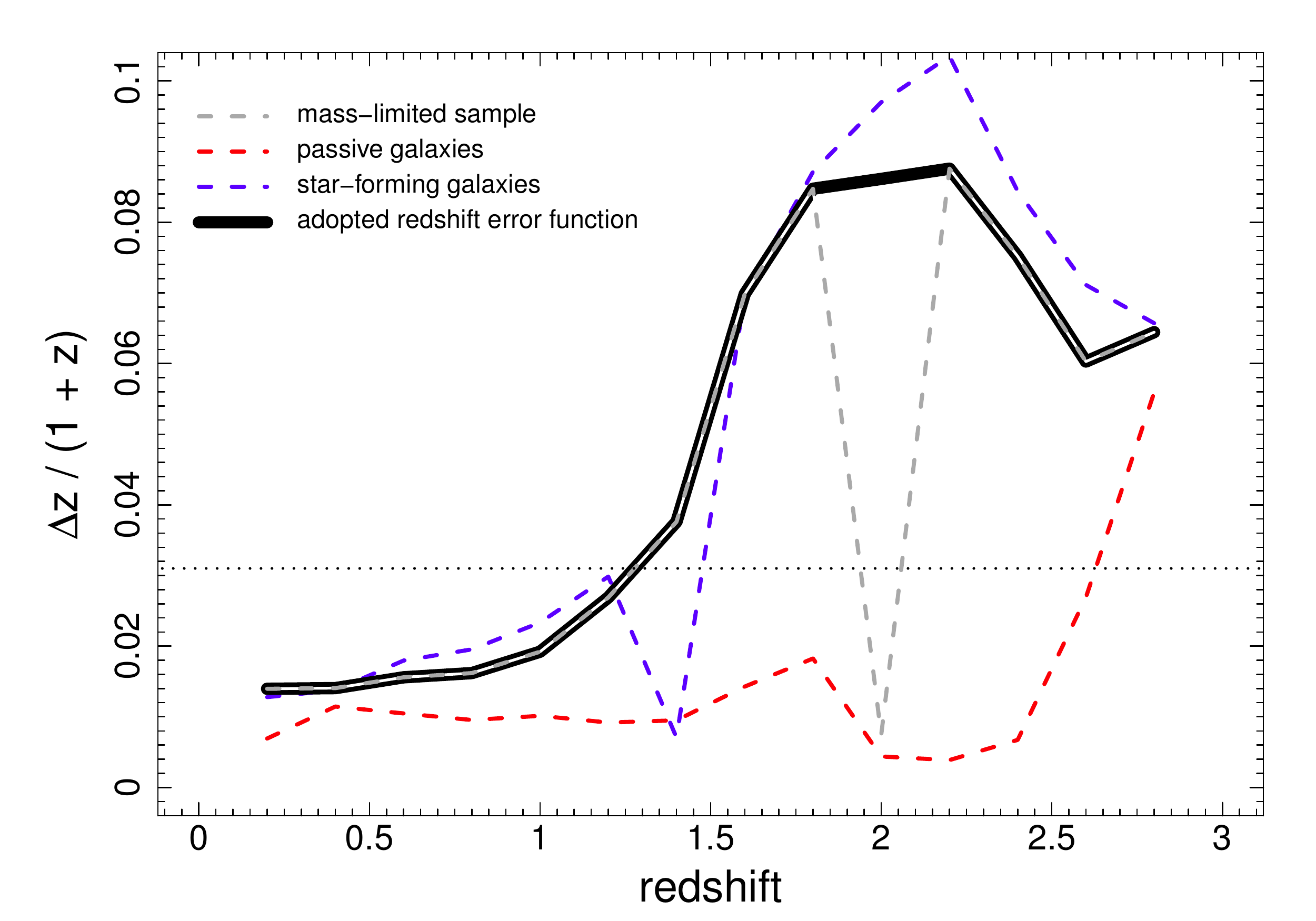}
\caption{Photometric redshift uncertainty as a function of redshift, following the method of \protect\cite{Quadri10}. The full mass-limited sample (grey dashed line) is shown alongside passive (red dashed line) and star-forming (blue dashed line) sub-samples (see section \ref{sample}). The redshift error function adopted in this work is shown by the solid heavy black line and the dotted horizontal line shows the uncertainty derived from the NMAD estimator applied to the spectroscopic sub-sample ($\Delta z/(1+z) = 0.031$). A colour version of this and all other figures are available in the on-line journal.}
\label{zerr}
\end{figure}

\begin{figure}
\includegraphics[angle=0, width=240pt]{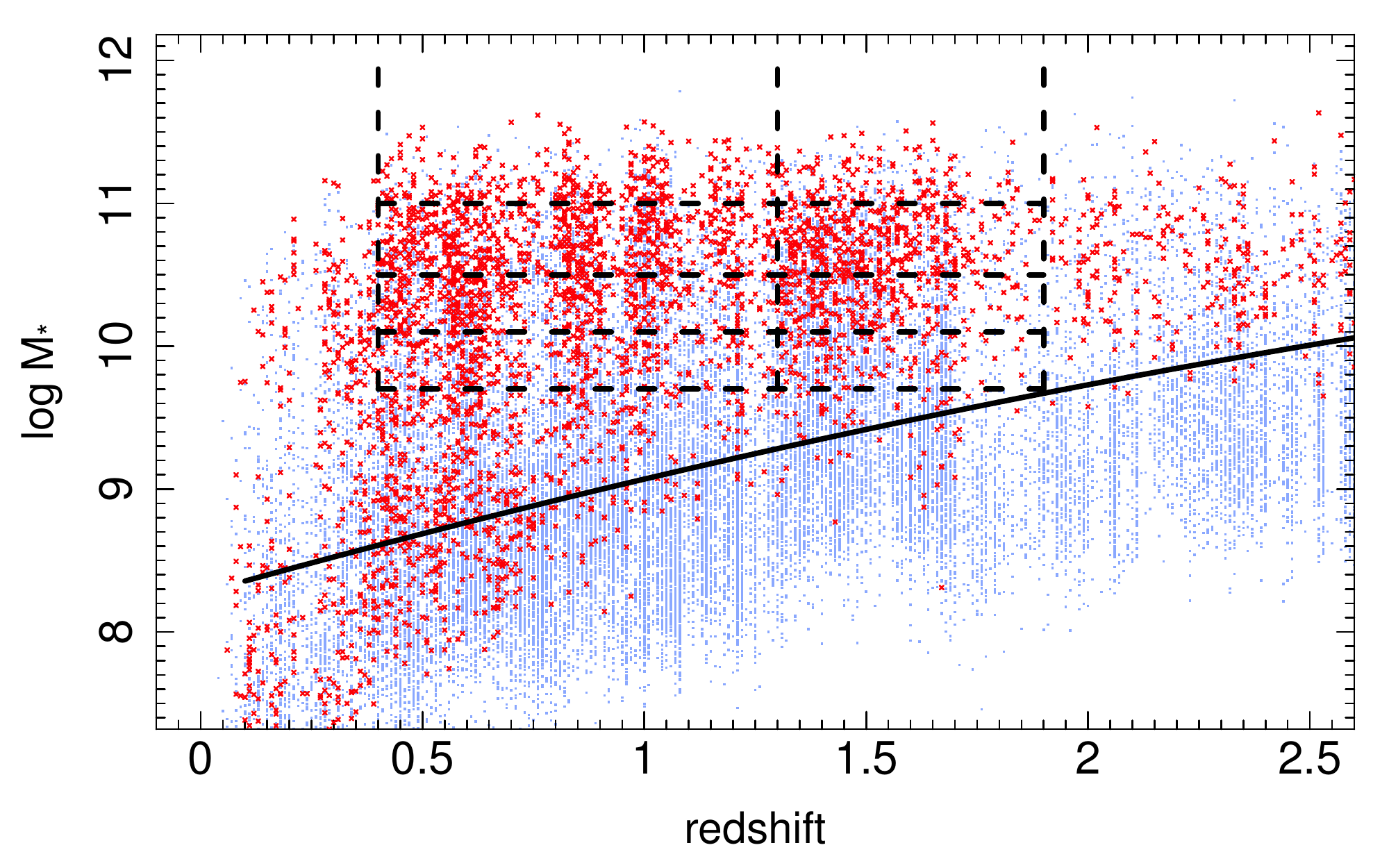}
\caption{Log (stellar mass) vs redshift for a random sub-sample of one third of our dataset. Light blue points are the ${\rm K}<24.3~$(AB) galaxy sample. The passive subset are shown by red crosses. The selection boundaries used for central and satellite populations are shown by the thick dashed lines. Finally, the solid curve shows the stellar mass completeness limit computed in \protect\cite{Hartley13}.}
\label{mz}
\end{figure}

Obtaining spectroscopic redshifts for substantial numbers of faint, non-star-forming galaxies at high redshift is prohibitively expensive with even the most sensitive multi-object spectrographs. The use of photometric redshifts in this study is therefore mandatory and the reliability of those redshifts is of crucial importance. To derive photometric redshifts (phot-z) we use the {\tt EAZY} package \citep[v1.0, ][]{Brammer08} using the default set of principal component templates with the addition of a slightly dust-reddened star-forming template. This extra template was constructed from {\tt EAZY}'s bluest template with ${\rm A_v}=0.1$ following the SMC extinction law \citep{Prevot84}. Motivated by inspection of the available $z>2$ rest-frame UV spectra, this slightly reddened template compensates for the deeper absorption lines and variety in dust extinction laws at high-z. We furthermore adapted the photometric zero-points in each band to improve agreement between the predicted and measured photometry for the sub-sample with secure spectroscopic redshifts. The phot-z of an object was taken as the maximum likelihood redshift, after applying {\tt EAZY}'s default K-band apparent magnitude prior.

The resulting photometric redshifts were tested in two ways. Firstly, we used the UDSz spectroscopic data (ESO Large Programme 180.A-0776, Almaini et al., in prep.) and archival redshifts taken from the literature \citep{Simpson12} to test a bright (${\rm K}<23$) sub-set of galaxies. In \cite{Hartley13} we found a dispersion of $\Delta z / (1+z) \sim 0.031$ using the NMAD estimator, after excluding AGN and catastrophic outliers. Secondly, we used the approach outlined in \cite{Quadri10}. This approach relies on the fact that galaxies with small separations on the sky are more likely to be physically associated (and therefore have the same redshift) than galaxies with large separations. The distribution of redshift differences, $|z_1 - z_2|$, for unassociated galaxies can be found by randomising galaxy positions. If the phot-z are perfect, then we would expect a sharply peaked excess over the randomised distribution at $|z_1 - z_2| = 0$. In reality there is a width to this excess peak, which reveals the statistical photometric redshift error distribution. We represent this distribution by fitting a simple Gaussian.

To estimate the uncertainty as a function of redshift we defined a primary galaxy sample within an interval of $\delta z_{{\rm phot}} \pm 0.2$ about each redshift within $0.2<z<2.8$, in steps of $dz=0.2$. These primary galaxies were then cross correlated with the whole sample to estimate the uncertainty. In Figure \ref{zerr} we show redshift uncertainties recovered in this way as a function of redshift for three samples: the full mass-complete (grey dashed line), passive (red dashed line) and star-forming (blue dashed line) samples (see section \ref{sample} for the definition of the latter two samples). At $z_{{\rm phot}}<1.3$, the photometric redshift uncertainties for all three samples are within the value derived from the spectroscopic sample. At higher redshift the redshift uncertainty for star-forming galaxies rises steeply and then begins to fall again at $z>2.2$ (at which point the Lyman break enters the u$^{\prime}$-band). It is worth noting that the redshift uncertainty for passive galaxies remains small out to $z\sim2.5$. More accurate photometric redshifts for high redshift passive galaxies (compared with similar redshift star-forming objects) have been reported in the literature previously \citep{Quadri12}. This can easily be understood due to the presence of a strong $4000{\rm \AA}$ break feature, and partly because the passive sample is biased towards brighter galaxies. Beyond $z=2.5$, the relatively small number of galaxies in the passive sub-sample which are above the stellar mass completeness limit means that the redshift uncertainties are themselves unreliable. 
 
A small number of redshift sub-samples were not fit well by a simple Gaussian. One of these sub-samples is the simple mass-limited sub-sample at $z\sim2$. The best-fit Gaussian suggests that the redshift uncertainty is extremely small at this redshift. The failure of an appropriate fit in this case is caused by a strong peak of pairs with $z_1 - z_2 = 0$. Through inspection of the star-forming sample at the same redshift (which dominates the mass-selected sample by number), interpolation of neighbouring data points should provide a reasonable estimate of the true uncertainty. The redshift dependent uncertainty we adopt for this work is then shown by the thick solid black line in Figure \ref{zerr}.

\subsection{Stellar masses, M$_*$}
\label{masses}

Stellar masses were determined by fitting the available photometry (u$^{\prime}$BVRi$^{\prime}$z$^{\prime}$JHK, and IRAC bands 1 and 2) to a large grid of synthetic SEDs, using the \cite{Bruzual03} stellar population synthesis models and a \cite{Chabrier03} initial mass function. Star-formation histories were assumed to be exponentially declining and parametrised by an age and e-folding time:

\begin{equation}
{\rm SFR}_{t=obs}={\rm SFR}_{t=form}\times e^{-\frac{t}{\tau}}.
\end{equation}

These parameters were allowed to vary within, $\tau= 0.01 - 10.0~{\rm Gyr}$ and $t= 0.001 - 13.7~{\rm Gyr}$. Templates that are older than the age of the universe at the redshift of the galaxy were forbidden. In addition dust extinction was included with values, ${\rm A_v} \le 5$, and metallicities, $0.0001 \le {\rm Z} \le 0.05$, were allowed. The stellar mass of a galaxy was found by scaling the best-fitting template such that the implied K-band flux was equal to the observed value. Further details of the method can be found in \cite{Hartley13} and \cite{Mortlock13}. From the best-fitting template we also recover rest-frame luminosities by integrating the appropriate filter over the synthetic SED. The distribution of our data in stellar mass and photometric redshift is shown in Figure \ref{mz}, where we plot a random sample of one third of our data. Star-forming galaxies are represented by blue points, while the small red crosses are passive galaxies according to our rest-frame colours and multi-wavelength data (Section \ref{sample}). We show the $95\%$ stellar-mass completeness limit computed in \cite{Hartley13} by the solid lines (M$_{{\rm lim}} = 8.27 + 0.87z - 0.07z^2$), and delimit the samples used in this work with dashed lines.

\subsection{Sample selection}
\label{sample}

To construct our galaxy samples, the UDS catalogue was first trimmed to a magnitude limited sample. For this stage we used ${\rm K}<24.3$ to ensure high completeness and to account for slight variations in depth across the image. Stars were removed using a combination of observed-frame colours where stars form a tight sequence (namely in B-z, z-K space, a K-excess diagram and a J-K versus K plot), and the difference in K-band magnitudes measured within $0.7{\arcsec}$ and $1.5{\arcsec}$ apertures. Cross-talk artefacts \citep{Dye06} are also a potential source of contamination, mimicing high-redshift passive galaxies in their colours. Fortunately the cross-talk pattern is relatively well known. The positions of cross-talk artefacts were predicted from the locations of stars and cross-correlated with the galaxy catalogue. Sources with colours indicative of being cross-talk were also added to the list of potential artefacts. This cross-talk catalogue was then exhaustively checked by eye.

The separation of our sample into passive and star-forming sub-samples follows the method described in \cite{Hartley13}. A two-colour technique (the UVJ selection), first used by \cite{Wuyts07} and subsequently refined by \cite{Williams09}, was used to separate the catalogue into passive and star-forming samples. Selection is performed in the rest-frame, using the U-V and V-J colours, from which we are able to broadly separate red passive objects from red dusty galaxies. We apply the selection boundaries from \cite{Williams09}, extending their selection for galaxies at $1<z<2$ to higher redshift as we do not expect strong colour evolution at $z>2$. We then refined these sub-samples by removing X-ray or radio-detected objects. Objects selected as passive by the UVJ selection, but with sSFR$>7.43\times10^{-11} {\rm yr}^{-1}$ (implied by either $24{\rm \mu m}$ flux or the SED template fit) were re-assigned as star-forming galaxies (see \citealt{Hartley13} for further details).

\subsection{Central galaxy definition}
\label{censat}

Currently, and for the foreseeable future, it is not possible to obtain reliable spectroscopic redshifts for significant numbers of low-mass passive galaxies at intermediate or high redshift ($z>1$). Such samples would be required if we are to identify individual pairs of central and satellite galaxies, as \cite{Phillips14} do. Our approach is instead to use the deep photometry and statistical power of the data at our disposal to find the average satellite surface density profiles around $\M^{\ast}$ galaxies. 

We began by defining a `central' galaxy sample in the following way. Galaxies with stellar mass $10.5<{\rm log} (\M_*/\M_{\odot})<11.0$ of the required type (passive, star-forming or purely mass-selected) and in the required redshift range were considered potential centrals. This stellar mass range is chosen to broadly cover star-forming galaxies which are likely to soon quench off the main sequence and the most recently quenched passive central galaxies, while maintaining roughly similar numbers in the star-forming and passive sub-samples. We use $\M \sim \M^{\ast}$ synonymously with this mass range, though most recent determinations of the characteristic Schechter function mass are towards the upper end of this range (e.g., \citealt{Ilbert13, Muzzin13}; Mortlock et al., in prep.). The value of $\M^{\ast}$ for star-forming galaxies does not evolve strongly with redshift \citep[e.g.][]{Ilbert10}, so we use the same mass interval at all redshifts. 

If any potential central had another galaxy from the parent catalogue that was more massive by $0.3~$dex (or greater) within $450~{\rm kpc}$ (projected) and $\pm\sqrt{2}~\sigma_z (1+z)$ in redshift, then it was rejected. These limits are somewhat arbitrary and were motivated by the desire to avoid contamination whilst retaining the greatest number of objects possible. In particular, $450~{\rm kpc}$ is comfortably greater than the radii of the typical halos at $z\sim1$ that we expect to host the central galaxies in this work (${\rm M_{halo}}\sim 10^{12} - 10^{13}~{\rm M}_{\odot}$, \citealt{Mo02,Hartley13}). Meanwhile $0.3~$dex is the expected uncertainty in log stellar mass, including uncertainties in photometric redshift and mass to light ratio\footnote{We note that these values leave the possibility of a small amount of contamination by high-mass satellite galaxies in our central sample. However, we have verified that using more conservative parameters for radius or redshift does not affect our conclusions (though the uncertainties become larger due to the smaller number of objects used).}. The value of $\sigma$ is redshift dependent and shown in Figure \ref{zerr}. The factor $\sqrt{2}$ is to take account of the fact that each object has an uncertainty on its photometric redshift. The central galaxy candidates are among the brighter objects at their redshift, and so will have more accurate photometric redshifts than Figure \ref{zerr} suggests. The candidates that survived this cut are henceforth referred to as central galaxies.

Other similar works \citep[e.g.][]{Tal13} prefer a rejection criterion strictly requiring the galaxy to be the most massive object within the radius and redshift tolerance allowed (i.e. without the $0.3~$dex flexibility we use). In cases where there are two (or more) similarly massive galaxies in close proximity it is not always clear which should be considered the central galaxy, or even whether defining a single object as the `central' galaxy has any meaning. Furthermore, uncertainties associated with stellar mass determination could lead to the intrinsically higher-mass galaxy being rejected. We introduced the $0.3~$dex stellar mass flexibility to avoid misidentifying centrals in this way and to allow similar mass galaxies within a single halo an equal standing in our analysis. To prevent satellite galaxies in such `dumb-bell'-like halos being over-represented, we also introduced a weight to each central galaxy. 

The central galaxies were given a weight, initially equal to $1/(1+n_{{\rm neighb}})$, where $n_{{\rm neighb}}$ is the number of objects with stellar mass $\pm 0.3~{\rm dex}$ of the central galaxy mass, and which satisfy the radius ($450~$kpc) and redshift ($\pm \sqrt{2}~\sigma$) conditions. However, even within the stellar mass interval $10.5<{\rm log} (\M_*/\M_{\odot})<11.0$, the stellar mass distributions of star-forming and passive galaxies are different, and this difference could introduce biases into the analysis. We therefore re-weighted the galaxies to remove this bias as so: weighted histograms of stellar mass, in bins of $d{\rm log}(\M_*/\M_{\odot})=0.05$, normalised by the sample size, were constructed for the intended central sample (passive or star-forming). An identically constructed but purely mass-selected sample at the same redshifts was then used to produce a stellar mass histogram in the same way. The weights of the target sample (passive or star-forming) were adjusted using a simple bin-by-bin rescaling, such that the weighted stellar-mass histograms of the target sample and mass-selected sample were equal. A quantified example of the weighting procedure used during this work is provided in the Appendix.

\section{Radial satellite profile measurement}
\label{rad_measurement}

The counting of satellite galaxies was performed in concentric annuli around the central galaxy. We used logarithmically spaced radii with width $\Delta {\rm logR} = 0.2$, starting with an inner radius, $\R=1.6~$kpc. We took account of bad regions and inaccessible areas of the image, due to e.g. bright stars, by dividing the counts for an object by the ratio of unmasked area to expected area in each annulus. Each object in the parent catalogue was considered as a possible satellite of a central galaxy if it lay within $1~{\rm Mpc}$ radius\footnote{Scales larger than $350~{\rm kpc}$ are not used in the analysis, but provide a further check on how far in radius we can reliably measure.} (projected, using the angular diameter distance at the redshift of the central), and within $1.5\times\sigma_{z}(1+z)$ in redshift of the central (again, using the redshift of the central to define the redshift uncertainty). We make the assumption that the redshift uncertainty is dominated by the satellite galaxies, and that the phot-z of the central is essentially correct. 

Choosing the allowable range in $\Delta z$ is a balance between counting a high fraction of the true satellites and avoiding contamination by fore and background objects. The photometric redshift uncertainties as a function of redshift for each of our samples are shown in Figure \ref{zerr}. As shown, passive galaxies have more reliable phot-zs on average. Using $1.5\sigma$ in our redshift tolerance means we miss only $\sim10\%$ of satellites, with the vast majority of them likely to be lower-mass star-forming objects. If the phot-z errors are distributed according to a Gaussian, as assumed for our uncertainty estimation, then it would be reasonable to simply multiply the measured number of star-forming galaxies to account for these potentially missing satellites. However, whilst it is not an unreasonable approximation, it is unlikely that the redshift errors are truly Gaussian. Moreover, catastrophic outliers, whilst few at only $4\%$, would not be represented by this correction. We therefore elect to present the results as measured and not correct the satellite counts. We do not expect this systematic error to have an important impact upon our conclusions, but we highlight the possible effects where relevant during our analysis.

If a candidate satellite is physically associated with the central then its estimated stellar mass could be incorrect. It's true redshift should be that of the central galaxy, but the stellar mass is computed at the phot-z of the satellite and not the redshift of the central. Keeping their mass to light ratios fixed, we therefore scale the masses of candidate satellites using the difference in luminosity distance between the phot-z of the central and satellite candidate. They were included in the satellite count if their scaled stellar mass fell into one of the required ranges, which we define as $9.7<{\rm log}(\M_{*}/\M_{\odot})<10.1$ and $10.1<{\rm log}(\M_{*}/\M_{\odot})<10.5$. The lower limit corresponds to the stellar mass completeness limit at our upper redshift limit, $z=1.9$. It is worth noting that the masses of the satellites we are investigating are not particularly small. For instance, the Large Magellanic cloud is below the stellar mass cut of our lower-mass sample \citep{James11}. Nevertheless, the star-formation properties of these relatively high-mass satellites can still contain vital information about the role of environment in galaxy evolution. The satellite counts within an annulus are multiplied by the central galaxy weight, and summed over the central population. To obtain the number of satellites per central, we divide the overall count by the sum of weights of the central sample.

Although measured, the satellite counts in radii, $\R<10~$kpc are unreliable. At $\R<10~{\rm kpc}$ the separation between the central and satellite galaxies is only $1-2\arcsec$ and so blending could become a serious problem (the FWHM of the K-band detection image is $\sim0.7\arcsec$). Furthermore, because we choose our central galaxy rejection criterion to be within $450~$kpc, there is a risk of substantial contamination in the satellite counts at large radii. Formally, we could expect some contamination to relatively small scales ($\R\sim100~$kpc) if we are unlucky enough to have a very massive galaxy just outside the $450~$kpc limit. However, we see no evidence for significant contamination at these scales, and the background correction (see section \ref{background}) should account for additional satellites counted in this way. Nevertheless, to be cautious we consider points at $\R>350~$kpc as unreliable. These ranges in separation are not included in our analysis.

\subsection{Background correction}
\label{background}

The counts of satellite galaxies in concentric annuli comprise three contributing sources. These are:

\begin{enumerate}
\item Physically associated satellite galaxies (our intended signal).
\item An uncorrelated background (and foreground) from galaxies that happen to lie along of sight with photometric redshifts within $1.5~\times~\sigma_{z}~(1+z)$ of the central object.
\item A correlated background arising from the fact that dark matter halos and their hosted galaxies are not randomly distributed in space (2-halo clustering contribution). 
\end{enumerate}

The second and third contributions are contaminants in our attempt to measure the first. The third source is the more challenging to remove as it depends upon the intrinsic clustering of the central sample. Fortunately, on sufficiently small scales it is sub-dominant to genuine satellite counts, as can be seen in halo occupation decompositions of clustering data \citep[e.g.][]{Zheng07}. Nevertheless, as we wish to examine the radial distribution of satellite galaxies we must attempt to remove this contribution in addition to the more common uncorrelated background correction. 

This issue was looked at in detail by \cite{Chen06}. Using mock galaxy samples they investigated four background correction schemes for radial satellite profiles, paying particular attention to the 2-halo clustering contribution. Their conclusion was that this contamination is best removed by taking the background correction annuli around a centre placed within a correlation length (r$_0$) of the central galaxy. For small enough fields this condition is automatically satisfied by the constraint of the field size on random galaxy placement. In our case, we have a large enough field that we must explicitly consider the correlated contamination. The nearer to the original central galaxy the correction is estimated, the closer the estimate of the satellite profile gets to the true profile. Needless to say however, if the correction is taken too close to the original object then the counts will be over corrected by genuine satellites being mistaken for contaminants.

For each central galaxy (the `parent' central galaxy) we constructed a random catalogue of $20$ objects, around which we count satellite galaxies in the exact same way as the real central galaxies. The number of randoms was chosen to ensure that the uncertainties are dominated by the real galaxy counts and not the background estimation. These random objects inherit all of the relevant characteristics of their parent galaxy (e.g. redshift, stellar mass). The minimum correlation length of our samples is r$_0\sim4~{\rm h}^{-1}{\rm Mpc}$ \citep{Hartley10}. We place the correction objects randomly in an annulus $1-2~{\rm Mpc}$ from the original galaxy. This choice allows us to reliably study the profile of satellite galaxies to the typical halo virial radius, while keeping the random locations well within the minimum correlation length. Halos with masses much higher than ${\rm log}~(\M_{{\rm halo}}/\M_{\odot})=13$ may suffer some degree of over correction in the outskirts, but very high mass halos with a central galaxy of ${\rm log}~(\M_*/\M_{\odot})<11$ are expected to be rare \citep[e.g.][]{Birrer14}.

The weighting procedure we adopt for our central sample requires an additional layer of complexity for the random sample. We must apply the same rejection and weighting to the $20$ randoms as we do to the central galaxy sample, otherwise the background could be overestimated. In the case where a random is rejected (due to a galaxy with mass, log (M$_*/\M_{\odot}) > 0.3$ more massive than the random lying within the search cylinder), it is replaced by another randomly chosen position. Otherwise, the randoms are weighted according to the number of similar mass galaxies that lie within the radius and redshift tolerances given in section \ref{sample}, exactly as we do for the central galaxies. The $20$ random galaxies must, between them, contribute the same weight to the background estimate as the parent galaxy does to the satellite count. The weights of these $20$ random galaxies are therefore re-normalised such that the sum of their weights is equal to the weight of their parent central galaxy. This aspect of the method is also covered by the example in the Appendix.

\begin{figure}
\includegraphics[angle=0, width=240pt]{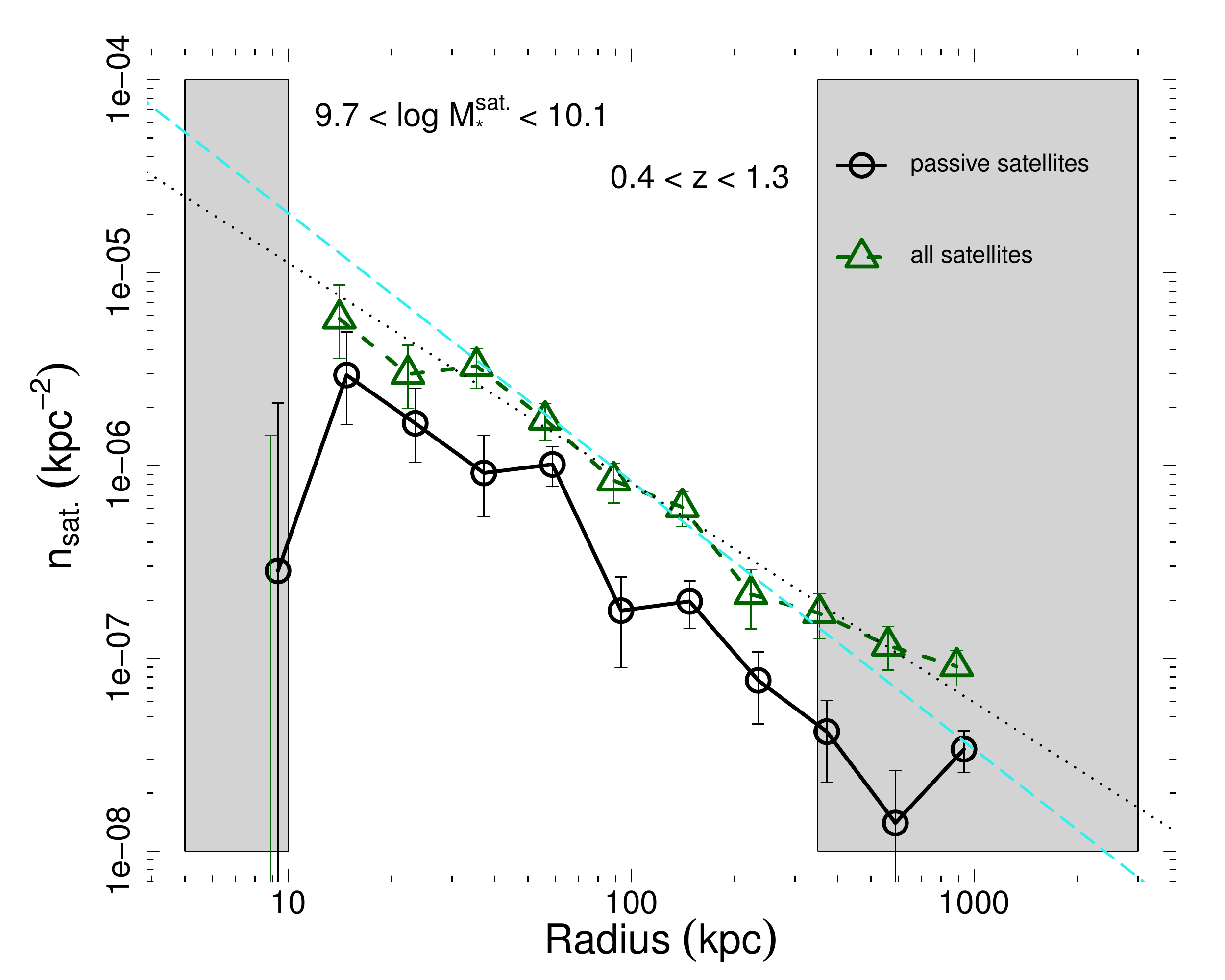}
\includegraphics[angle=0, width=240pt]{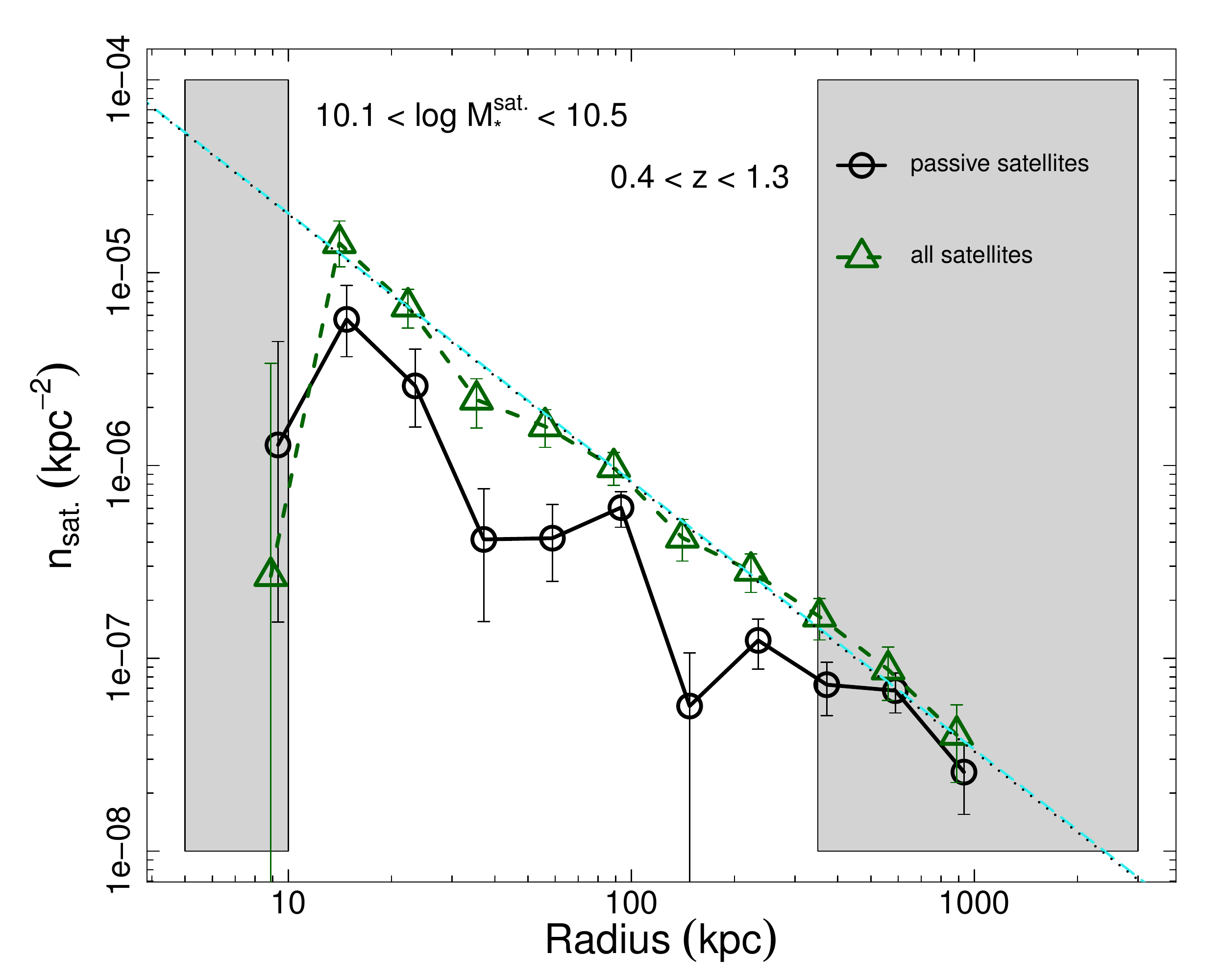}
\caption{Surface number density of satellite galaxies around $\M\sim \M^*$ central galaxies in the redshift interval $0.4<z<1.3$, as a function of projected radius. The projected surface density profile of mass-selected satellite galaxies (green dashed lines with triangles) and the passive subset (black solid lines with circles, offset slightly for clarity) are shown. The stellar mass limits of the satellite samples are $9.7<{\rm log}~(\M_*/\M_{\odot})<10.1$ (upper panel) and $10.1<{\rm log}~(\M_*/\M_{\odot})<10.5$ (lower panel). The dotted line shows the best-fit power law for the profile of mass-selected satellites. The parameters of all power-law fits are given in Table \ref{pltable}. The cyan dashed line is repeated in all the profile plots in this work to assist comparison of the different samples. It is the power-law fit for high mass galaxies at low-z. Uncertainties throughout this work are computed from Poisson statistics.}
\label{massfig_lowz}
\end{figure}

\begin{figure}
\includegraphics[angle=0, width=240pt]{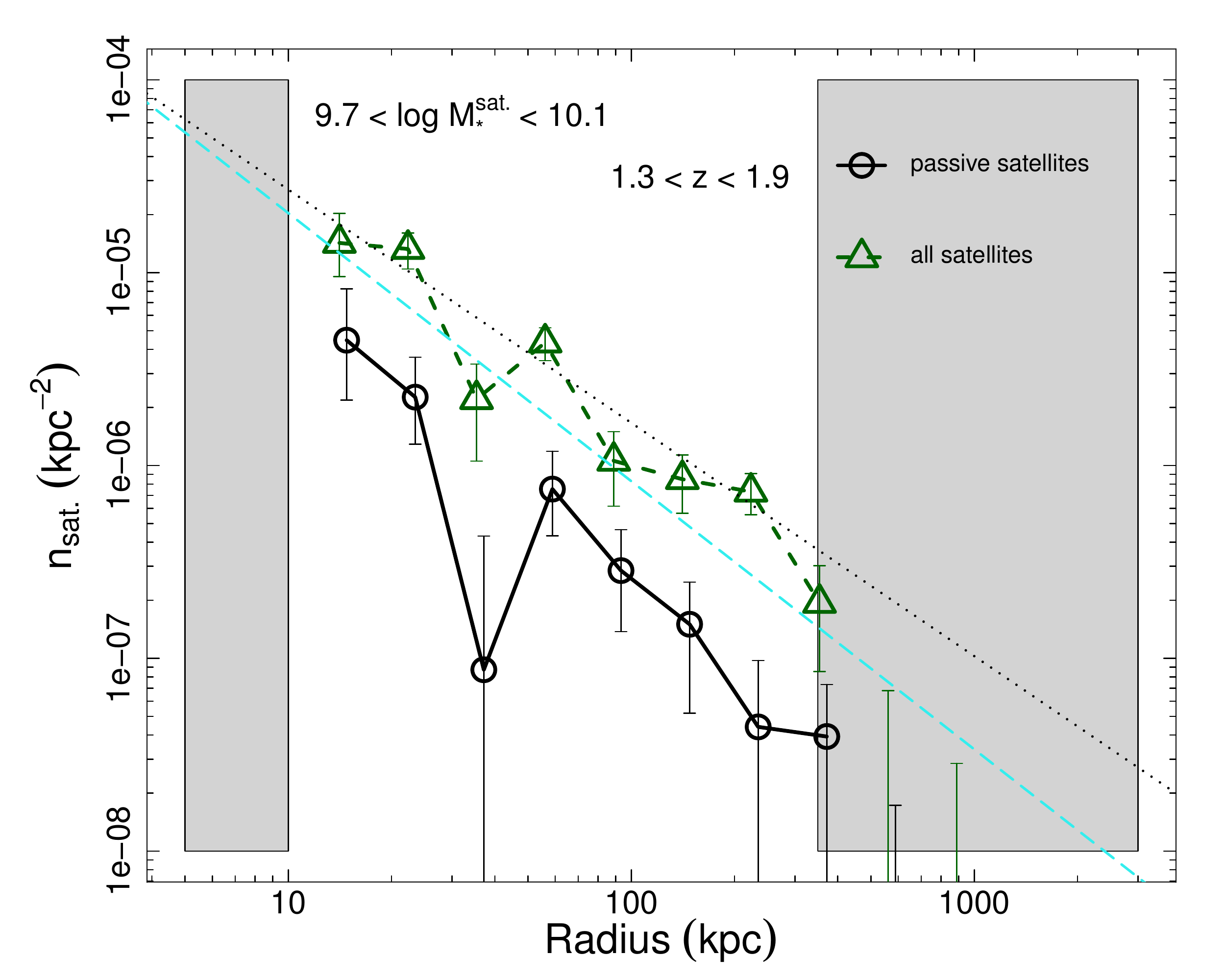}
\includegraphics[angle=0, width=240pt]{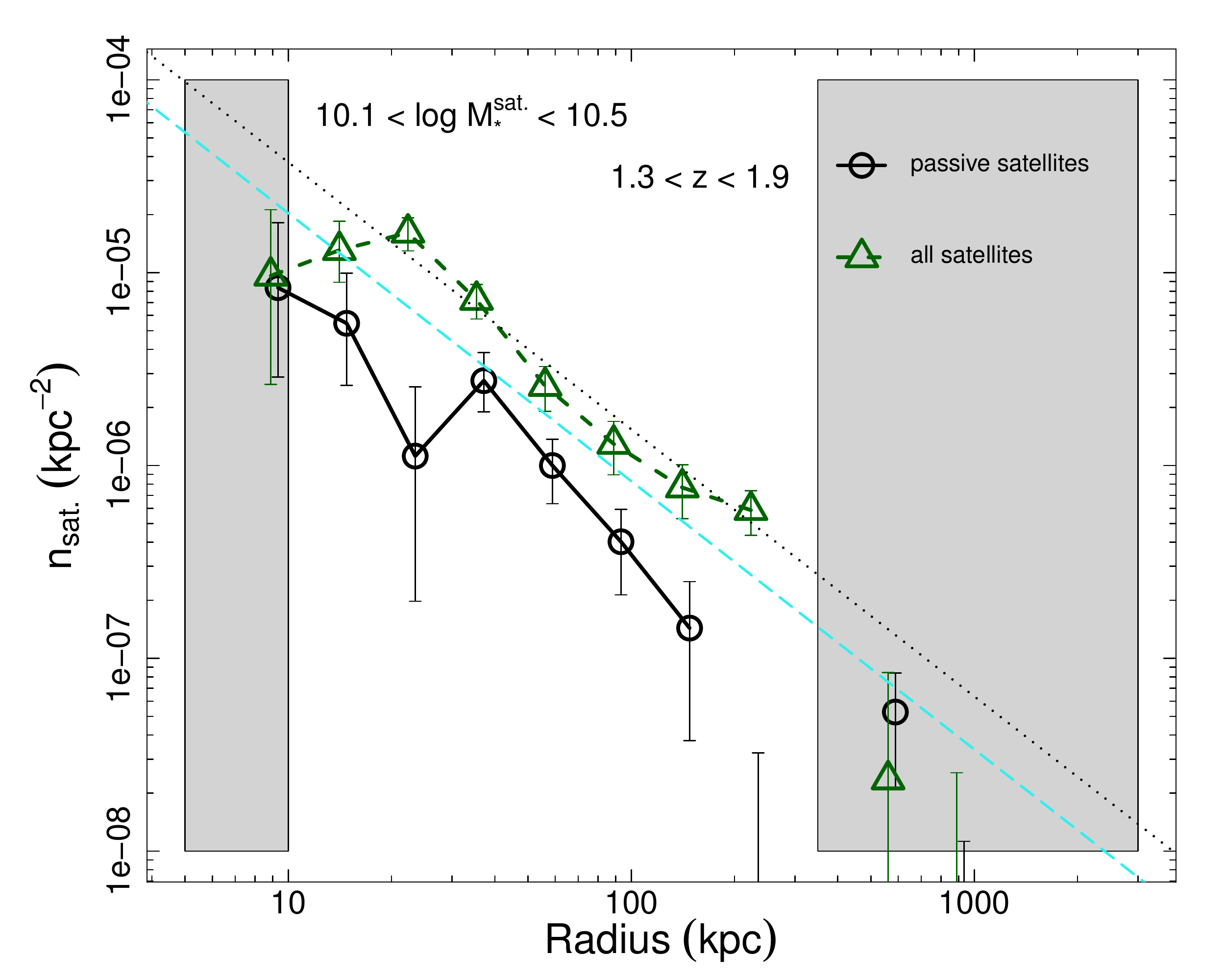}
\caption{Surface number density profiles for $\M\sim \M^*$ central galaxies in the redshift interval $1.3<z<1.9$. All lines and symbols have the same meanings as in Figure \ref{massfig_lowz}.}
\label{massfig_highz}
\end{figure}

\section{Satellite profiles around $\M^{\ast}$ galaxies}
\label{prof_mass}

\begin{table*}
\caption{Power-law fitting parameters for the various samples used in this work. $^a$ key for galaxy types: m - simple mass selection; p - passive subset of the mass-selected sample; sf - star-forming subset of the mass-selected sample. $^b$ $68\%$ confidence interval of the marginalised distribution. $^c$ Normalisation at $1~$kpc. $^{\ast}$ indicates a best fit at the limit of the parameter space explored. $^{\dagger}$ a simple power law is a poor representation of the radial surface-number-density dependence for passive satellites around star-forming centrals, but the values are included for completeness.}
\begin{tabular}[h]{|c|c|c|c|c|c|c|c|c|}
Redshift & Central type$^a$ & $N_{{\rm central}}$ & Satellite mass & Sat. type$^a$ & Slope & $\sigma_{{\rm Slope}}^b$ & log (Norm.)$^c$ & $\sigma_{{\rm log (Norm.)}}^b$\\
\hline
{\bf Central mass:} & \multicolumn{3}{l}{$10.5 < {\rm log}(\M_*/\M_{\odot}) < 11.0$} \\
\\
$0.4 < z < 1.3$ & m & 2697 & $9.7 < {\rm log}(\M_*/\M_{\odot}) < 10.1$ &  m & -1.15 & 0.11 & -3.80 & 0.67 \\
$0.4 < z < 1.3$ & p & 1491 & $9.7 < {\rm log}(\M_*/\M_{\odot}) < 10.1$ &  m & -1.22 & 0.12 & -3.57 & 0.48 \\
$0.4 < z < 1.3$ & sf & 1206 & $9.7 < {\rm log}(\M_*/\M_{\odot}) < 10.1$ &  m & -0.94 & 0.3 & -4.44 & 1.315 \\
$0.4 < z < 1.3$ & m & 2697 & $9.7 < {\rm log}(\M_*/\M_{\odot}) < 10.1$ &  p & -1.34 & 0.17 & -3.89 & 0.52 \\
$0.4 < z < 1.3$ & p & 1491 & $9.7 < {\rm log}(\M_*/\M_{\odot}) < 10.1$ &  p & -1.39 & 0.165 & -3.61 & 0.345 \\
$0.4 < z < 1.3$ & sf & 1206 & $9.7 < {\rm log}(\M_*/\M_{\odot}) < 10.1$ &  p & -1.01$^{\dagger}$ & 0.61 & -4.84 & 1.565 \\
$0.4 < z < 1.3$ & m & 2697 & $10.1 < {\rm log}(\M_*/\M_{\odot}) < 10.5$ &  m & -1.38 & 0.115 & -3.34 & 0.21 \\
$0.4 < z < 1.3$ & p & 1491 & $10.1 < {\rm log}(\M_*/\M_{\odot}) < 10.5$ &  m & -1.27 & 0.135 & -3.47 & 0.375 \\
$0.4 < z < 1.3$ & sf & 1206 & $10.1 < {\rm log}(\M_*/\M_{\odot}) < 10.5$ &  m & -1.72 & 0.295 & -2.89 & 0.665 \\
$0.4 < z < 1.3$ & m & 2697 & $10.1 < {\rm log}(\M_*/\M_{\odot}) < 10.5$ &  p & -1.38 & 0.235 & -3.83 & 0.495 \\
$0.4 < z < 1.3$ & p & 1491 & $10.1 < {\rm log}(\M_*/\M_{\odot}) < 10.5$ &  p & -1.58 & 0.27 & -3.39 & 0.48 \\
$0.4 < z < 1.3$ & sf & 1206 & $10.1 < {\rm log}(\M_*/\M_{\odot}) < 10.5$ &  p & -0.50$^{\ast \dagger}$ & 0.755 & -5.88 & 2.22 \\
$1.3 < z < 1.9$ & m & 1449 & $9.7 < {\rm log}(\M_*/\M_{\odot}) < 10.1$ &  m & -1.16 & 0.15 & -3.50 & 0.515 \\
$1.3 < z < 1.9$ & p & 603 & $9.7 < {\rm log}(\M_*/\M_{\odot}) < 10.1$ &  m & -1.05 & 0.2 & -3.66 & 0.77 \\
$1.3 < z < 1.9$ & sf & 846 & $9.7 < {\rm log}(\M_*/\M_{\odot}) < 10.1$ &  m & -1.30 & 0.235 & -3.29 & 0.425 \\
$1.3 < z < 1.9$ & m & 1449 & $9.7 < {\rm log}(\M_*/\M_{\odot}) < 10.1$ &  p & -1.40 & 0.515 & -3.92 & 0.86 \\
$1.3 < z < 1.9$ & p & 603 & $9.7 < {\rm log}(\M_*/\M_{\odot}) < 10.1$ &  p & -1.04 & 0.625 & -4.36 & 1.25 \\
$1.3 < z < 1.9$ & sf & 846 & $9.7 < {\rm log}(\M_*/\M_{\odot}) < 10.1$ &  p & -1.33$^{\dagger}$ & 0.77 & -4.23 & 1.505 \\
$1.3 < z < 1.9$ & m & 1449 & $10.1 < {\rm log}(\M_*/\M_{\odot}) < 10.5$ &  m & -1.36 & 0.12 & -3.11 & 0.215 \\
$1.3 < z < 1.9$ & p & 603 & $10.1 < {\rm log}(\M_*/\M_{\odot}) < 10.5$ &  m & -1.37 & 0.245 & -3.14 & 0.44 \\
$1.3 < z < 1.9$ & sf & 846 & $10.1 < {\rm log}(\M_*/\M_{\odot}) < 10.5$ &  m & -1.35 & 0.135 & -3.09 & 0.245 \\
$1.3 < z < 1.9$ & m & 1449 & $10.1 < {\rm log}(\M_*/\M_{\odot}) < 10.5$ &  p & -1.65 & 0.265 & -3.25 & 0.455 \\
$1.3 < z < 1.9$ & p & 603 & $10.1 < {\rm log}(\M_*/\M_{\odot}) < 10.5$ &  p & -1.36 & 0.29 & -3.52 & 0.525 \\
$1.3 < z < 1.9$ & sf & 846 & $10.1 < {\rm log}(\M_*/\M_{\odot}) < 10.5$ &  p & -1.86$^{\dagger}$ & 0.58 & -3.02 & 1.16 \\
\hline
{\bf Central mass:} & \multicolumn{3}{l}{${\rm log}(\M_*/\M_{\odot}) > 11.0$} \\
\\
$0.4 < z < 1.3$ & m & 1050 & $9.7 < {\rm log}(\M_*/\M_{\odot}) < 10.1$ &  m & -1.13 & 0.075 & -3.28 & 0.415 \\
$0.4 < z < 1.3$ & p & 791 & $9.7 < {\rm log}(\M_*/\M_{\odot}) < 10.1$ &  m & -1.18 & 0.08 & -3.17 & 0.31 \\
$0.4 < z < 1.3$ & sf & 259 & $9.7 < {\rm log}(\M_*/\M_{\odot}) < 10.1$ &  m & -1.16 & 0.195 & -3.31 & 0.48 \\
$0.4 < z < 1.3$ & m & 1050 & $9.7 < {\rm log}(\M_*/\M_{\odot}) < 10.1$ &  p & -1.26 & 0.11 & -3.41 & 0.35 \\
$0.4 < z < 1.3$ & p & 791 & $9.7 < {\rm log}(\M_*/\M_{\odot}) < 10.1$ &  p & -1.31 & 0.12 & -3.26 & 0.255 \\
$0.4 < z < 1.3$ & sf & 259 & $9.7 < {\rm log}(\M_*/\M_{\odot}) < 10.1$ &  p & -1.09$^{\dagger}$ & 0.535 & -4.08 & 1.175 \\
$0.4 < z < 1.3$ & m & 1050 & $10.1 < {\rm log}(\M_*/\M_{\odot}) < 10.5$ &  m & -1.23 & 0.095 & -3.20 & 0.27 \\
$0.4 < z < 1.3$ & p & 791 & $10.1 < {\rm log}(\M_*/\M_{\odot}) < 10.5$ &  m & -1.24 & 0.105 & -3.13 & 0.23 \\
$0.4 < z < 1.3$ & sf & 259 & $10.1 < {\rm log}(\M_*/\M_{\odot}) < 10.5$ &  m & -0.92 & 0.34 & -4.02 & 1.115 \\
$0.4 < z < 1.3$ & m & 1050 & $10.1 < {\rm log}(\M_*/\M_{\odot}) < 10.5$ &  p & -1.26 & 0.105 & -3.33 & 0.31 \\
$0.4 < z < 1.3$ & p & 791 & $10.1 < {\rm log}(\M_*/\M_{\odot}) < 10.5$ &  p & -1.38 & 0.11 & -3.05 & 0.21 \\
$0.4 < z < 1.3$ & sf & 259 & $10.1 < {\rm log}(\M_*/\M_{\odot}) < 10.5$ &  p & -0.50$^{\ast \dagger}$ & 0.19 & -5.15 & 1.93 \\
$1.3 < z < 1.9$ & m & 663 & $9.7 < {\rm log}(\M_*/\M_{\odot}) < 10.1$ &  m & -1.01 & 0.11 & -3.48 & 0.69 \\
$1.3 < z < 1.9$ & p & 384 & $9.7 < {\rm log}(\M_*/\M_{\odot}) < 10.1$ &  m & -1.03 & 0.145 & -3.37 & 0.63 \\
$1.3 < z < 1.9$ & sf & 279 & $9.7 < {\rm log}(\M_*/\M_{\odot}) < 10.1$ &  m & -1.02 & 0.21 & -3.56 & 0.765 \\
$1.3 < z < 1.9$ & m & 663 & $9.7 < {\rm log}(\M_*/\M_{\odot}) < 10.1$ &  p & -1.32 & 0.185 & -3.41 & 0.365 \\
$1.3 < z < 1.9$ & p & 384 & $9.7 < {\rm log}(\M_*/\M_{\odot}) < 10.1$ &  p & -1.55 & 0.175 & -2.82 & 0.43 \\
$1.3 < z < 1.9$ & sf & 279 & $9.7 < {\rm log}(\M_*/\M_{\odot}) < 10.1$ &  p & -0.50$^{\ast \dagger}$ & 0.915 & -5.60 & 2.24 \\
$1.3 < z < 1.9$ & m & 663 & $10.1 < {\rm log}(\M_*/\M_{\odot}) < 10.5$ &  m & -1.34 & 0.145 & -3.06 & 0.26 \\
$1.3 < z < 1.9$ & p & 384 & $10.1 < {\rm log}(\M_*/\M_{\odot}) < 10.5$ &  m & -1.28 & 0.175 & -3.05 & 0.335 \\
$1.3 < z < 1.9$ & sf & 279 & $10.1 < {\rm log}(\M_*/\M_{\odot}) < 10.5$ &  m & -1.76 & 0.405 & -2.49 & 0.94 \\
$1.3 < z < 1.9$ & m & 663 & $10.1 < {\rm log}(\M_*/\M_{\odot}) < 10.5$ &  p & -1.54 & 0.205 & -3.12 & 0.35 \\
$1.3 < z < 1.9$ & p & 384 & $10.1 < {\rm log}(\M_*/\M_{\odot}) < 10.5$ &  p & -1.92 & 0.275 & -2.10 & 1.27 \\
$1.3 < z < 1.9$ & sf & 279 & $10.1 < {\rm log}(\M_*/\M_{\odot}) < 10.5$ &  p & -1.72$^{\dagger}$ & 0.53 & -2.91 & 0.935 \\
\label{pltable}
\end{tabular}
\end{table*}

\begin{figure}
\includegraphics[angle=0, width=240pt]{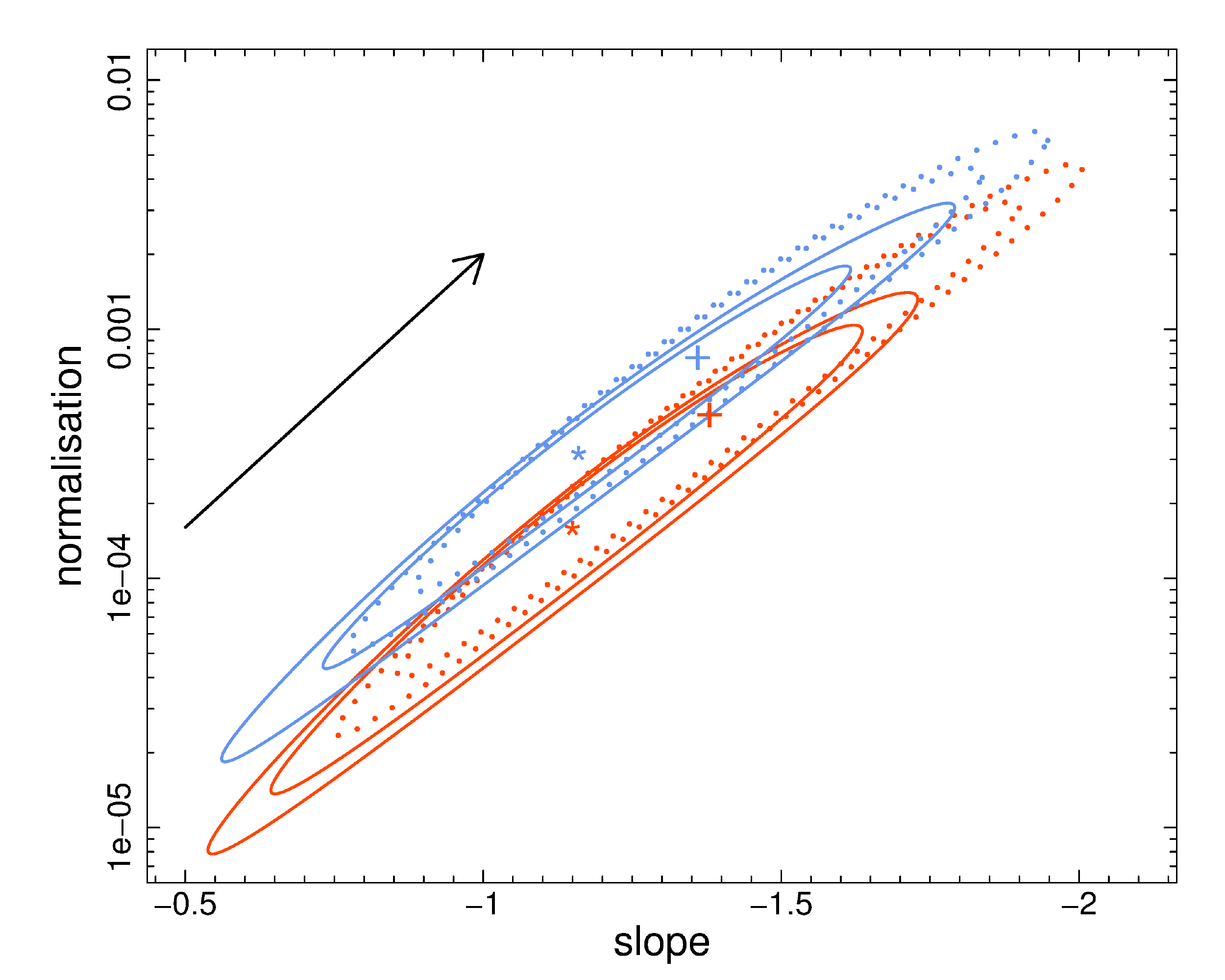}
\caption{Confidence contours ($1$ and $2\sigma$) for the power-law fits to satellite profiles of mass-selected centrals. Solid contours correspond to our lower-mass satellite galaxies, while dotted lines are for higher mass. The lower pair (in red) are for the lower redshift interval and the upper pair for the higher redshift range ($1.3<z<1.9$). The black arrow shows the direction of the degeneracy expected between the fitting parameters for a constant number of satellites within $\R<350~$kpc.}
\label{confplot}
\end{figure}

We begin our analysis with a simple stellar-mass selected central galaxy sample in two redshift intervals: $0.4<z<1.3$ and $1.3<z<1.9$. The surveyed volumes probed by these two intervals are roughly equal, but there are twice as many central galaxies in our stellar mass range in the lower redshift interval. The boundary between these two samples corresponds to the redshift at which the photometric redshift uncertainty rises above the global value of $\Delta z /(1.+z) = 0.031$. For the satellite population we consider two ranges in stellar mass, $9.7 < {\rm log} (\M_*/\M_{\odot}) < 10.1$ and $10.1 < {\rm log} (\M_*/\M_{\odot})< 10.5$, where the lower limit of the low-mass satellite sample is determined by the stellar-mass completeness limit at $z=1.9$ (see section \ref{masses}). We perform separate measurements for stellar-mass-selected satellite samples and for the passive subset of these satellites (primarily based on the addition of a rest-frame UVJ criterion, see section \ref{sample}). 

In Figure \ref{massfig_lowz} we show the radial galaxy surface-number-density profiles of satellites around $\M \sim \M^\ast$ central galaxies within the redshift interval $0.4<z<1.3$ for lower-mass (upper panel) and higher-mass (lower panel) satellites. Green triangles with joining dashed lines represent the profile of the simple mass-selected satellites, while black circles and solid lines show the results for passive satellites. Throughout this work uncertainties on the measurements are taken from Poisson statistics, assuming that they are dominated by the counts around the real central galaxies. The grey regions indicate radii where the profiles are likely to be poorly estimated (see Section \ref{rad_measurement}). Finally, we show, by the dotted lines, the best fit two-parameter power laws, 
\begin{equation}
n_{{\rm sat.}}=\alpha~\R^{\beta}, 
\label{plfit}
\end{equation}
to the stellar-mass selected satellites of each mass interval between $10<\R({\rm kpc})<350$. 

\subsection{Power-law fitting results}
\label{fitting}

Single component power laws provide a good description of the data and the fitting parameters for each sample are given in Table \ref{pltable}, together with marginalised uncertainties. It is usually assumed in halo occupation models \citep[e.g.][]{Cooray02} that satellite galaxies are distributed according to an NFW profile \citep{Navarro95}, reflecting the underlying dark matter distribution. The average number of satellites that a central galaxy in our sample has within the adopted stellar mass intervals is $\lesssim 1$. It is interesting to ask whether the averaged distribution does indeed follow an NFW profile in this case. Our data show no clear evidence for a departure from a power-law profile, as might have been expected for a projected NFW distribution \citep{Bartelmann96}. However, it is extremely difficult to differentiate a projected NFW profile from a simple power law in most current data sets \citep{Chen06}, and so we measure simple power laws for ease of comparison. 

One perhaps surprising result of our data is the higher normalisation of the power-law fit (taken at $\R=1~$kpc) for higher mass satellites. Integrating the two fits overs our reliable range of radii ($10 < {\rm R(kpc)} < 350$), we find that the average number of low and high-mass satellites are almost equal, well within the measurement uncertainties. Such behaviour is somewhat counter-intuitive because less massive galaxies are more numerous in general. However, we should not take this similarity in number too literally. Firstly, as we mentioned in Section \ref{rad_measurement}, we expect to miss around $10\%$ of the star-forming satellites. These missing objects are likely to be predominantly low-mass satellites due to the smaller passive fraction at low masses, and the lower signal-to-noise photometry of low-mass galaxies. Secondly, the two intervals in satellite mass that we consider are only twice the expected uncertainty in the log of the mass to light ratio from SED fitting. We would therefore expect a significant Eddington bias in the relative numbers of higher and lower-mass satellites. Finally, it is not clear that we should expect the relative numbers of satellites to exactly follow the broader galaxy stellar mass function.

Examining the slopes of the power-law fits, we find that low-mass satellite profiles are flatter than those of high mass satellites, albeit at marginal significance. For the two fits to equation \ref{plfit} shown in Figure \ref{massfig_lowz} the values for $\beta$ are, $\beta=-1.15\pm0.11$ and $-1.38\pm0.12$ for low-mass and high-mass satellites respectively. The flatter surface-number-density profiles of low-mass satellites perhaps indicate that low mass satellites are being disrupted when they approach the central galaxies. In fact, the difference in slope is almost entirely driven by a slight deficit of low mass satellites at small separations, which would be expected if satellite disruption were important. An alternative explanation for the observed difference in profile slopes could be the effects of dynamical friction. The deceleration due to dynamical friction is satellite-mass dependent \citep{Chandrasekhar43,Boylan08}, and over long timescales should result in mass segregation, where more massive galaxies have smaller radial separations. This behaviour is observed in galaxy clusters \citep{Old13,Wu13}, but it is speculative to apply it to the systems we are studying. 

In Figure \ref{massfig_highz} we show the results for the higher redshift ($1.3<z<1.9$) mass-selected central galaxies. All lines and symbols have the same meanings as in Figure \ref{massfig_lowz}. These results are similar to those at lower redshift, but there are subtle differences. The satellite distributions around the higher-redshift centrals have a slightly higher average normalisation: the average number of satellites around low redshift centrals is $\sim 0.31\pm 0.03$, i.e. around one in three centrals hosts a satellite of the masses studied in this work. Meanwhile, at high redshift the average number of satellites is $\sim 0.44\pm 0.04$, i.e. $\sim50\%$ higher. We would expect the average host halo mass to correlate with the average number of satellites, and this could therefore be taken as evidence that galaxies of mass $\M\sim \M^{\ast}$ are found in slightly more massive halos at higher redshift. This behaviour would be consistent with other studies \citep[e.g.][]{Moster10, Behroozi13, Hartley13}. However, because the average number of satellites per central galaxy is less than unity, and our satellites have masses within a factor of $10$ of the central, it is not obvious that we can infer a difference in halo masses from these numbers alone.

To better understand our measurements we inspect the confidence contours in our two fitting parameters from equation \ref{plfit}. These are shown in Figure \ref{confplot}. Solid contours show the $1$ and $2\sigma$ intervals for low-mass satellites, with the corresponding maximum likelihoods given by the asterisk symbols. Dotted contours with plus marks are for high mass satellites. The lower pair of contours, shown in red, are the results for low-redshift samples, while high-redshifts are shown in pale blue. There is a clear degeneracy in the fitting parameters which shows that the difference in slope and normalisation between high and low-mass satellites is actually not significant at either redshift. Far greater statistics would be required to differentiate the slopes of the two satellite profiles. We show, by the black arrow, the interdependence of slope and normalisation for a fixed integrated number of satellites per central for $\R<350~$kpc. Whilst it is unsurprising that this arrow follows the degeneracy in the confidence contours, it demonstrates that the integrated number of galaxies within some radius (the vector perpendicular to the arrow) is well constrained by the data.

\subsection{Profiles of passive satellites}
\label{passprof}

Turning to the passive satellite profiles, we find that the power-law slopes are systematically steeper than for purely mass-selected satellites of the same mass and redshift. We caution that in the majority of cases the difference is not significant, but it is consistent across all mass-selected central samples. This behaviour is qualitatively similar to the well-known morphology-density relation \citep{Dressler80} in much more massive systems, and is possible evidence for an environmental impact on the star-formation properties of the satellites.

A further subtle trend in our measurements is that the difference in normalisation between low-mass passive and mass-selected satellites is greater at higher redshift. i.e. the passive fraction of low-mass satellites is lower around the high-redshift galaxies. This is despite the fact that higher redshift centrals host more satellites and may therefore be hosted by higher-mass halos. This behaviour is reminiscent of the well known redshift evolution of galaxy populations in much more massive systems \citep{Butcher78}, and is also known to be present in group environments \citep[e.g.][]{Lin14}.

There could be a number of reasons for this apparent trend. Firstly, it may simply be that the passive fraction in the field at these masses is higher at low redshift, and that the satellite population reflects this difference. We may also be seeing the effects of the extended structure known to exist in the UDS field at $z\sim0.62$ \citep{vanBreukelen06}. An extended super-cluster-like structure could be responsible for quenching galaxies well beyond its virial radius \citep[e.g.][]{Wetzel14} and it is possible that our central galaxy selection is not perfect in such a dense region. Even if neither of these explanations are responsible, the region near a super-cluster is certainly a peculiar region of the Universe and we may be seeing the influence of formation bias. Alternatively, this difference might arise purely due to the time interval spanned. Low mass halos may be capable of quenching satellites (albeit slowly), and we are simply seeing the result of an extended evolution timescale. 

Finally, there is the intriguing possibility that the phenomenon of galactic conformity which has been observed at low redshift \citep{Weinmann06,Ross09,Kauffmann13,Phillips14}, and introduced in section \ref{intro}, may provide a very natural explanation. The fraction of central galaxies with mass, $\M\sim \M^{\ast}$, that are passive is weakly dependent on redshift \citep{Hartley13, Knobel13}. If galactic conformity is present at high redshift also, then the difference in the fraction of passive satellites may reflect the evolving {\em central} passive fraction. To investigate this possibility and find the origin of galactic conformity we now consider the passive and star-forming central galaxy populations separately.

\subsection{Satellite profiles around passive and star-forming galaxies}
\label{prof_passsf}

\begin{figure*}
\noindent\begin{minipage}{180mm}
\begin{tabular}[htbp]{@{}ll@{}}
\includegraphics[angle=0, width=240pt]{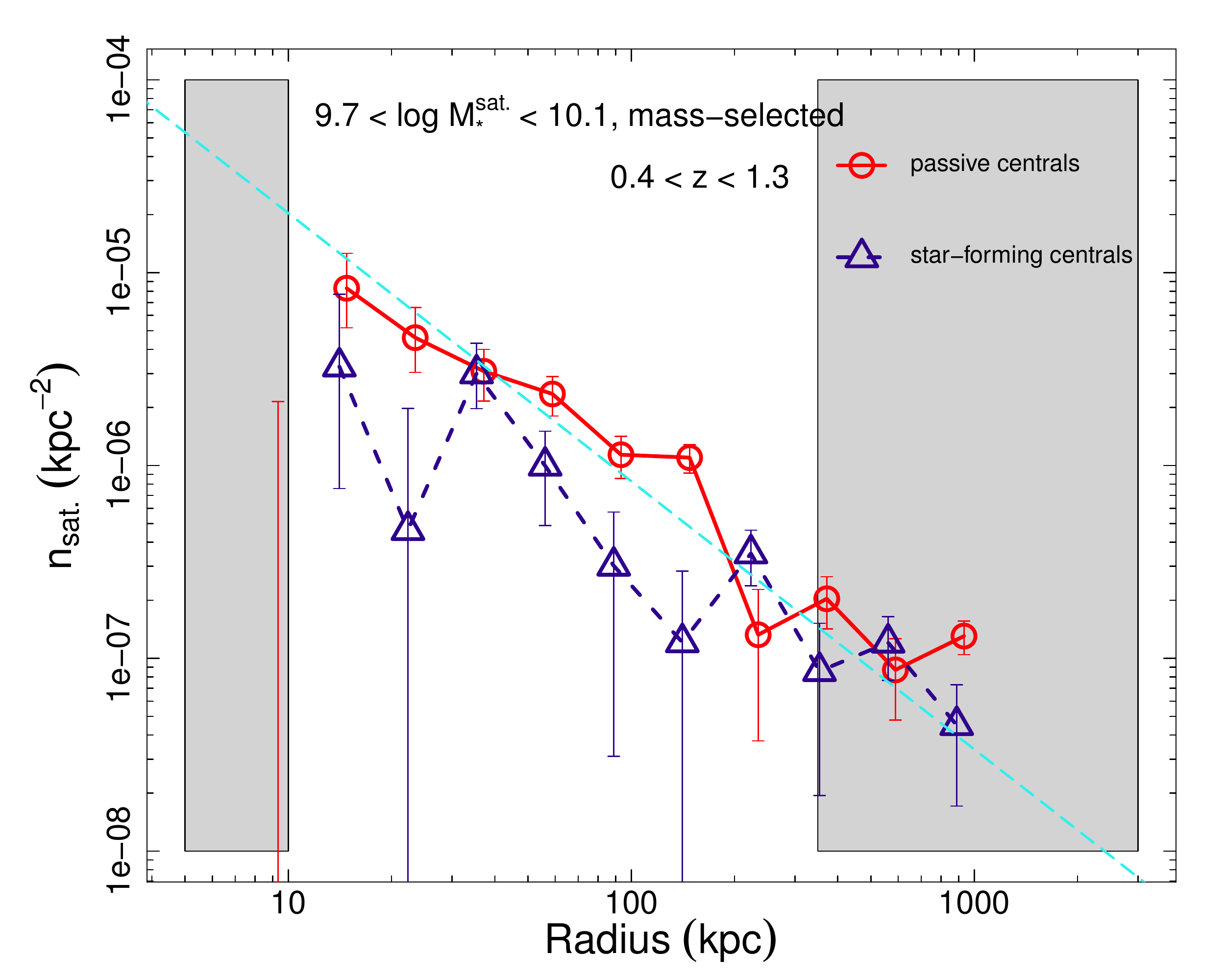} &
\includegraphics[angle=0, width=240pt]{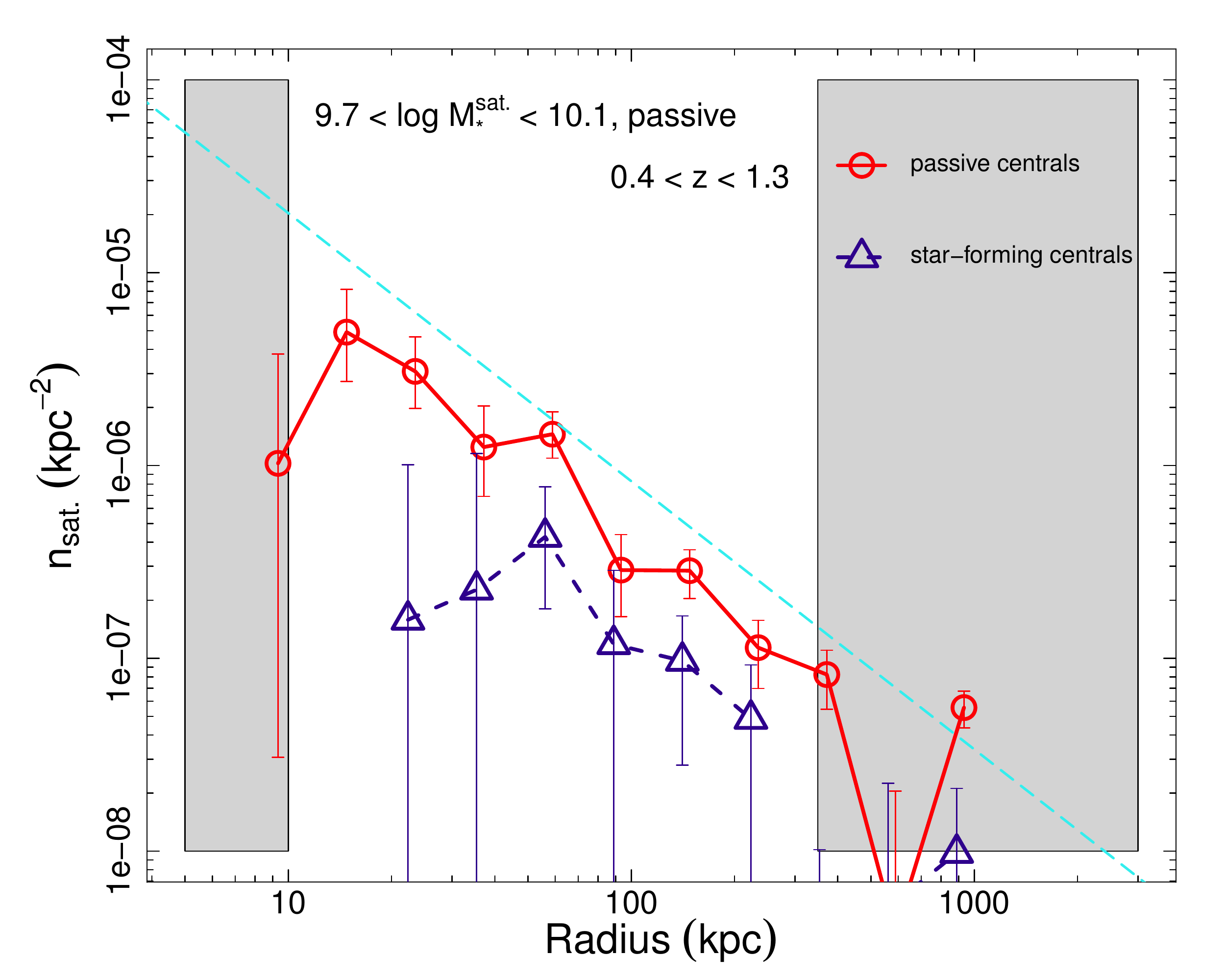}\\
\includegraphics[angle=0, width=240pt]{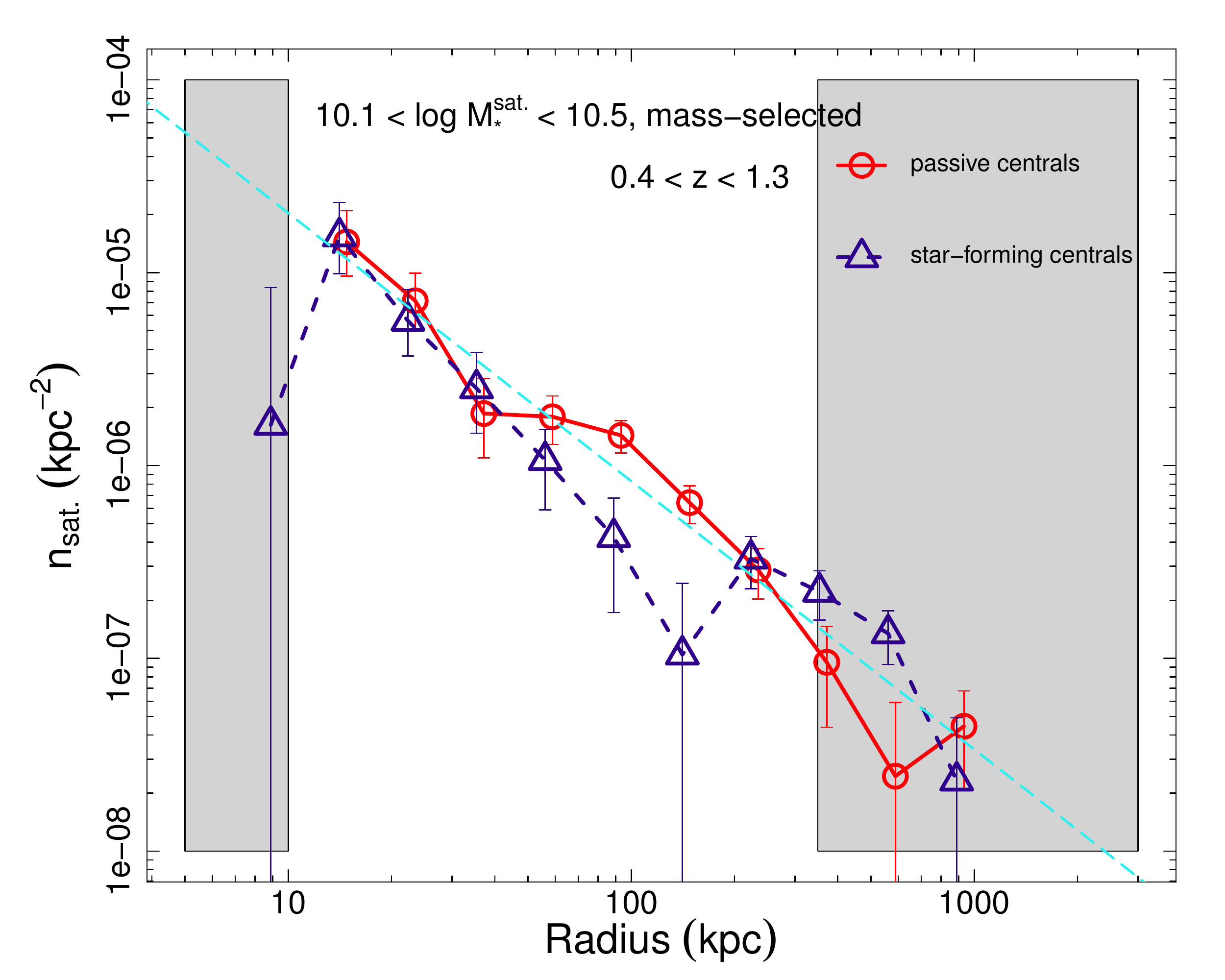} &
\includegraphics[angle=0, width=240pt]{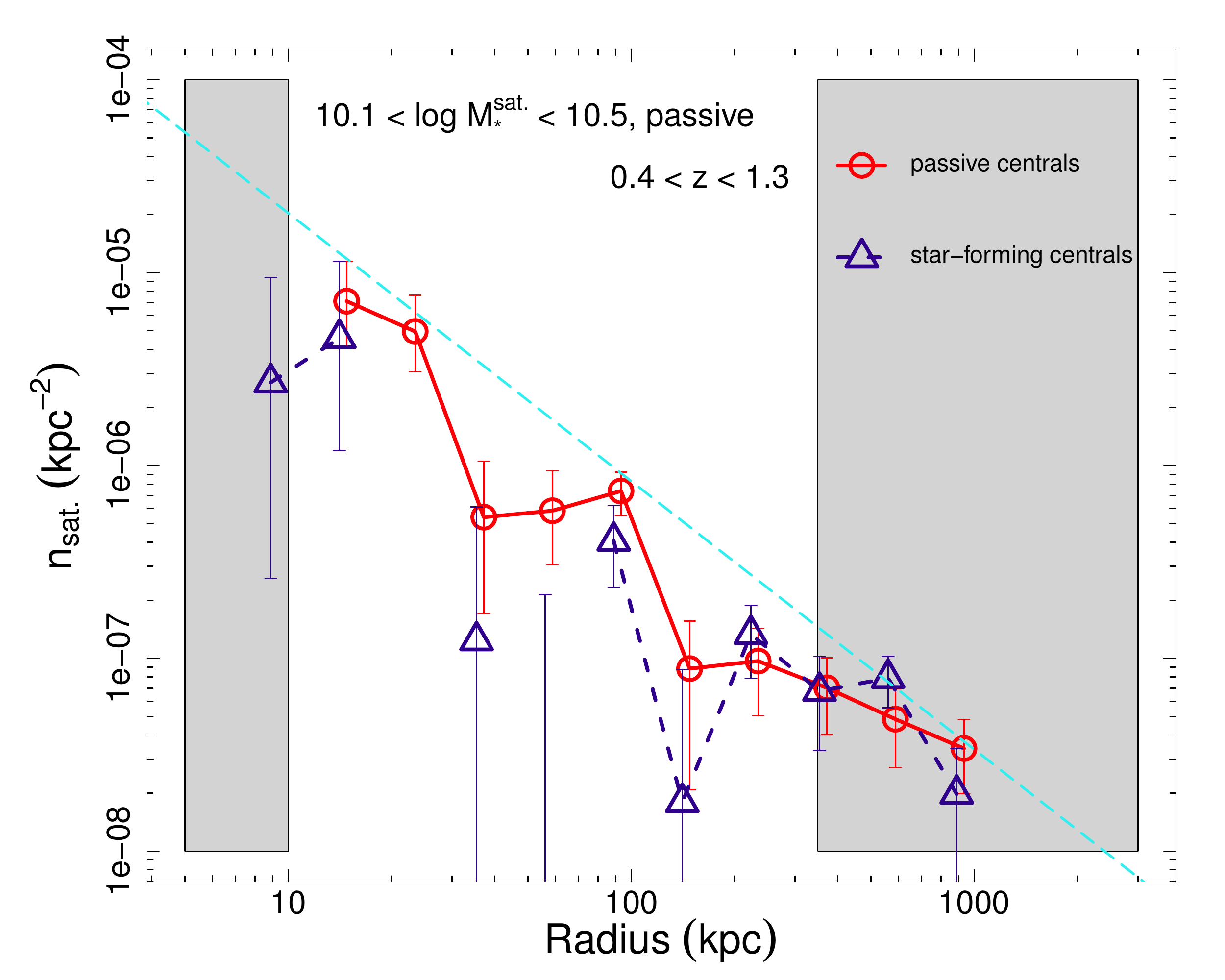}\\
\end{tabular}
\caption{Satellite galaxy projected surface density profiles around $\M\sim \M^*$ star-forming and passive central galaxies. Profiles for passive (red solid lines with circles) and star-forming (blue dashed lines with triangles) central samples at $0.4<z<1.3$ were computed for low-mass ($9.7<{\rm log}(\M_*/\M_{\odot})<10.1$, upper panels) and high-mass ($10.1<{\rm log}(\M_*/\M_{\odot})<10.5$, lower panels) satellites. Left hand panels show the profiles of mass-selected satellites, while right-hand panels show the profiles of passive satellites only.}
\label{lowz}
\end{minipage}
\end{figure*}

\begin{figure*}
\noindent\begin{minipage}{180mm}
\begin{tabular}[htbp]{@{}ll@{}}
\includegraphics[angle=0, width=240pt]{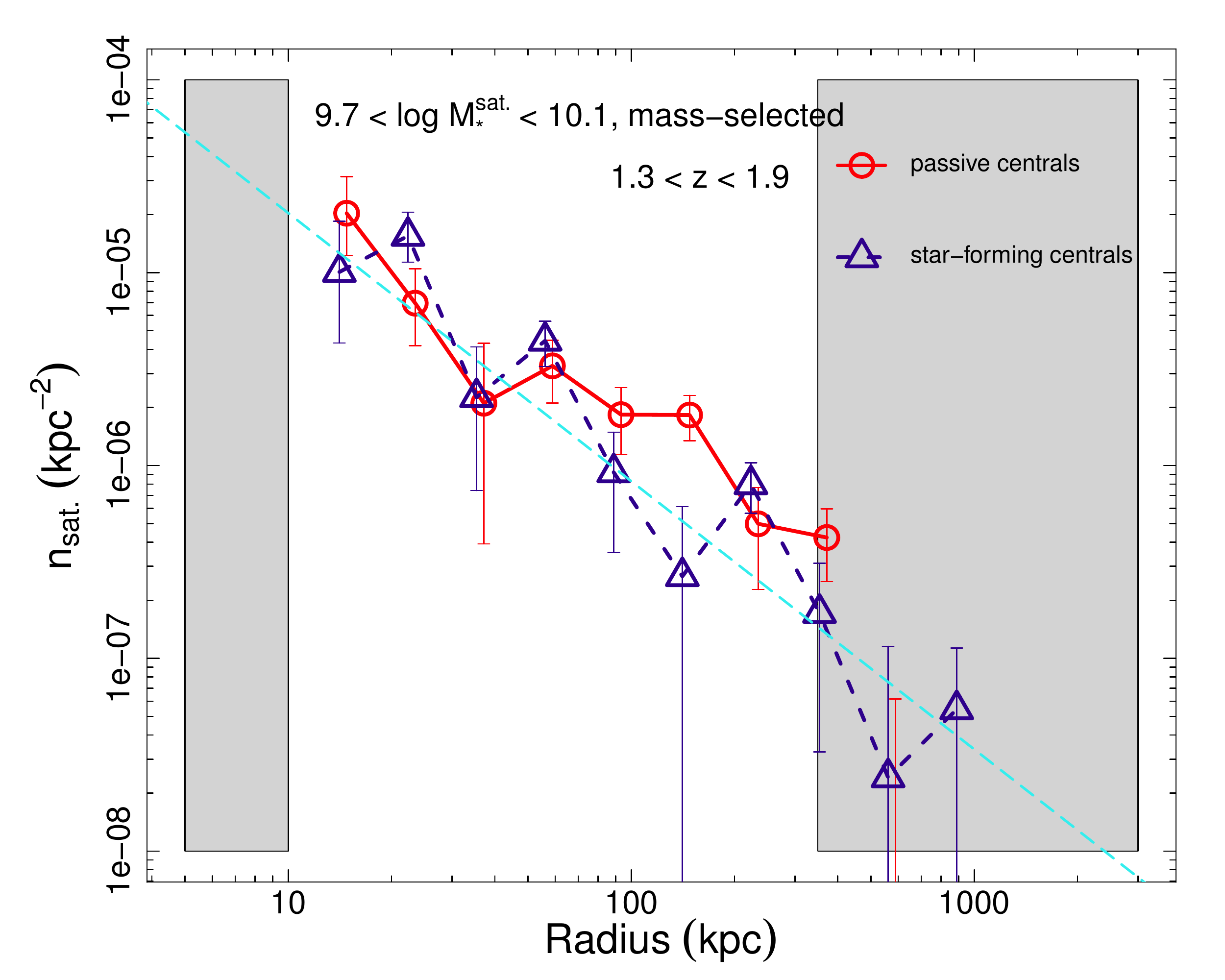} &
\includegraphics[angle=0, width=240pt]{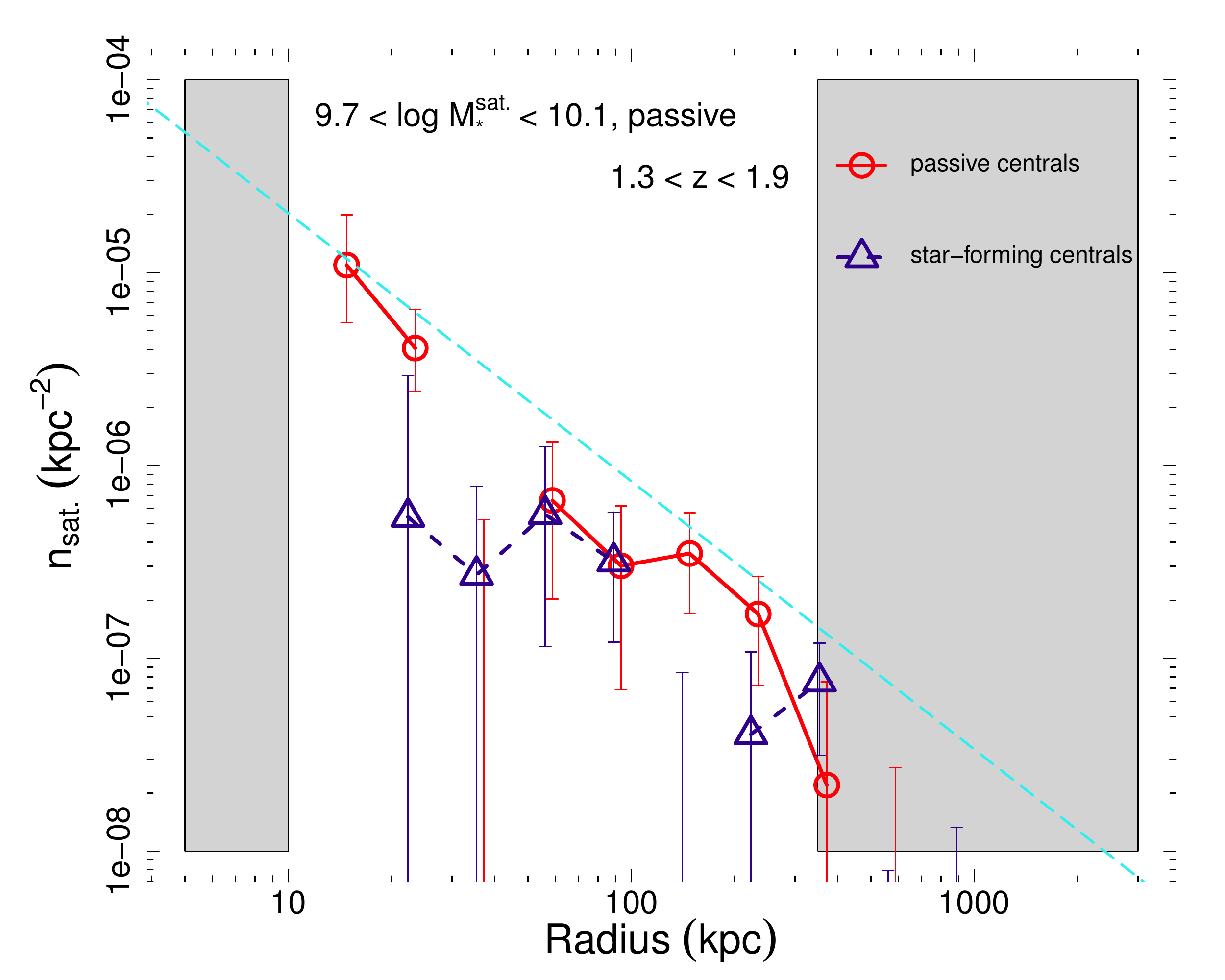}\\
\includegraphics[angle=0, width=240pt]{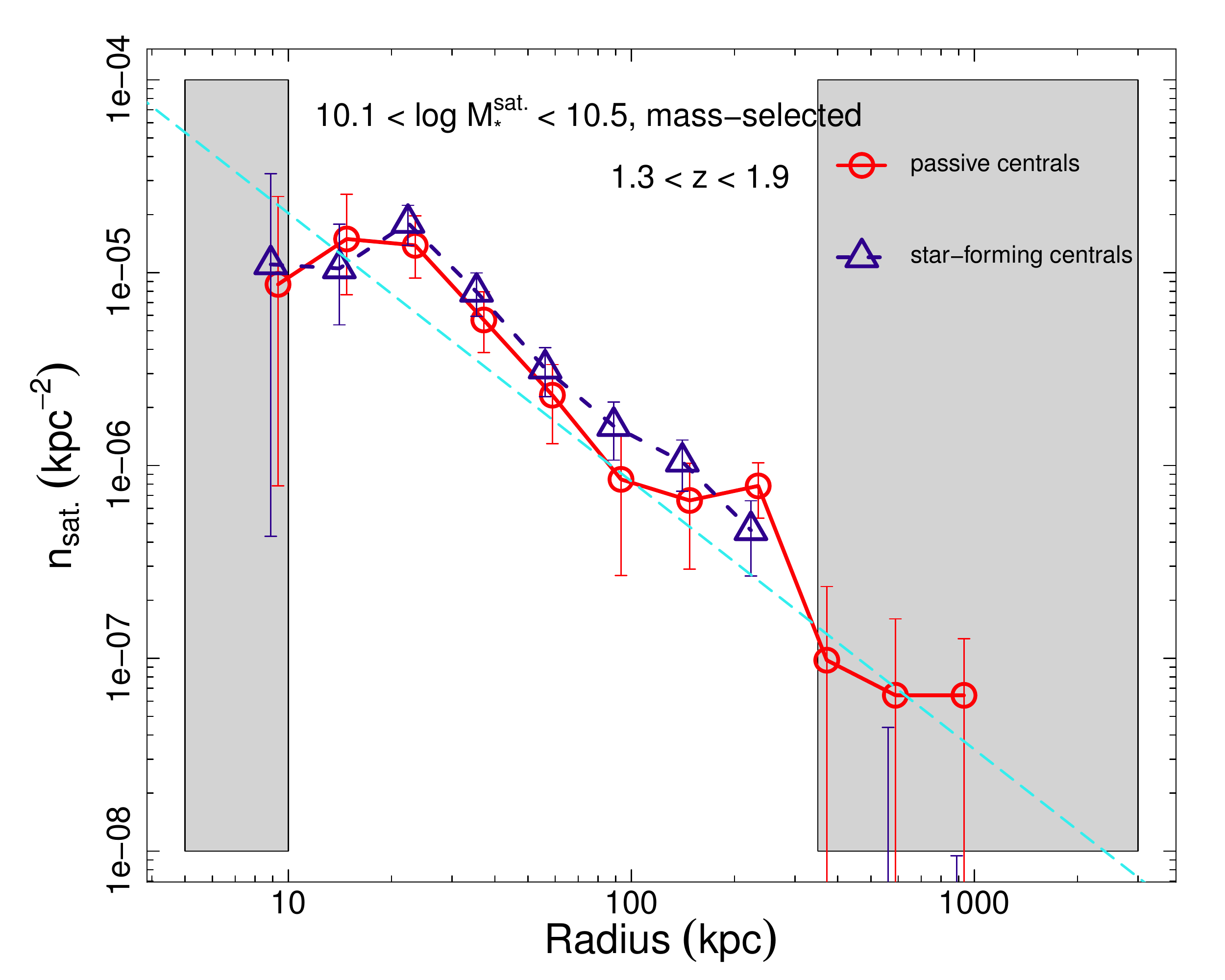} &
\includegraphics[angle=0, width=240pt]{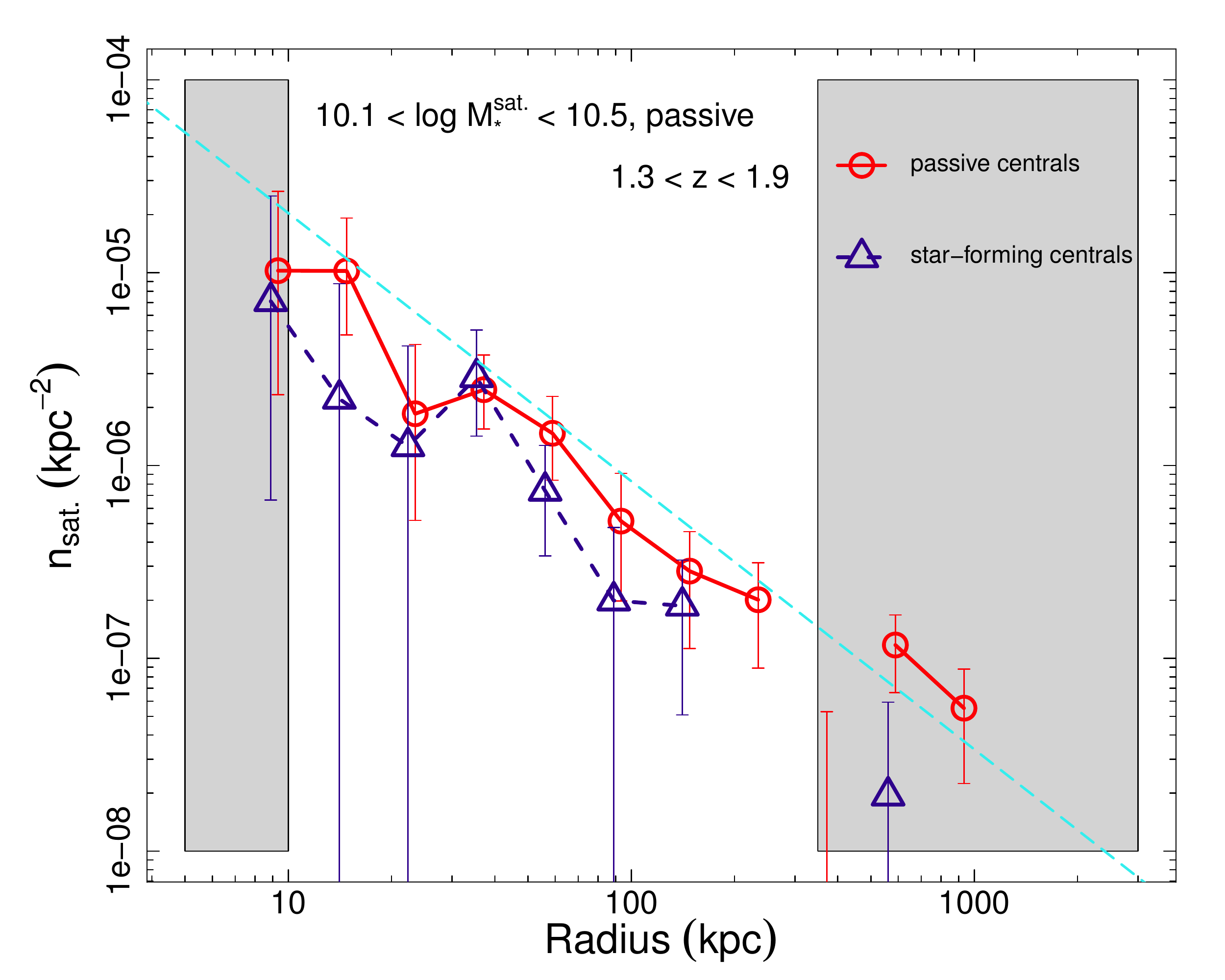}\\
\end{tabular}
\caption{Same as Figure \ref{lowz}, but for redshifts $1.3 < z < 1.9$.}
\label{highz}
\end{minipage}
\end{figure*}

Our central galaxy samples were split into passive and star-forming galaxies and stellar-mass matched with the parent sample as described in section \ref{sample}. The surface-number-density profiles of mass-selected and passive satellites, in the same two stellar mass intervals we used in the previous section, are shown in Figures \ref{lowz} and \ref{highz} for redshifts $0.4<z<1.3$ and $1.3<z<1.9$ respectively. In each Figure the left hand panels show stellar-mass selected satellites, while the right hand panels are for the passive satellite subset. Upper panels are the results for low-mass satellites, with higher-mass satellites shown in the lower panels. In each panel the satellite profile around passive central galaxies is plotted with red circles joined by solid lines. Star-forming centrals are displayed by blue triangles with dashed lines. As before, measurement uncertainties are computed from Poisson statistics.

Passive central galaxies in the lower redshift range appear to have a higher number of satellites on average, particularly at radii, $\R>50~{\rm kpc}$ and for lower-mass satellites. However, in our high-redshift interval the profiles of mass-selected satellites of passive and star-forming centrals are indistinguishable. As remarked previously, it is not clear that the number of satellites alone can be used as evidence for a difference in halo mass for our particular samples. Indeed, from the results of galaxy clustering we would expect to find a difference in the halo masses of star-forming and passive galaxies in both redshift intervals \citep[e.g.][]{Ross09,Williams09,Coupon12,Hartley13}. We return to the question of a potential halo mass difference in section \ref{disc}.

Looking at the profiles of passive satellites we find behaviour that is consistent across our two redshift intervals. In each of the redshift intervals studied, and for both stellar mass ranges of satellites, there are a greater number of passive satellites around passive central galaxies than star-forming centrals. This trend is most obvious in the measurements of low-mass satellites at low redshift, but can also be seen in the other samples. Moreover, there are several radial bins with less than zero passive satellite galaxies (after background correction) around star-forming centrals. These simple observations suggest that galactic conformity may well exist out to high redshift ($z\sim2$), but in order to confirm this suspicion we need to estimate the passive fraction of satellites around the two central samples.

\subsection{Literature comparison}
\label{litcomp}

The subject of satellite number density profiles has received a great deal of attention in recent years. This has in part been motivated by the missing satellites problem \citep[e.g.][]{Moore99}, but also due to its importance in understanding the role of environment in ending star-formation in galaxies and the frequency of galaxy mergers. In much of the literature it has been found that the radial profiles of satellite distributions are adequately fit by simple power laws. 

Determinations of power-law slopes vary between those that are similar to the inner slope of a projected NFW profile (at $\beta=-1$ \citealt{Navarro95,Prescott11,Guo12,Jiang12,Nierenberg12}), to much steeper slopes approaching $-2$ \citep{Chen06,Tollerud11,Watson12}. Claims have been made for dependencies on the luminosity (or stellar mass) and colour of the central galaxy \citep{Lares11,Prescott11,Guo12,Wang14} and also of the satellites \citep{Ann08,Guo12,Wang14}. Furthermore, variation in methodology, and in particular the interloper or background correction, can cause significant changes in the slope \citep{Chen06}. Finally, as the majority of studies rely on co-adding the satellites of samples of central galaxies (as does our work presented here), the question of whether one should scale the radial dimension by some physically-motivated quantity (such as the virial radius) could also become important \citep{Guo12}. 

Perhaps due to differences in methodology, results vary considerably between different studies; even at low-z, where spectroscopic data sets are plentiful and galaxy group catalogues can be constructed with confidence. It seems unlikely that a consensus on the precise satellite profile around central galaxies will be reached in the near future. Our intention in this section is to draw together the results of the most directly comparable works from low-redshift, and the small number of high-redshift efforts, to compare broad trends rather than detailed values. The few studies that have presented results at both low and intermediate redshift (e.g. \citealt{Nierenberg12}) find no clear redshift evolution of the satellite profile slope. We will therefore compare the trends in our lower-redshift interval with those at $z\sim0$. 

Only a single study reports a dependence of the satellite profile on satellite luminosity or mass. \cite{Guo12} find that fainter satellites are more centrally concentrated around the primary galaxy when a constant physical radius is used to count satellites. However, when the radii are scaled to units of $R_{200}$ (the radius enclosing a density $200$ times the mean density of the Universe), then this dependence disappears. Their findings indicate that a spurious difference in slope can arise when the satellites in halos of different masses are naively co-added, which is the chosen method in most studies, including ours. Other observational estimates of the satellite profile at different masses find either a weak or insignificant difference \citep{Chen06,Nierenberg12,Wang14}. We find slightly steeper slopes for higher-mass satellites, i.e. in the opposite sense to the dependence found by \cite{Guo12}. For a given central sample the difference is not significant. However, it is consistent across the majority of our samples.

The dependence of the profile slope on central galaxy mass or luminosity is less consistent across the literature. A number of authors find steeper satellite profile slopes around lower-mass central galaxies \citep{Lares11,Prescott11,Guo12,Wang14}. Others, meanwhile, find no significant difference \citep[][this work]{Chen06,Jiang12,Nierenberg12}. The central galaxy mass dependence (or lack thereof) may be linked to a dependence on central galaxy colour, due to the fact that red colours dominate at the high mass end of the galaxy population. Indeed, the authors that report a central galaxy {\em mass} dependence on the satellite profile slope also find a dependence on central galaxy {\em colour}, with shallower slopes found around red galaxies \citep{Lares11,Prescott11,Guo12}. We find no consistent or significant behaviour (similar to e.g. \citealt{Chen08}). If (central galaxy) stellar mass is the more fundamental quantity in predicting the satellite distribution slope, then this absence of any notable behaviour is entirely expected because we stellar mass-match our passive and star-forming central samples.

The final aspect of interest in the profile slopes at low-z is the satellite colour dependence. Here again the literature is split, between those that report a satellite-colour dependency (in the sense that red satellites have a steeper radial distribution; \citealt{Chen08,Guo12,Wang14}), and those that find no significant difference \citep{Ann08,Lares11}. In the studies that find a slope dependence, it is more apparent around higher mass central galaxies and is perhaps a manifestation of a more physical interpretation: a radially-dependent satellite red fraction caused by star-formation quenching close to the primary galaxy \citep[e.g.][]{Prescott11}. Our results are consistent with this scenario, in that passive satellite galaxies exhibit steeper slopes than the full mass-selected satellite samples. However, the statistical precision of our measurements does not warrant a firmer conclusion. 

At higher redshift ($z>1$) there are currently just two comparable studies to our knowledge, \cite{Tal13} and Kawinwanichakij et al. (in prep.). \cite{Tal13} concentrate on high-mass central galaxies and use a mass ratio to define their satellite samples, rather than fixed cuts. Our method is otherwise very similar to theirs, and so a direct comparison with our high-mass central samples is reasonable. Fitting their high-redshift data with equation \ref{plfit} (they do not find any redshift dependence on slope or normalisation) we recover a slope of $\alpha=-1.08 \pm 0.08$, which is consistent with our measurement of lower-mass satellites. Combining our two satellite samples, we furthermore find that the normalisation for our high-mass centrals is very similar to the value we derive from the \cite{Tal13} measurements. Using a similar central stellar mass selection to our main sample, but extending the satellite sample to lower masses, Kawinwanichakij et al. (in prep.) find a slope, $\alpha = -1.28 \pm 0.08$. This value is slightly steeper than our equivalent measurement of $\alpha = -1.16 \pm 0.15$, but nevertheless consistent.

\section{Satellite passive fractions}
\label{passfrac}

\begin{figure}
\includegraphics[angle=0, width=240pt]{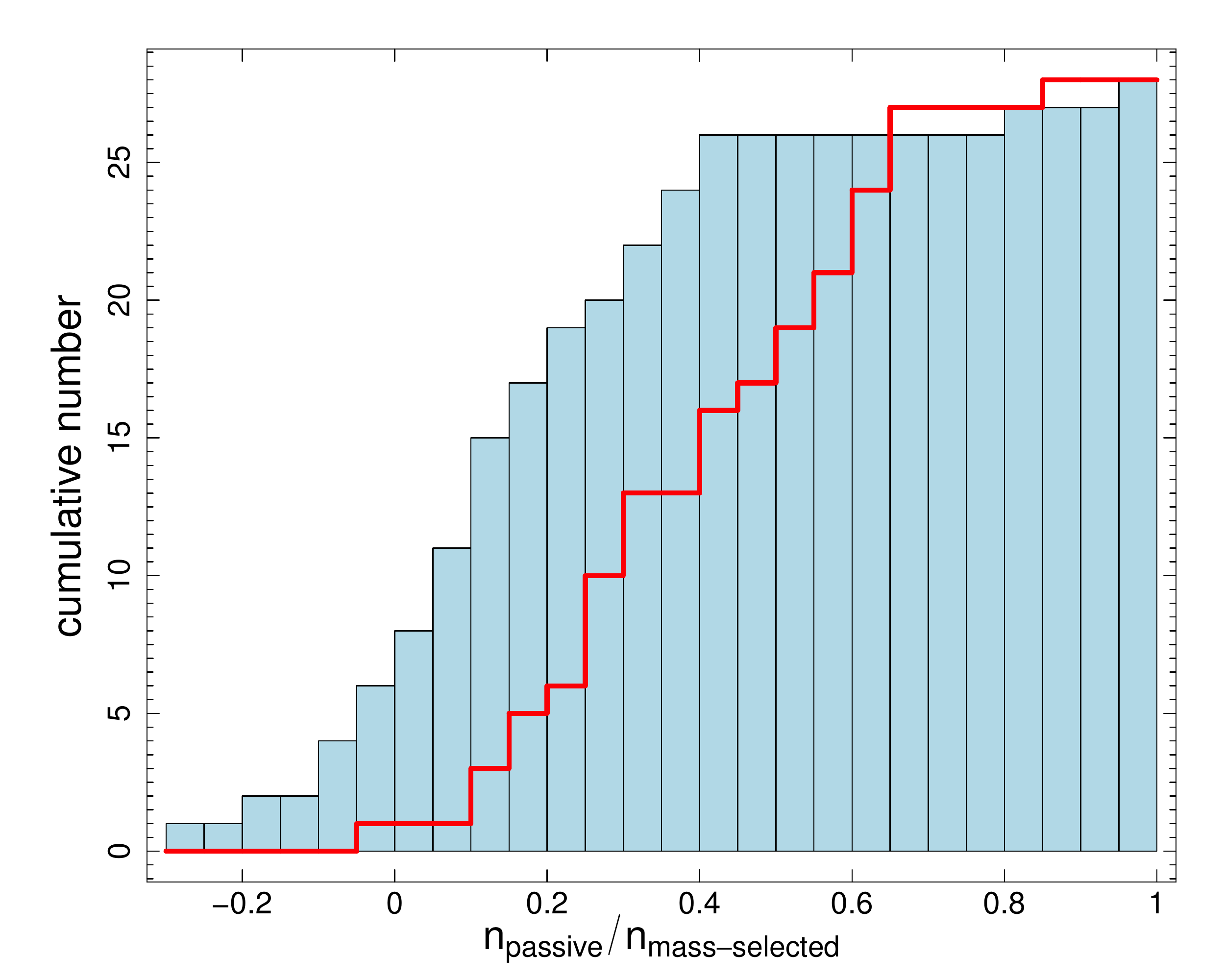}
\caption{Distributions of the ratio of passive and mass-selected satellite galaxy surface densities (passive fraction). The light blue filled histogram is the distribution for central galaxies that are star forming and the red line shows the passive centrals. The measurements for both redshifts and satellite masses are combined, with each radial bin (Figures \ref{lowz} and \ref{highz}) in the range $10 < {\rm R (kpc)} < 350$ contributing a single count. Performing a K-S test, we find that the distributions are inconsistent with having being drawn from the same distribution at $>3\sigma$ significance.}
\label{pfhist}
\end{figure}

We count and background-correct the mass-selected and passive satellites separately, and so do not have a formal measure of the fraction of satellites that are passive. However, the ratio of the two measurements is a suitable estimator for the passive fraction and we therefore utilise this quantity throughout the following analysis. Due to the methodology this ratio may exceed unity, or fall below zero in the cases where a negative number of passive satellites is recorded. In Figure \ref{pfhist} we plot cumulative histograms of the bin-by-bin passive to mass-selected surface number density ratio. We have seven radial bins, two redshift intervals and two satellite mass ranges, leading to $28$ independent estimates for each central sample. The results for star-forming centrals are shown by the light blue filled histogram, and the passive centrals by the solid red line. There is a marked difference between the two distributions, with the passive satellite fraction around passive central galaxies clearly shifted to higher values. Performing a two-sample K-S test on these distributions, we find that they are inconsistent with having been drawn from the same parent distribution at $3\sigma$ significance. We therefore confirm the existence of galactic conformity in our sample.

The elevated abundance of passive satellites around passive central galaxies can be taken as evidence of environmental processing (though there are alternative explanations). In order to quantify this effect we now compare this passive fraction with that from the wider field. The dependence of colour and passivity on local environment has been studied in detail in this field previously \citep{Chuter11,Quadri12}, and a colour-density relation identified to at least $z\sim1.5$. In addition to the field measurement, \cite{Quadri12} estimated the environmental quenching efficiency, $\epsilon$, in a  $z=1.62$ cluster candidate in the UDS field, finding $\epsilon \sim 0.4$. The environmental quenching efficiency is defined, following \cite{Peng10}, as the fraction of satellite galaxies that are passive, but that would otherwise have been star-forming had they been field galaxies, i.e. 
\begin{equation}
\epsilon = \frac{f_{p,sat} - f_{p,field}}{1 - f_{p,field}}. 
\label{efficiency}
\end{equation}
Here $f_{p,sat}$ is the passive satellite fraction and $f_{p,field}$ is the passive fraction of the field population. We define the field as all galaxies from our parent sample, within the stellar mass and redshift intervals of the satellite sample we wish to compare with. In this way we find the following passive fractions for field galaxies: $\sim 20\%$ (low-z, low mass); $\sim 35\%$ (low-z, high mass); $\sim 10\%$ (high-z, low mass); and $\sim 25\%$ (high-z, high mass). Other values found in the literature for the passive fraction of field galaxies are similar to those we find here \citep{Brammer11, Hartley13, Knobel13, Muzzin13, Tinker13, Lin14}, but with a degree of variation due to differing definitions for passivity and field-to-field variance. Because the field comprises galaxies in dense environments as well as isolated objects in our definition, it will provide only a lower limit of the true environmental quenching efficiency. 

At $\R=150~{\rm kpc}$ the satellite passive fractions around passive centrals at low redshift are $\sim 35\%$ for each mass range of satellites (possibly enhanced by the extended structure in the field). At higher redshift, the passive fraction of low-mass satellites is $\sim 20\%$ rising to $\sim 25\%$ at higher satellite masses. Passive fractions computed at $\R=50~{\rm kpc}$ are similar but slightly higher on average. From these indicative passive fractions we find $\epsilon \sim 0.1 - 0.2$  from the low-mass satellites, but minimal environmental quenching efficiency in higher mass satellites. This very rough calculation provides only an indication of the range of values for $\epsilon$ because projection effects mean that we are mixing a range of physical environments. Furthermore, the uncertainties are large and reducing the data to a single quenching efficiency does not take advantage of all the information that we have at our disposal. In the following subsection we examine how the passive fraction of satellites depends on separation from the central galaxy.

\subsection{Radial dependence of the passive fraction}
\label{pfrac}

\begin{figure*}
\noindent\begin{minipage}{180mm}
\begin{tabular}[htbp]{@{}ll@{}}
\includegraphics[angle=0, width=240pt]{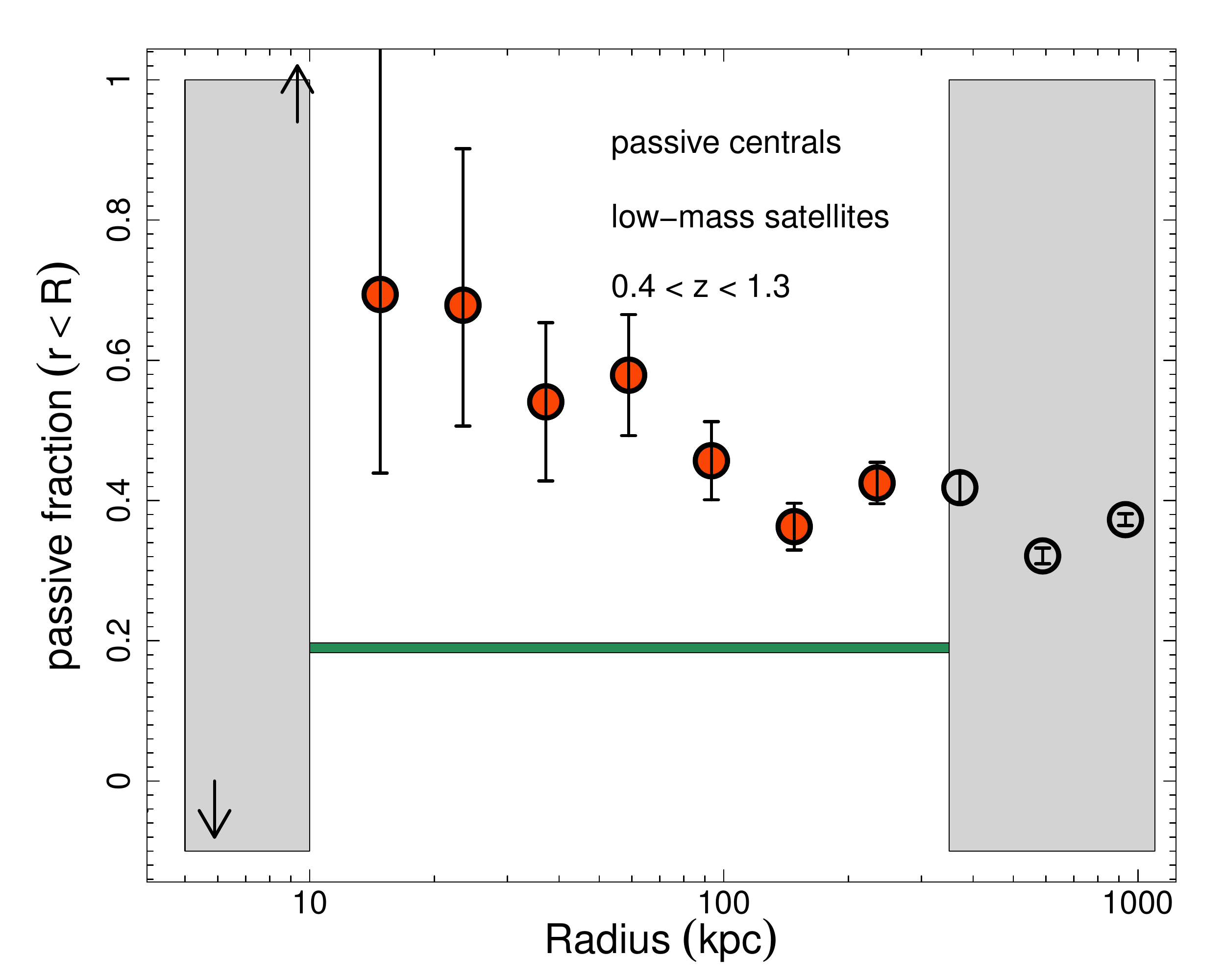} &
\includegraphics[angle=0, width=240pt]{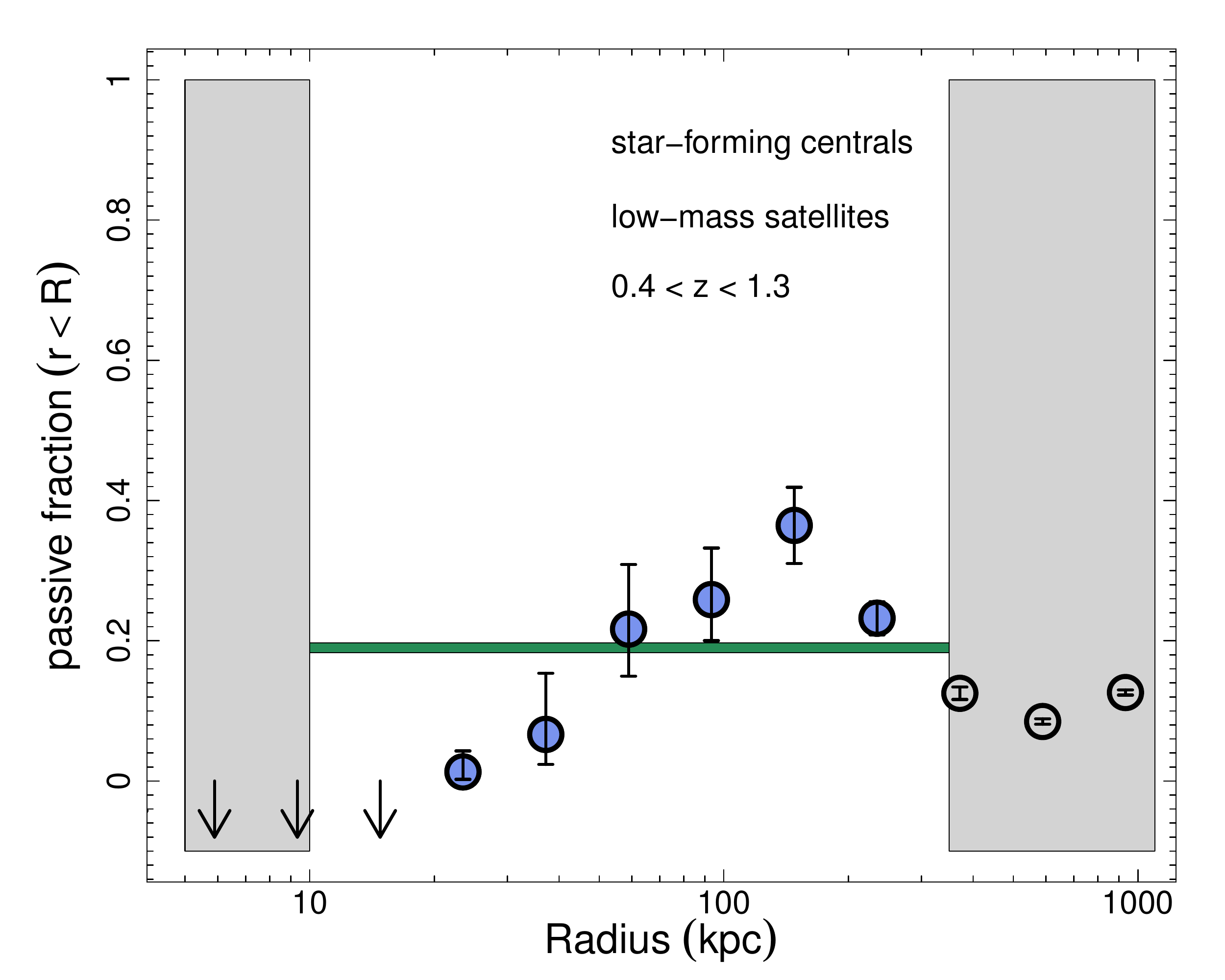}\\
\includegraphics[angle=0, width=240pt]{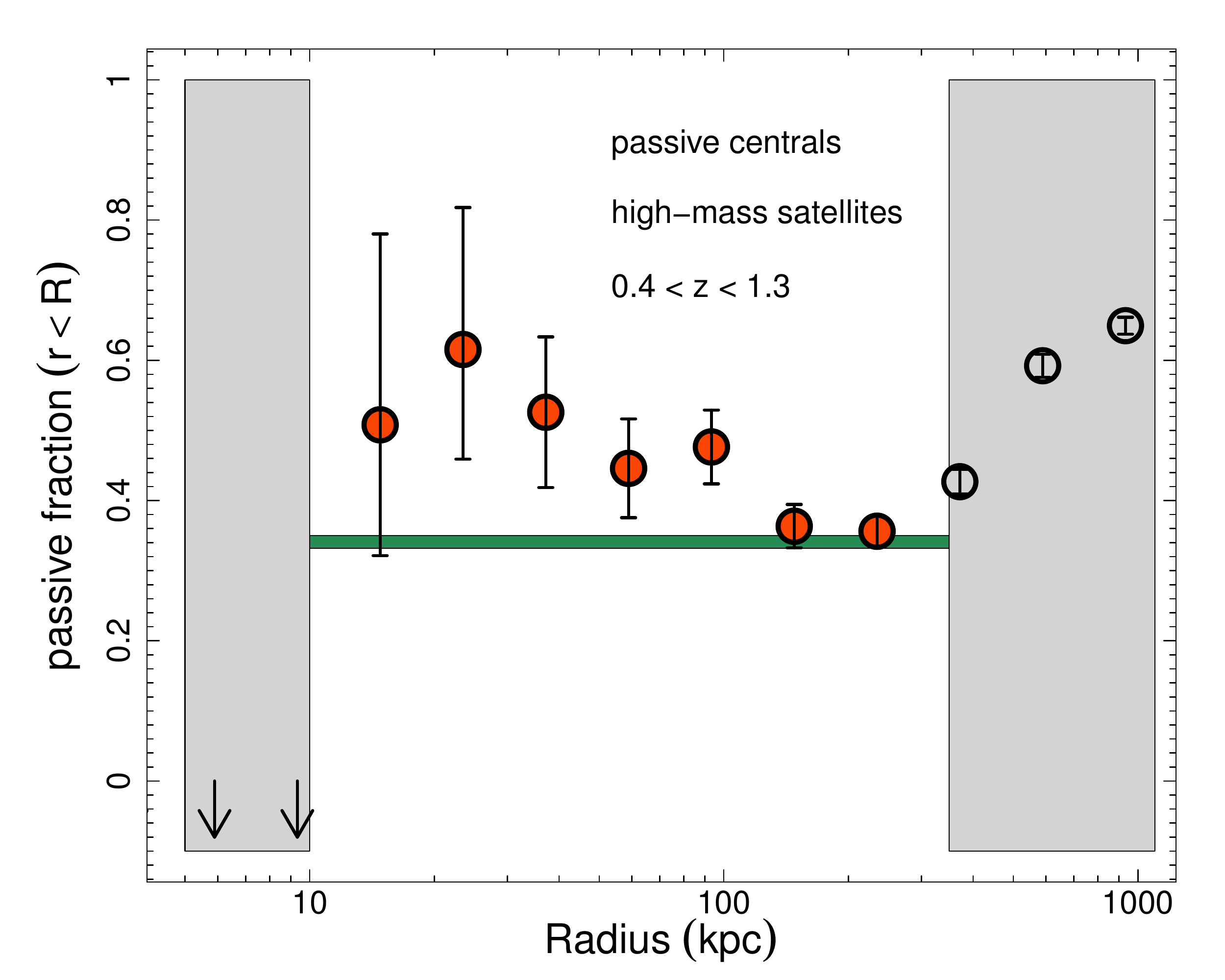} &
\includegraphics[angle=0, width=240pt]{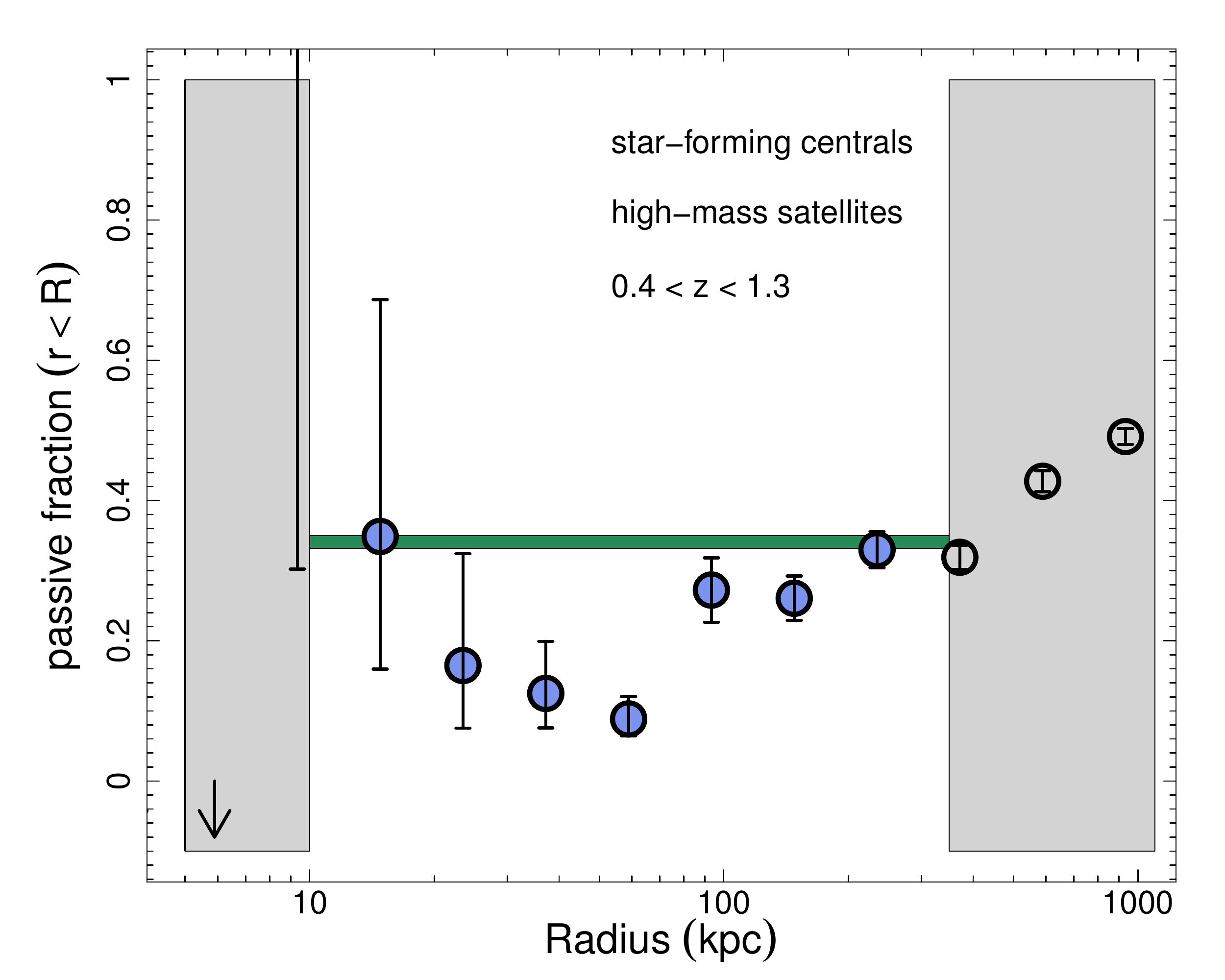}\\
\end{tabular}
\caption{The ratio of passive satellite galaxies to mass-selected satellites (our proxy for passive fraction) within $r<\R$, as a function of radius. Passive central $\M\sim \M^*$ galaxies in the redshift interval $0.4<z<1.3$ are represented by the left-hand panels, while measurements for star-forming centrals are given in the right-hand panels. Upper panels are for low-mass ($9.7<{\rm log}(\M_*/\M_{\odot})<10.1$) satellites, while the lower panels represent high-mass ($10.1<{\rm log}(\M_*/\M_{\odot})<10.5$) satellites. The green filled rectangles show the passive fractions and $1\sigma$ uncertainties in the field, for galaxies of the same stellar mass as the satellites. Because mass-selected and passive satellites are background-corrected independently, some data points lie above unity or below zero. These are represented by an errorbar alone, or an arrow if the uncertainty also lies beyond the plotting region.}
\label{lowz_frac}
\end{minipage}
\end{figure*}

\begin{figure*}
\noindent\begin{minipage}{180mm}
\begin{tabular}[htbp]{@{}ll@{}}
\includegraphics[angle=0, width=240pt]{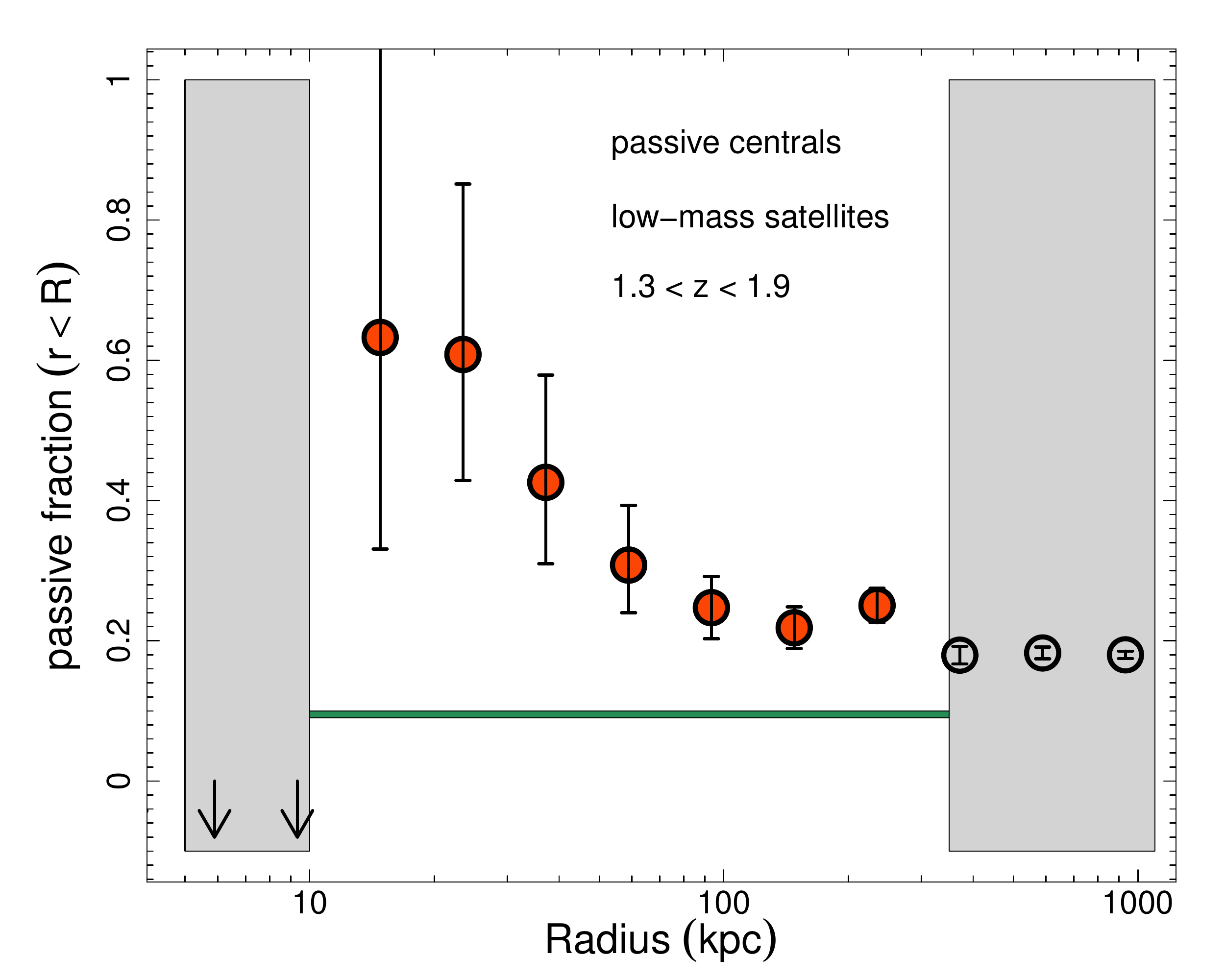} &
\includegraphics[angle=0, width=240pt]{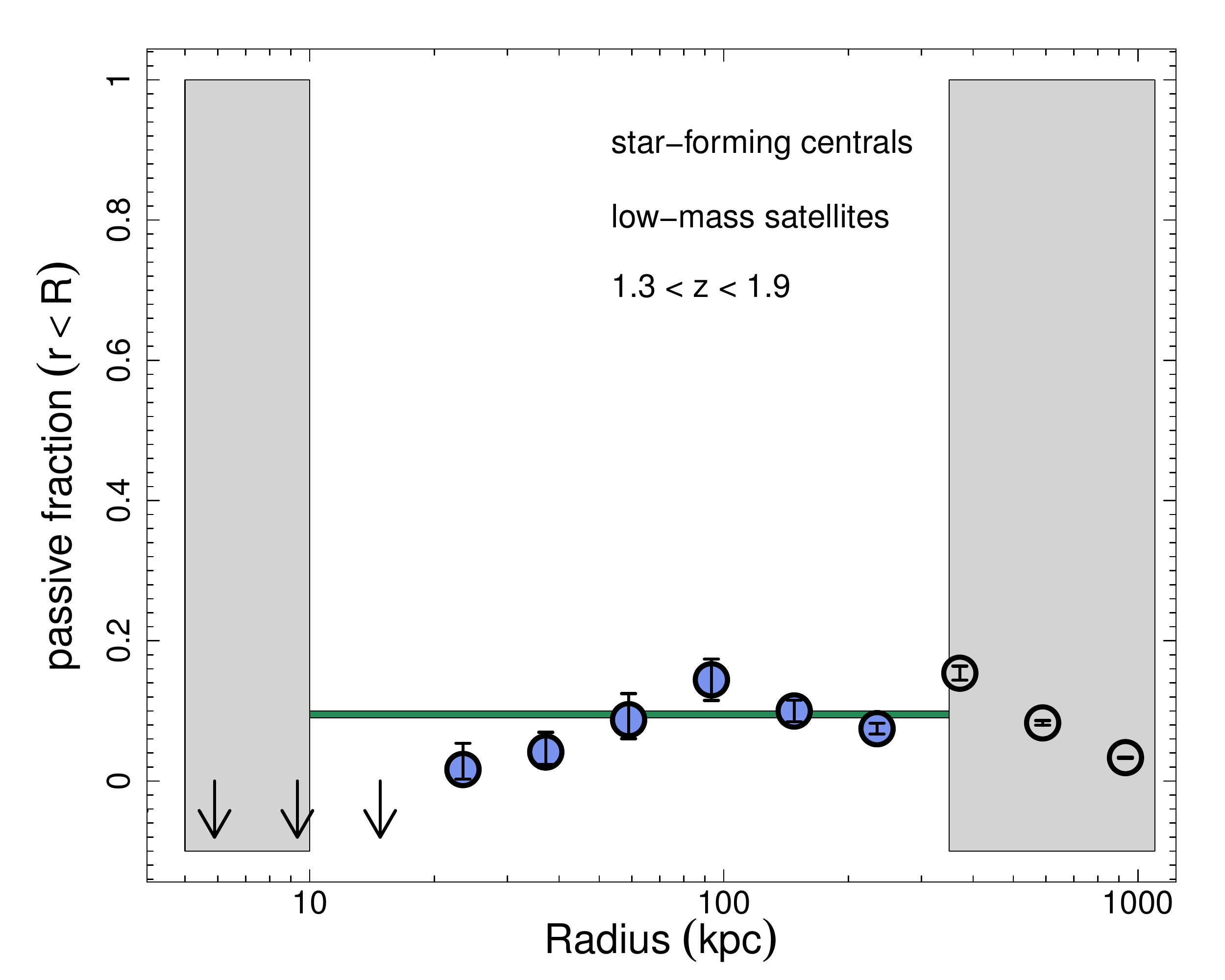}\\
\includegraphics[angle=0, width=240pt]{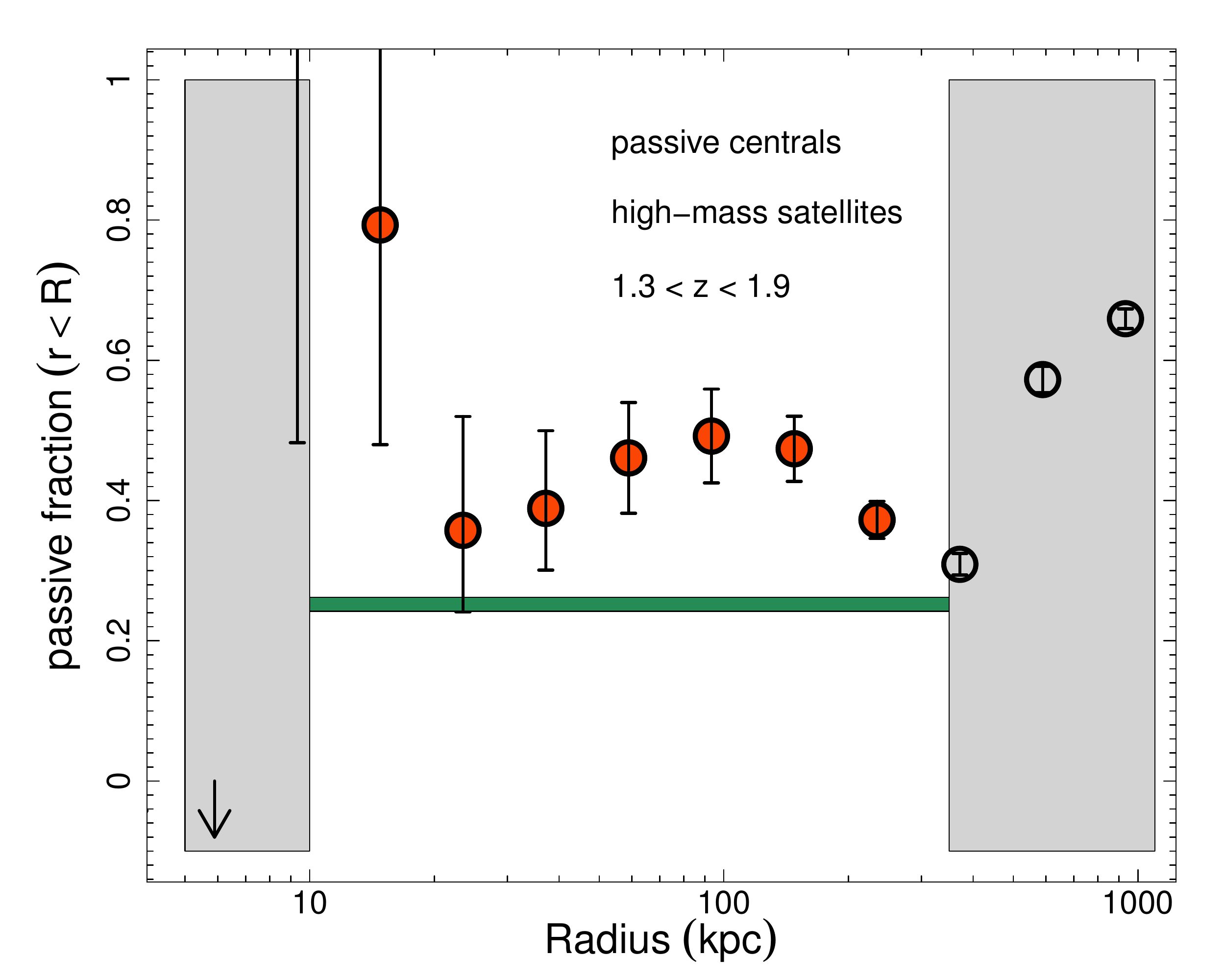} &
\includegraphics[angle=0, width=240pt]{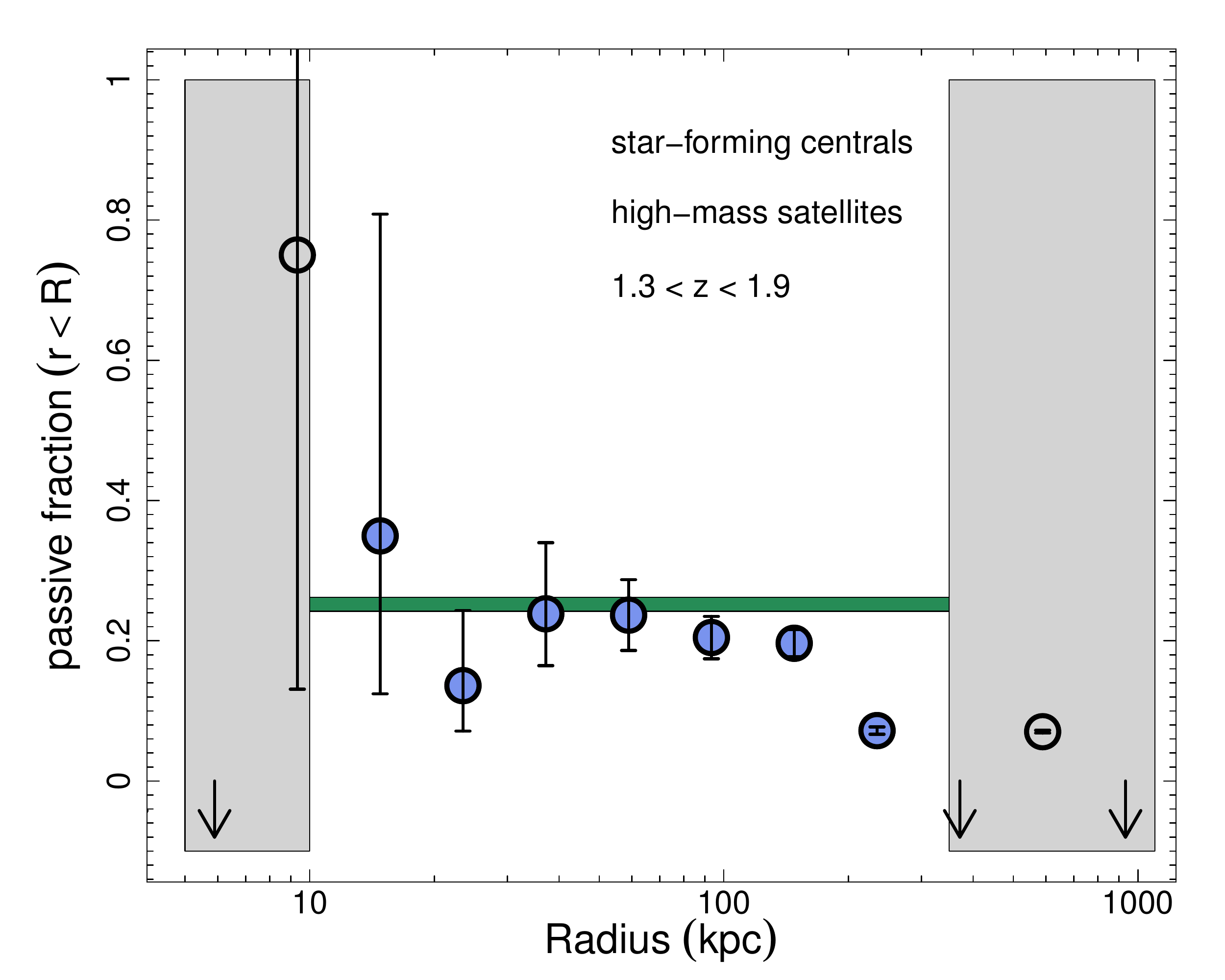}\\
\end{tabular}
\caption{Same as Figure \ref{lowz_frac}, but for the redshift interval $1.3<z<1.9$.}
\label{highz_frac}
\end{minipage}
\end{figure*}

The radial dependence of the influence of environment on satellite galaxies can encode important clues as to the physical process (or processes) at work. We showed in Section \ref{fitting} that the cumulative number of satellites within some radius, $\R$, is better constrained than either the slope or normalisation of the power-law fit. We therefore compute cumulative satellite profiles, and use these to find the ratio of passive to total satellites within some stellar mass interval and radius. 

The results of this cumulative ratio approach are shown in Figures \ref{lowz_frac} and \ref{highz_frac} for the two redshift ranges we have studied in this work. In each figure the left-hand panels are the data for passive central galaxies, while the right-hand panels display the results for star-forming galaxies. Filled circles (red for passive centrals and blue for star-forming samples) show our results, whilst the open circles are data points at radii that are expected to be unreliable (see Section \ref{rad_measurement}). The upper panels show measurements for low-mass satellites ($9.7<{\rm log}(\M_*/\M_{\odot})<10.1$), while the lower panels are our higher-mass sample ($10.1<{\rm log}(\M_*/\M_{\odot})<10.5$). 

The green rectangles in each panel show the $1\sigma$ confidence region of the passive fraction for field galaxies at the appropriate redshift and of equal stellar masses to the satellites. We show the same rectangular region for both passive and star-forming central samples. However, the fact that the passive fraction at fixed stellar mass evolves with redshift (albeit rather weakly for high-mass galaxies) can introduce a small bias in the interpretation. The passive fraction of satellites around our passive central galaxies should be compared with a lower redshift field sample, and hence a slightly higher passive fraction. We have computed to magnitude of this bias using the photometric redshift distributions of our central samples, finding $\Delta {\rm p_{frac}} = 0.02$ in the low-redshift interval and negligible contribution in the high-redshift interval. This bias is therefore at most comparable to the Poisson uncertainty already included and has no impact upon our conclusions.

As described above, no environmental information is used when computing field passive galaxy fractions. Some field galaxies will have been quenched due to their environment and therefore a data set which lies on the green rectangle could be regarded as evidence for low-level environmental quenching of satellites. However, we recall (Section \ref{rad_measurement}) that we are missing a small fraction of star-forming satellite galaxies. The sense of this bias is opposite to that caused by the field definition above. It is impossible to determine which effect is greater without extensive spectroscopy. However, the $z\sim0.62$ cluster notwithstanding, we expect the field passive fraction to be dominated by isolated galaxies, and the precise degree of the net bias does not have an important impact upon our conclusions. We therefore ask that the reader simply bare these biases in mind during the following discussion.

Some of the trends that we highlighted in the previous section are immediately obvious in these plots. In particular, passive central galaxies have a higher fraction of passive satellite galaxies in both of the redshift ranges studied. Viewed with respect to the field population the differences become more apparent. Across almost all radii, in each stellar mass and redshift range, the satellite passive fractions around passive central objects lie systematically and significantly above the field value. For star-forming centrals, meanwhile, the passive fractions are consistent with the satellites having been drawn directly from the field sample. Although we cannot rule out some low-level environmental quenching of the satellites of star-forming centrals, these results are similar to those found by \cite{Phillips14} for pairs of galaxies in the SDSS. The implication of these results is that the only environments capable of impacting the star-forming properties of modest-mass satellite galaxies are those in which the central galaxy has already ended its major episode of star-formation. Given the uncertainties involved in identifying satellite populations with only photometric redshifts (even on a statistical basis), a statement this strong is perhaps unwarranted. We therefore await further studies to more accurately quantify the impact of environmental processes on the satellites of star-forming galaxies.

\subsubsection{Redshift evolution}

One of our intentions at the outset of this work was to use the redshift dependence of galactic conformity to constrain the physical origins behind its manifestation. Considering the results of satellites around passive central galaxies to begin with, the passive fraction of low-mass satellites is very similar in both redshift intervals. If we take account of the variation in the field passive fraction and instead compute the environmental quenching efficiency as a function of radius (Equation \ref{efficiency}), we find that the two are statistically indistinguishable. 

For higher-mass satellites, the same statement is true for all but the two largest radial separations (which differ at $\sim3\sigma$), which are in the sense that a stronger conformity signal is found at higher redshift. In these two cases the actual passive fractions are similar at each redshift. The disparity arises due to the high field passive fraction at low redshift and the small statistical uncertainties. However, we have neglected to include sources of systematic uncertainty in our analysis, such as sample variance. As we described above, the quenching efficiency we calculate in the lower redshift interval may be biased in the presence of strong large-scale structure variance. We therefore do not find any convincing evidence for an evolution of the galactic conformity signal with redshift from the passive central galaxies. 

The results for the star-forming central galaxies are very similar: a minority of points show evidence for a statistical difference between the two redshift intervals. However, the redshift dependence is in the opposite sense for low and high-mass satellites. Moreover, there is only a single such point that implies a quenching efficiency above that of the field. The other points that are statistically different between the two redshift intervals are either consistent with or below the field passive fraction. We therefore conclude that there is no clear evidence for redshift evolution from the star-forming central galaxies either. This is not to say that we confirm an absence of {\em any} evolution, but any redshift dependence must be relatively minor.

\subsubsection{Radial dependence}

A final feature of our results is the difference in radial dependence of the passive satellite fraction for the two central samples. Around passive central galaxies, the fraction of passive satellites appears to rise at small radii, with most ($\gtrsim 60\%$) satellites within $20-30~{\rm kpc}$ being passive. The same is not true for the satellites of star-forming objects, where the equivalent passive fraction for low-mass satellites is essentially zero in both redshift intervals. The apparent radial dependence of satellite passive fraction around passive central galaxies is, however, driven largely by the very uncertain data points at small radii. We therefore cannot rule out a flat passive fraction across the range of radii we study. However, because the ratios we use as an estimator of the passive fractions are based on the cumulative galaxy counts, any gradient that we find must be shallower than the intrinsic (differential) gradient. This is simply because the high fraction of passive satellites at small radii would be included and contribute to the passive fraction at larger radii. 

The apparent radial dependence appears much more strongly in the results of low-mass satellites. In fact, to uncover the influence of environment it is prudent to study lower-mass systems where possible. The intrinsic passive galaxy fraction (`mass quenched' in the nomenclature of \citealt{Peng10} and other recent works) is smaller at low stellar mass \citep{Geha12}, and so variations in the efficiency of environment quenching are more readily apparent. Although our lower-mass satellites are not of particularly low mass (a galaxy within our low-mass satellite interval at $z=1.5$ may well evolve into a Milky Way-mass galaxy or greater by the present day), it is certainly these objects at both redshifts that drive the impression of a radially-dependent environmental quenching efficiency. Moreover, a radial dependence of the passive fraction, even at fixed stellar mass, has been reported previously in group-mass halos \citep{Prescott11, Carollo14}. For these reasons we suggest that the gradient is likely to be real, but note that further investigation is required to confirm this finding. Overall, our results show that the `galactic conformity' correlation observed at low redshift exists to at least $z\sim2$ for $\M \sim \M^{\ast}$ galaxies, and appears increasingly prevalent at smaller central-satellite separations.

\section[]{Discussion - The origin of galaxy conformity}
\label{disc}

We have found that an observed correlation between the colour or star-formation properties of central and satellite galaxies at low redshift (so called `galactic conformity') is present to at least $z\sim2$. There are a number of possible explanations for a correlation between the central and satellite galaxy star-formation properties. Some of these take the form of a common origin of quenching, while others are more coincidental. In this section we discuss some of the more likely origins of galaxy conformity and, drawing additional insight from the literature, we propose one possible description of the principal route for galaxy quenching.

\subsection{Halo mass as the primary driver of galactic conformity}

\begin{figure*}
\noindent\begin{minipage}{180mm}
\begin{tabular}[htbp]{@{}ll@{}}
\includegraphics[angle=0, width=240pt]{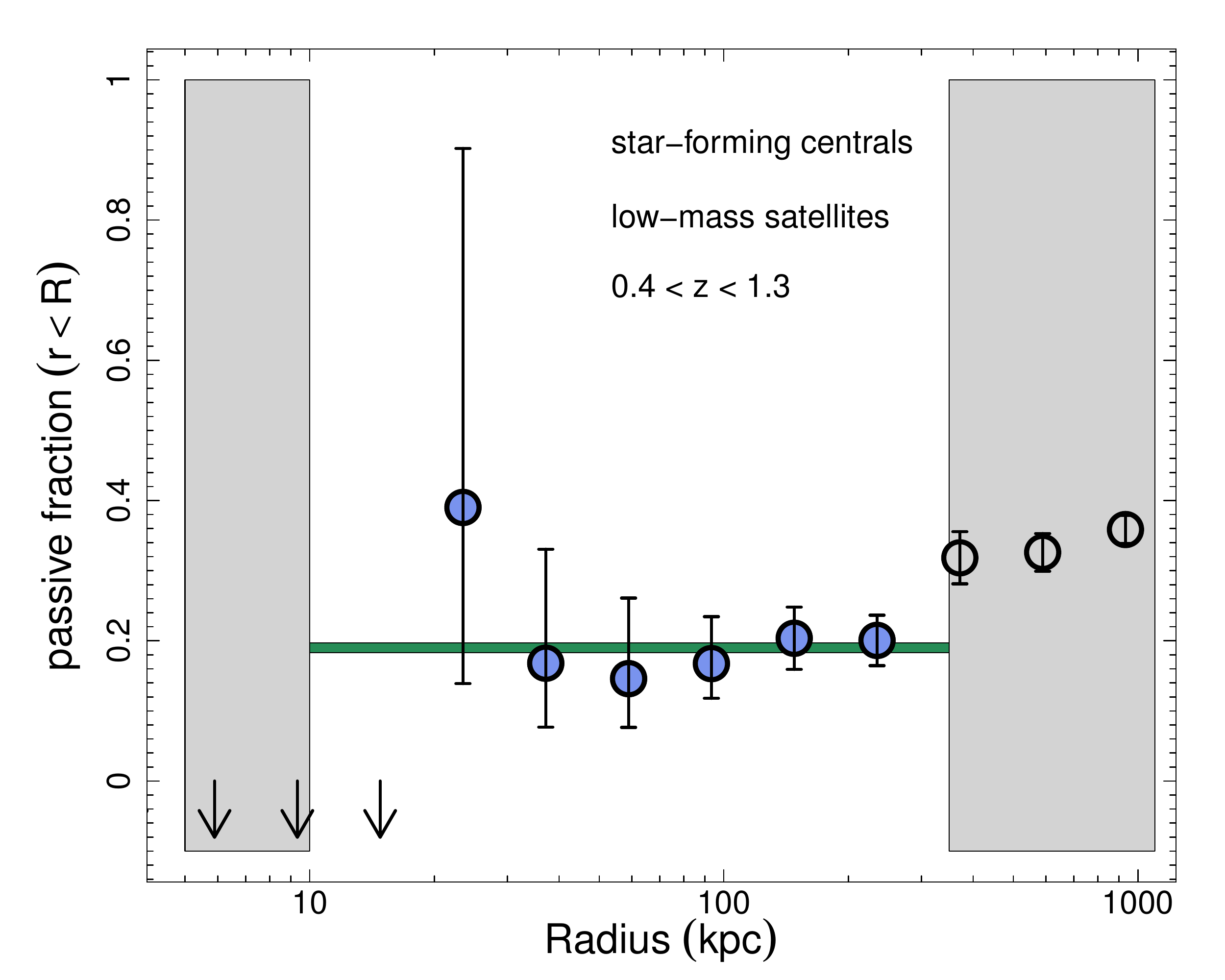} &
\includegraphics[angle=0, width=240pt]{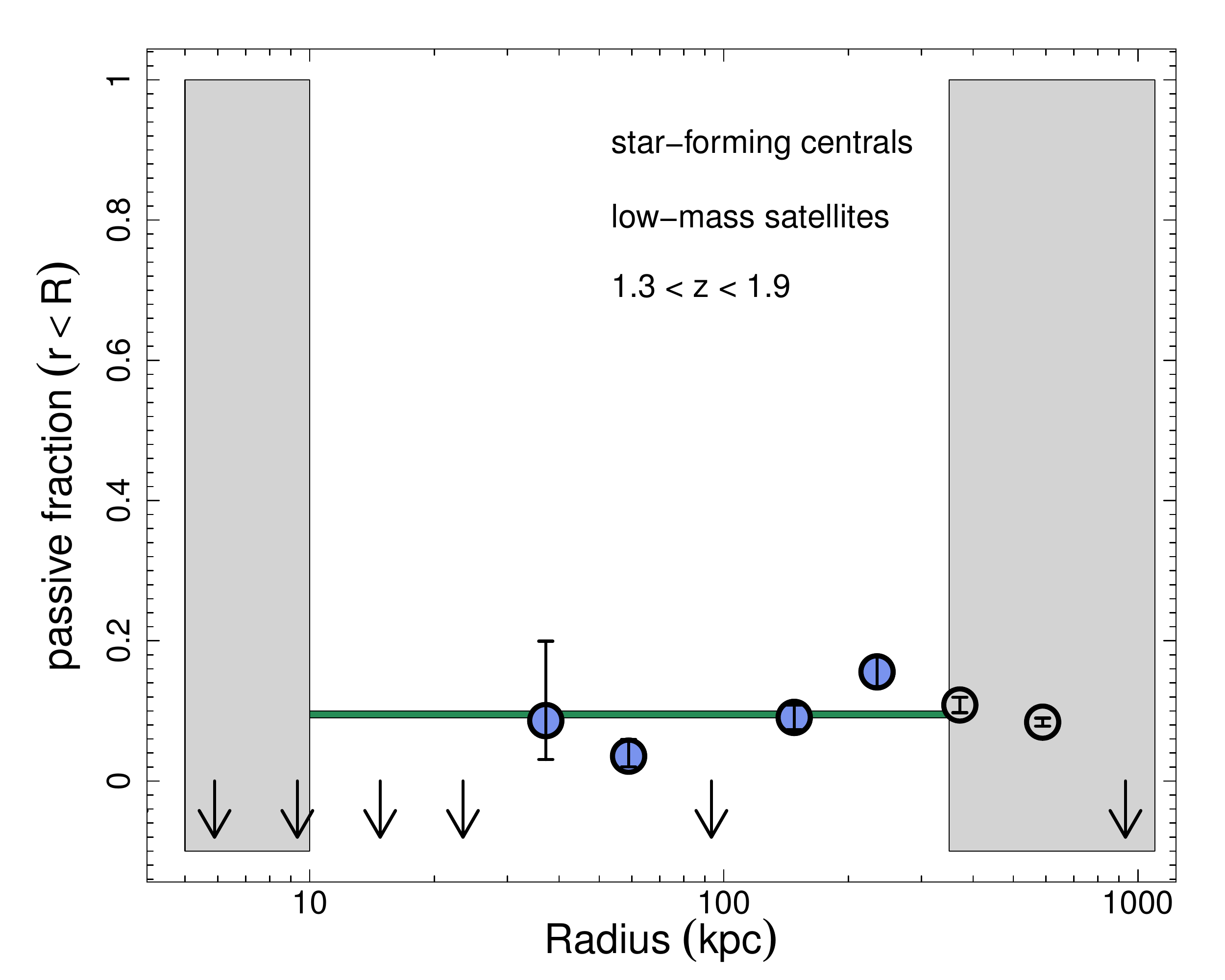}\\
\includegraphics[angle=0, width=240pt]{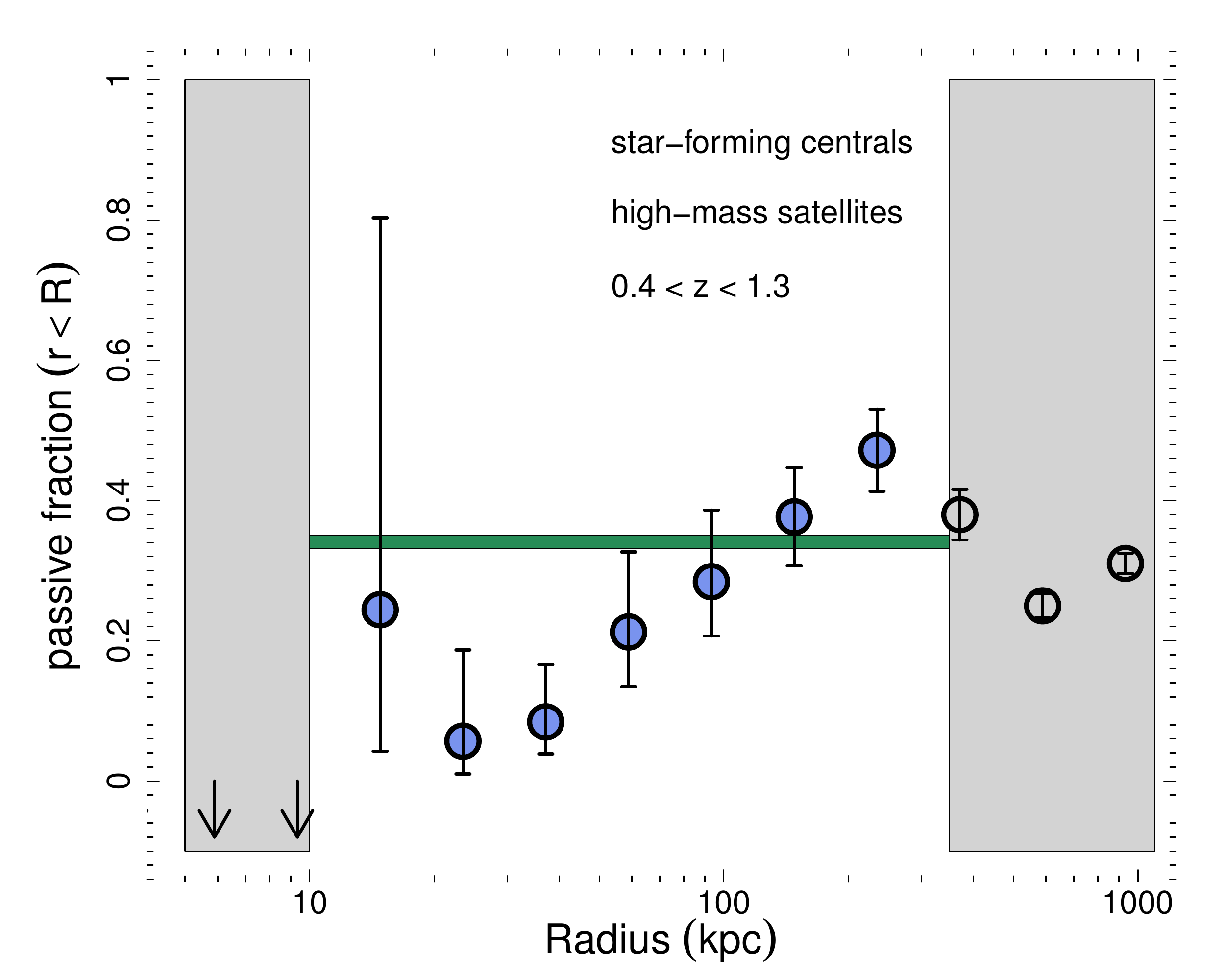} &
\includegraphics[angle=0, width=240pt]{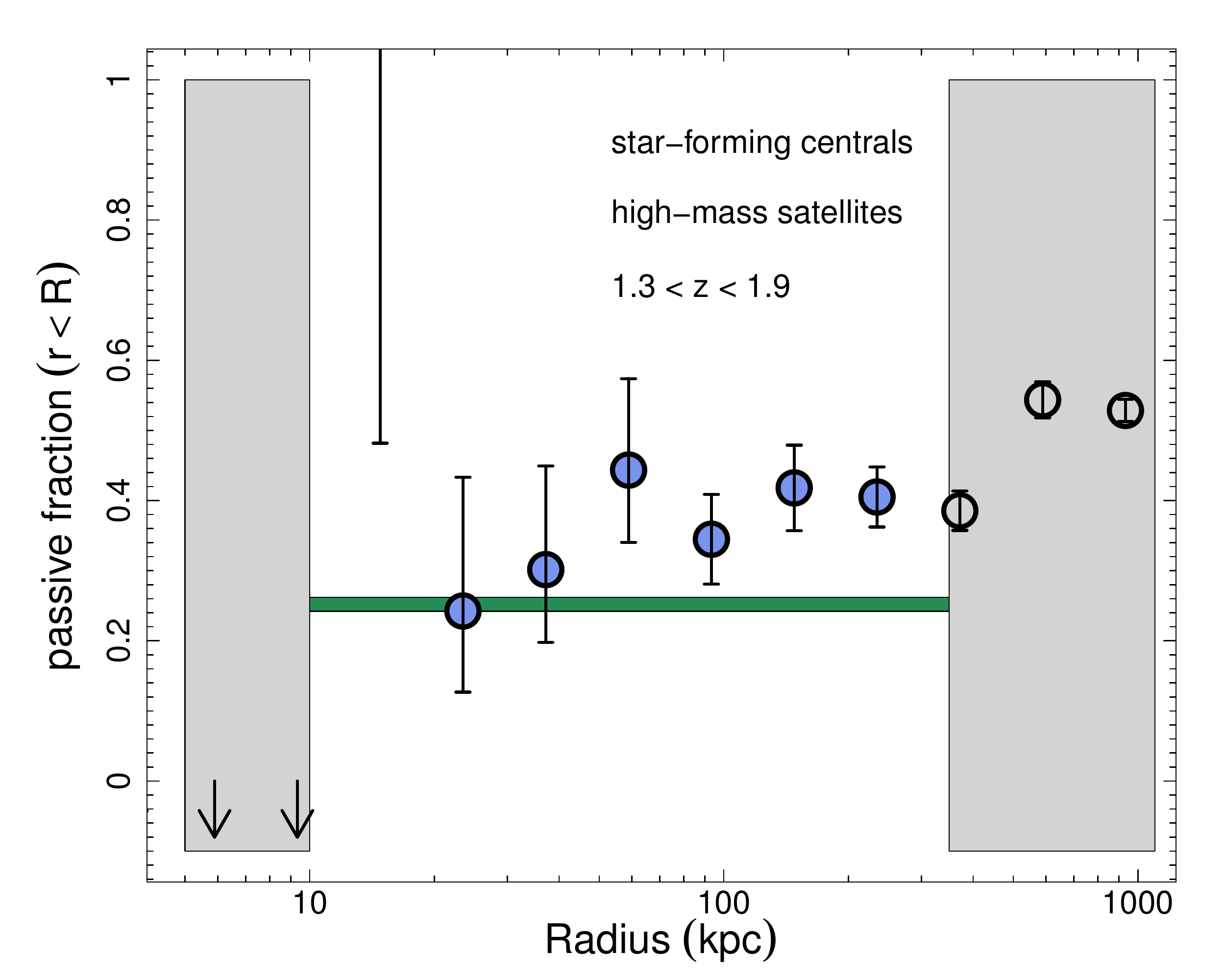}\\
\end{tabular}
\caption{Ratio of passive satellites to mass-selected satellites for higher-mass (${\rm log}(\M^*/\M_{\odot} > 11$) star-forming centrals. Symbols and lines have the same meanings as in Figure \ref{lowz_frac}.}
\label{highmass}
\end{minipage}
\end{figure*}

Perhaps the most obvious explanation for galactic conformity is that passive central galaxies (and their satellites) are hosted by more massive halos than star-forming central galaxies of similar stellar mass. The majority of studies investigating galactic conformity have concluded that halo mass is {\em not} responsible. However, given that the majority of satellite and central  galaxy quenching mechanisms are expected to correlate with halo mass (and the relative simplicity of the arguments), it is worth re-visiting this question with our data.

At high masses, dark matter halos are thought to be able to sustain a halo of hot gas which can efficiently prevent cooling \citep{White78,Dekel06, Croton06, Cen11}. This halo quenching proceeds by gravitational and shock-induced heating of newly accreted gas to the virial temperature of the halo; thereby delaying the conversion of the gas into stars by the time it takes for the gas to cool. For high mass halos, ${\rm log}(\M_{{\rm halo}}/\M_{\odot}) \gtrsim 12.5$, the cooling time of gas at the virial temperature approaches the Hubble time, and the gas can be maintained at high temperature by a modest amount of radio AGN feedback \citep{Croton06}. 

In \cite{Hartley13} we found that galaxy host halo masses inferred from clustering statistics were entirely consistent with this scenario, with passive galaxy samples out to redshifts, $z\sim3$, typically hosted by high mass halos (${\rm log}(\M_{\rm halo})/\M_{\odot} \gtrsim12.5$), irrespective of stellar mass. The hot gaseous halo surrounding passive objects could also create an environment which is able to efficiently process satellite galaxies, either by strangulation \citep{Larson80} or ram pressure stripping \citep{Gunn72}. In order to identify halo mass as the physical quantity responsible for the central-satellite SFR correlation, we must show two things: 1. that passive central galaxies are found in more massive dark matter halos than the star-forming galaxies that they are compared with; 2. that the difference is sufficient to give rise to the conformity effect. Although an attractive and simple explanation, we will argue that halo mass does not satisfy the second condition and is therefore unlikely to be the {\em sole} cause of galactic conformity.

As stated previously in section \ref{prof_mass}, it might reasonably be expected that the number of satellites of a given stellar mass would correlate with the overall halo mass. At high masses this is certainly the case, as richness has been found to be a good estimator for the total mass in clusters \citep[e.g.][]{Andreon10}. Indeed we find that, at low redshift at least, passive central galaxies have a greater number of satellites. However, as the satellites in our sample are of fairly high stellar mass (${\rm log}(\M_*/\M_{\odot} > 9.7$), and the expected number count of satellites per central galaxy in our sample is less than unity, we cannot rely on this quantity alone to estimate halo masses. An illustration of the difficulty in inferring halo mass from satellite counts is provided by the phenomenon of fossil groups. These are groups in which the magnitude difference between the brightest and second-brightest object is $\Delta m > 2$ \citep{Ponman94}. The lack of relatively massive satellites is related to the formation history of the dark matter halo, rather than its mass \citep{Dariush07}. 

Clustering results routinely find that passive or red galaxies at all redshifts are hosted, on average, by more massive halos than the equivalent-stellar-mass star-forming objects \citep[e.g.][]{Norberg02,Zehavi05,Hartley13}. However, the results from 2-point correlation function measurements include both central galaxies and satellites. To recover the typical halo mass of only the central galaxies requires the use of a halo occupation distribution model \citep[HOD, see ][for a review]{Cooray02}. Modelling the measurements of star-forming and passive samples separately is fraught with uncertainties and necessitates potentially important assumptions (such as the satellite number-density profile of each sample within halos, and the probability of a central being passive as a function of halo mass). Most high mass (${\rm log} (\M_*/\M_{\odot}) > 10.5$) galaxies at $z>1$ are likely to be the central object in their halo, and the simpler assumption that each halo hosts one such galaxy is perhaps reasonable for these samples. Nevertheless, there have been a few notable efforts to interpret the results of passive galaxy clustering through an HOD framework from the local universe to $z\sim2$ \citep{Ross09,Tinker10,Coupon12}. These studies typically find a $50\%$ occupation for red (passive) galaxies at halo masses, ${\rm log}(\M_{{\rm halo}}/\M_{\odot}) > 12$ (the value is luminosity dependent), with most results showing red centrals are hosted by more massive halos than similar stellar mass blue central galaxies. The exception is \cite{Tinker10}, who enforce a fixed central fraction at high halo masses of $\sim25\%$. Perhaps the most reliable results are those at low redshift by \cite{Ross09}, who explicitly identified and modelled galactic conformity in their measurements. Given our results, this is clearly a necessity for future halo occupation modelling of passive and star-forming samples at any redshift.

From a physical perspective, finding a higher average halo mass for passive central galaxies than for similar stellar mass star-forming central galaxies is almost inevitable. While a central galaxy is forming stars, it is natural to expect that the stellar and dark matter masses increase in such a way that a correlation between these quantities is maintained. This is indeed what we observe at low redshift \citep[e.g.][]{Mandelbaum06,More11}. However, once a galaxy becomes quenched, the smooth accumulation of dark matter halo mass and (central galaxy) stellar mass become decoupled. The stellar mass ceases to grow appreciably, while the dark matter mass continues to accumulate, producing lower stellar-to-halo mass ratios (SHMR, i.e. higher halo mass at fixed stellar mass) for massive galaxies \citep[e.g.][]{Tinker05,Mandelbaum06,Foucaud10,Moster10,Behroozi13,Rodriguez13}. Although a broad correlation is maintained via hierarchical assembly, the scatter in SHMR also increases for passive central galaxies with respect to similar-mass star-forming objects. This difference in halo mass ranges is illustrated in the models of \cite{Birrer14}, and is easy to understand by considering the progenitors of each population at the time when the observed passive centrals were last star-forming. Passive galaxies have, by definition, added little stellar mass over the $\sim$Gyr prior to observation (unless dry merging is substantially more prevalent than we currently believe). In contrast, a Gyr is approximately the stellar mass doubling time for a star-forming galaxy on the `main sequence' at redshift, $z\sim2$ \citep[e.g.][]{Daddi07}. When a passive galaxy at this redshift was quenched, therefore, its stellar mass could have been twice that of the star-forming objects that it is later compared with. It would be a surprise, then, if passive central galaxies did not, on average, have higher host halo masses than similar stellar mass star-forming galaxies. Nevertheless, the difference in {\em average} halo mass between the two population is not expected to be large (a factor of $\sim2$ or so at the stellar masses we study here, \citealt{More11}). 

The above arguments suggest (but do not prove) that passive central galaxies are indeed hosted by more massive dark matter halos. However, we must still attempt to show that it is the halo mass which is driving the correlation in the star-formation properties of centrals and satellites. To investigate a possible halo-mass effect we need to compare the satellites of passive and star-forming galaxies at constant halo mass. Unfortunately, accurate halo mass estimates for $\M^{\ast}$ galaxies at high redshift are difficult to obtain, even for small samples. Instead we must use an indirect method to infer the halo mass-dependent behaviour. To do so we use the existence of the correlation between star-forming central stellar mass and host halo mass. From our parent sample we select high-mass star-forming galaxies with stellar mass, ${\rm log} (\M_*/\M_{\odot}) > 11$. These galaxies are, on average, a factor of $\sim3$ more massive than the star-forming sample we considered previously, and should therefore, at least approximately, be hosted by similar mass dark matter halos to our passive ($10.5 < {\rm log} (\M_*/\M_{\odot}) < 11$) sample\footnote{As an rough indication, we find that the cumulative number of satellites are very similar for the high-mass star-forming and $\M^{\ast}$ mass passive centrals.}. This new high-mass sample is treated in exactly the same way as our main samples and the results should therefore reveal the extent to which halo mass is responsible for galactic conformity.

In Figure \ref{highmass} we show the radial dependence of the satellite passive fraction for these high mass star-forming centrals in both redshift ranges and satellite samples. The results clearly mimic those of their $\M \sim \M^{\ast}$ counterparts, in Figures \ref{lowz_frac} and \ref{highz_frac}. With the possible exception of high-mass satellites at high redshift, the satellites of high-mass star-forming galaxies are consistent with being drawn from the field population. There is no clear evidence for the sort of environmental quenching that is occurring in the satellites of passive centrals. Whilst we cannot unequivocally rule out host halo mass being the only necessary ingredient for satellite quenching, we would need to assume an implausible degree of scatter to the stellar-to-halo mass ratio. Else, a rather unphysically sharp transition with halo mass would need to occur which is not supported by halo occupation models \citep{Ross09}. 

The question of whether halo mass is the principle driver of conformity has been a contentious issue in the literature. When \cite{Weinmann06} identified the correlation in colour between central galaxies and their satellites, they suggested that halo mass was {\em not} an important factor. This seemed to be supported by \cite{Ross09}, who found that they needed to apply conformity `to the maximum extent allowed' in order to produce a reasonable halo model fit to their data. However, \cite{Wang12} found that they could qualitatively reproduce the behaviour of conformity in the SDSS by a semi-analytic (S-A) model \citep{Guo11} in which quenching processes were only dependent on halo mass. Following this, \cite{Kauffmann13} performed a detailed investigation of the mass and radial dependence of conformity in the SDSS with comparison to the same S-A model, finding several discrepancies. 

It is well known that the fraction of quenched satellites is typically too high in S-A models of galaxy formation \citep{Weinmann11,Wang12,Wang14}. It may be that the quantitative disagreement between the model and observational data is nothing more than the result of over-efficient quenching. However, it is equally possible than some aspect of the physics is either missing or not correctly modelled, and that this missing physics is the true cause of galactic conformity. \cite{Kauffmann13} identify a few of the possible missing process, suggesting that some form of pre-heating \citep{Valageas99} could be necessary for low-mass centrals, and the effects of gas accretion important at high masses. In the following section we suggest an alternative origin for galactic conformity in central galaxies of $\M\sim \M^{\ast}$ and above.

\subsection{A simple picture of coincident quenching}
\label{model}

With the evidence seemingly weak for a {\em purely} halo-mass driven origin of conformity, we must now consider other possible origins of the correlation between central and satellite star-formation activities. We can consider two broad concepts: a physical connection between the star-formation properties of central and satellites, or a circumstantial correlation. The latter we might infer as being due to formation bias, where galaxy evolution is accelerated in regions of large-scale overdensity. 

At fixed mass, the dark matter halos (and the galaxies that they host) collapse earlier, on average, in denser regions of the Universe. This could in turn lead to a higher passive fraction by our observational definition. Recently it was shown by \cite{Hearin14} that formation bias can give rise to a large-scale (2-halo) conformity signal (identified in SDSS data by \citealt{Kauffmann13}) over radii $\R>1 {\rm Mpc}$. Although \cite{Hearin14} argued that this 2-halo conformity is created by the properties of {\em central} galaxies, they also point out that a number of these galaxies will ultimately become satellites around other quenched galaxies. Such effects could play an important role in establishing intra-halo galactic conformity \citep{Yang06}, and may in part explain the occurrence of an apparent `pre-processing' of cluster galaxies \citep[e.g.][]{Kodama01}. 

The most massive galaxies were the first to complete their major episode of star formation, with lower mass objects successively forming a greater proportion of their stars at lower redshifts \citep{Cowie96,Kodama04}. This trend became known as `downsizing', but can be re-cast in terms of formation bias: the most massive galaxies formed earlier, when typical sSFRs were higher, and very quickly reached the characteristic quenching mass \citep{Peng10}. If, in regions of the Universe dominated by passive central galaxies, low-mass galaxies formed earlier than typical for their mass, then we may well expect downsizing to be weaker for these objects. However, we currently have no reason to believe that downsizing would be absent entirely in these regions. If formation bias is entirely responsible for intra-halo galactic conformity, we would therefore expect it to weaken at higher redshift. We have not been able to measure any evolution in conformity thus far, but cannot rule out this possibility. Modelling the evolution of galactic conformity in the context of formation bias will be the subject of future work. In the remainder of this sub-section we will explore the alternative possibility of a physical relationship between the star-forming properties of central and satellite galaxies.

To search for a physical connection, we first consider the processes most likely to be involved in quenching central galaxies. A very compelling observation of the galaxy population is the coincidence of suppressed star-formation rates and a bulge-like morphology \citep{Blanton03, Bell11}. Bulge formation is currently poorly understood and may be entirely unrelated to the process(es) that end star-formation in galaxies. However, arguments to explain this correlation by a physical connection can be classed as either of violent origin, simultaneously producing a bulge and quenching a galaxy \citep[e.g.][]{Hopkins06}, or a more sedate process, by which the presence of a bulge stabilises the gaseous disk against fragmentation \citep{Toomre64,Martig09}. Because of its dependence upon the Toomre stability criterion, Q \citep{Toomre64}, the latter type of morphology-related quenching is sometimes referred to as `Q-quenching'. However, how a bulge-like morphology of a central galaxy could influence its satellites' star-formation activity is not clear. We suggest, therefore, that whatever role Q-quenching plays, it is not important in establishing galactic conformity. 

Merging could be responsible for a rapid-timescale quenching pathway, and is perhaps required to explain the paucity of intermediate sSFR objects \citep[see][for some recent arguments]{Lin14,Omand14,Schawinski14}. This violent bulge-forming process is expected to result in very strong galaxy-wide outflows. Whether these outflows are powered by a central super-massive black hole or a star-burst phase is not particularly important for this discussion. What is important is that galactic conformity requires that the quenching of the central galaxy be coupled to the satellite galaxies or wider halo environment. Outflows are a ubiquitous phenomenon, even for `normal' star-forming galaxies at moderate to high redshift \citep{Weiner09, Bradshaw13, Bordoloi13}. Moreover, from their phenomenological model, \cite{Peng10} argued that the probability of a central galaxy being quenched over some time interval could be described by a simple proportionality to its star-formation rate. In turn, the star-formation rate is correlated with nuclear activity \citep[e.g.][]{Chen13}, meaning that both star-formation and central black hole feedback are prime candidates for quenching the host galaxy. These arguments are well-established for quenching the central galaxy of a dark matter halo, but do not {\em directly} influence the surrounding satellite galaxies.

There are a wide variety of candidate processes for quenching satellite galaxies, from those sensitive to the gaseous halo environment (e.g. strangulation \citealt{Larson80} or ram pressure stripping, \citealt{Gunn72}), to those that depend predominantly on the halo potential (tidal stripping), or on the number and orbits of other satellites (e.g. harassment). As we are searching for a mechanism that depends neither on halo mass nor the number of other satellites (as the mean number of satellites we can measure is $<1$), the likely candidates would seem to be those that require a dense inter-galactic medium. Furthermore, there is growing evidence that a galaxy's environment does not have a significant impact upon its morphology. \cite{Maltby12} found that even a cluster environment is insufficient to alter the outer disk properties of the member galaxies appreciably. Similarly, the morphological mix of quenched galaxies in group environments is found to be independent of group-centric radius and halo mass, with the implication that a passive galaxy's morphology is established {\em before} quenching occurs \citep{Carollo14}. This is not to say that there is exactly {\em zero} environmental effect on a galaxy's morphology as it is quenched, but gas stripping (if responsible) is likely to be fairly gentle.

The hot gas content (that is necessary to strip galaxies of their outer gas halos) in galaxy groups is observed to be low, of order a few percent at halo masses, ${\rm log}(\M_{{\rm halo}}/\M_{\odot}) \sim 13$, and to depend on halo mass roughly as $f_{gas} \propto \M_{{\rm halo}}^{0.2}$ \citep{Giodini09}. However, these studies also show that the scatter in the gas-fraction - halo-mass relation is substantial. Direct observations of hot gas fractions at lower halo masses are very few and confined to the local Universe. However, they are typically very low \citep{Li14}. Environmental processes that are sensitive to gas density, such as ram pressure stripping, would have limited ability to quench satellites of modest stellar mass in such poorly established gaseous halos.

Following a sustained galaxy-wide outflow, however, the hot gas halo may be changed substantially. Especially at high redshift, where cold gas masses within star-forming galaxies can reach $\sim 1/3$ of the stellar mass or higher \citep{Daddi10}. During this time it is possible that the halo environment becomes suitable for satellite-quenching processes (e.g. stripping). At the same time this environment could increase in its ability to shock heat newly accreted gas and therefore restrict fuel for star-formation in the central galaxy. If so, then the process could become self-sustaining, accumulating more gas in the halo environment, rather than it becoming locked away in stars and cold gas within galaxies. In this way, galactic conformity can be explained by the central galaxy activating a switch, whereupon the halo environment becomes responsible for preventing star-formation in both the central galaxy and its satellites. This mechanism could then reproduce the correlation between central and satellite star-formation properties without the need for a dichotomy at some halo-mass scale. 

For many years there has been vigorous debate about whether the dark matter (and hot gas) halo or some form of feedback is primarily responsible for shutting off star-formation in central galaxies. The emerging picture that we present, based on our results and the existing literature, is that actually both could be important, albeit in a less direct manner than in their proposed form. A step-by-step description of the possible process for this coincident quenching is as follows:

\begin{enumerate}
\item We begin with a central galaxy and its satellite galaxy, each forming stars at the typical inefficient rate from their existing supply of gas. The hot gas corona surrounding the central galaxy is poorly established, and does not cause much obstruction to infalling cold gas joining the disks of the two galaxies and being available for star-formation. Star-formation in the galaxies heats and expels some gas, but the density of the circum-galactic medium remains low and over time this gas may cool and rejoin a disk. In low-mass halos this gas may instead leave the halo entirely. The satellite galaxy is unperturbed by the low-density medium that it passes through.
\item The galaxy stellar masses and host halo mass grow over time. At some point the star-formation rate of the central galaxy climbs to the point at which the resulting outflow has an important impact upon its surroundings. At high redshift such high star-formation rates are a natural consequence of rapid gas accretion in massive galaxies. At low redshift a merger may be required. A fraction of the remaining gas in the central galaxy is removed from the galaxy by galaxy-scale winds. However, it is not removed from the now massive halo due to depth of the potential well. The gas is heated by the supernovae (or possibly AGN) as it is swept out of the galaxy, with strong shocks further heating the gas. The central galaxy is now relatively gas poor and the star-formation rate declines.
\item Infalling gas, which was previously the major source of star-forming material for the central galaxy, now encounters a hot dense medium as it descends through the halo. It experiences shock heating in addition to gravitational heating and gets added to the existing hot gas component. The supply of cold gas reaching the central galaxy is heavily suppressed with respect to the time before the major outflow. The star-formation rate remains low relative to unquenched galaxies of similar mass and a few hundred Myrs later, the central will be classified as a passive galaxy. Meanwhile, the satellite galaxy also feels the effect of the increased halo gas density, and its more loosely bound gas is efficiently stripped, again adding to the pool of hot gas in the circum-galactic medium. The remaining gas in the satellite is eventually used up and it too becomes a quenched object.
\item Although the circum-galactic gas cools over time it will not fully return to a state in which the central galaxy appears to be a normal star-forming object. The density of the gas corona is maintained by the addition of shock-heated infall and gas stripped from the satellite, which outpaces the rate at which the gas can cool onto the central. The cooling of gas is also regulated by the action of radio mode feedback from the central super-massive black hole.
\end{enumerate}

We note that a star-formation (or radio) outburst, if responsible for conformity, is unlikely to be acting independently. A high halo mass is still required in order to retain a sufficiently dense gas halo for stripping and shock heating of infalling gas to occur, as at low halo masses feedback could well remove the outflowing gas from the halo entirely. In normal star-forming galaxies the gas mass in an outflow is much lower than the available cold gas reservoir \citep{Bradshaw13}, and this gas removal would therefore not be capable of quenching either the central or satellite galaxy. Furthermore, a massive galaxy (and hence halo) is necessary because these are the only objects with cold gas masses sufficient to establish a dense hot gas halo. Moreover, this picture does not preclude the necessity for radio mode feedback to prevent large quantities of gas cooling onto the central object at late times \citep[e.g.][]{Fabian99, Best05, Croton06}. 

\subsection{Support from the literature}

An argument along similar lines to the one we present was laid out in a toy model by \cite{Rawlings04}. In their model it was postulated that a powerful radio outburst resulting from two super-massive black holes merging would heat the {\em common} cool gas halo of the merger remnant host and nearby galaxies. This single event could end star-formation in all galaxies fed from this common gas reservoir simultaneously, at scales up to $\R=1~{\rm Mpc}$. Their model actually predicted a correlation in star-formation activity between galaxies within a proto-cluster that we might now include under the umbrella term of `galactic conformity'. From present observational studies of conformity it is not possible to differentiate the \cite{Rawlings04} model from our own description. 

The results in recent years from sophisticated hydro-dynamical models lend a degree of support to our descriptive process. Within many simulations the baryonic budget of galaxies is dominated by gas that was accreted in the so-called `cold mode', i.e. gas that never experienced shock heating as it fell from the outskirts of the halo to the disk of the central galaxy \citep[e.g.][]{Keres09}. In dark matter halos with masses ${\rm log}(\M_{{\rm halo}}/\M_{\odot}) > 12$, the specific gas accretion rate drops sharply. However, \cite{vandeVoort11} find that even high-mass (${\rm log}(\M_{{\rm halo}}/\M_{\odot}) \sim 13$) halos are unable to {\em fully} prevent cold-mode gas accretion (and hence star formation) in their central galaxy. The details of how gas is accreted onto galaxies in their model depends sensitively on the feedback prescriptions used, particularly in sub-group to group-mass halos. If the central galaxy is instrumental is establishing a hot gaseous halo as we have argued, then the halo-averaged trends presented by \cite{vandeVoort11}, the relatively shallow rise in passive fraction with halo mass found from HOD modelling and galactic conformity may all be manifestations of this common physical cause.

Recent observations of the hot gas halos of modest-mass galaxies also lend some weight to our proposed mechanism. \cite{Li14} find that the hot gas coronae of intermediate-mass ($10.5 < {\rm log} (\M_*/\M_{\odot}) < 11$) star-forming galaxies are far fainter at X-ray wavelengths than expected from simple analytic models. Furthermore, they find a clear correlation between the star-formation rate of a galaxy and the luminosity of its X-ray halo. Enhanced star-formation does then appear to be able to create a denser hot gas halo. This correlation is all the more compelling when we recall the possible SFR-dependence on the quenching probability for central galaxies in \cite{Peng10}. A simple correlation from sparse data is far from being conclusive proof for our descriptive process, but encouraging nonetheless. We hope that further investigations of the X-ray emitting gas around normal galaxies will be forthcoming in the future.

\subsection{A remaining puzzle}
\label{puzzle}

We have presented one possible description of galaxy quenching, incorporating both central and satellite galaxies, that can explain the phenomenon of galactic conformity as well as other broad trends from the literature \citep[e.g.][]{Peng10, Hartley13}. However, there remain some basic observations that cannot be explained within this context. In particular, the abundance of non-star-forming dwarf galaxies that are satellites of our own galaxy and of M31, both of which are relatively massive star-forming galaxies. These dwarf satellites clearly do not fit into the picture of galactic conformity. Neither do they fit onto other extrapolations from trends found in intermediate-mass galaxies \citep{Wheeler14}. 

There have been suggestions that these sort of objects are formed in a different way from the broader galaxy population, perhaps as the result of tidal debris \citep[e.g.][and references therein]{Pawlowski13}. If true, it would certainly help reconcile this strange population with the emerging picture of galaxy quenching. An alternative explanation is that galaxies on such small scales could be vulnerable to events that are unable to quench more massive galaxies. One recent example is the suggestion that a large-scale shock front, driven by star-formation or an AGN, could {\em directly} strip these small objects of their gas supply, effectively preventing further star-formation \citep{Nayakshin13}. Whatever their origin, ultra-low-mass galaxies remain outside of the broad galaxy description for the time being.

\section[]{Conclusions}
\label{conc}

We have used the UKIDSS UDS, the deepest degree-scale near-infrared survey to date, to investigate the phenomenon of galactic conformity. We have explored the surface number-density profiles of satellites around $\M\sim \M^{\ast}$ galaxies up to $z\sim2$ within two satellite stellar mass intervals. These results were used to derive the radial dependence of the ratio of passive to mass-selected satellites. Our principal findings are as follows:

1. The power-law slopes of massive satellites are systematically steeper than those of lower-mass satellites. Although the slope and normalisation of our fits are degenerate, this behaviour is consistent across almost all samples and requires further study from independent data sets in order to fully understand the trend. The numbers of high and low-mass satellites are found to be similar. We would expect a greater number of low-mass satellites. We suggest that satellite disruption could explain these two results but this observation remains largely unresolved.

2. The number of satellites is greater (around twice as many) in our high redshift interval ($1.3<z<1.9$) than at lower redshifts ($0.4<z<1.3$). If we interpret the number of satellites as an indicator of halo mass, these results are consistent with evolutionary trends derived from clustering studies \citep[e.g.][]{Hartley13}.

3. The power-law slopes of passive satellites are consistently steeper than those of the mass-selected satellites. This behaviour is qualitatively similar to the morphology-density relation and hints at a possible environmental process transforming satellites at small separations.

4. In our low-redshift interval, passive central galaxies have a greater number of satellites than star-forming centrals. This again can be interpreted as a halo-mass effect, consistent with clustering studies. We find no difference in the number of satellites between passive and star-forming centrals at higher redshift. However, we caution that the halo assembly history can also influence the number of satellites that have masses within an order of magnitude of the central \citep{Dariush07}.
 
5. The ratio of passive satellites to mass-selected satellites (our estimator for passive fraction) at $\R\sim 150~$kpc around passive central galaxies implies an environmental quenching efficiency of $\epsilon \sim 0.1 - 0.2$ \citep{Peng10,Peng12,Kovac14}.

6. The correlation between the star-formation activity of a central galaxy and its satellites, known as `galactic conformity', exists up to $z\sim2$ - a measurement that has only previously been made at $z\sim 0$. Star-forming centrals have a satellite population that is consistent with having been drawn directly from the field. Passive centrals, meanwhile, have an enhanced passive satellite population.

7. There is an apparent radial dependence to the passive fraction, rising with decreasing separation from the central galaxy. The small number of galaxies at small separations prevents us from making a definitive statement, however any measured slope is likely to be shallower than the true dependence.

8. We perform the same radially-dependent passive fraction measurements for high-mass (${\rm log} (\M_*/\M_{\odot}) > 11.$) star-forming central galaxies. These results are very similar to those of lower-mass star-forming galaxies and we therefore argue that the host halo mass is not exclusively the cause of galactic conformity. 

9. We suggest a possible origin for galactic conformity, whereby gas expelled from the central galaxy due to star-formation or an active nucleus establishes the conditions necessary to quench both the central and satellite galaxies. Once established this hot gas corona could be self-sustaining and does not require a rapid change in galaxy behaviour over a small interval in halo mass.

The study we have presented is at the limit of what we could achieve with the the UDS DR8. However, this work could be extended in the relatively near future. The final UKIDSS UDS data will reach $25.2$ (AB) in the K-band (i.e. 0.6 magnitudes deeper than the data used in this work), with a quarter of the field a further 0.3 magnitudes deeper from the forthcoming UDS+ survey. Furthermore, the Cosmic Assembly Near-infrared Deep Extragalactic Legacy Survey (CANDELS) observations have now been completed. These data should allow us to push our study to higher redshift, where the first substantial populations of passive galaxies are found, and to lower satellite masses.

Testing our picture for the origin of conformity will be no easy task. Observations of the hot gas within modest-mass halos could be the most sensitive test, but are extremely challenging and can only currently be achieved at low redshift. As a first step, we intend to perform simple modelling of the processes we have described, place them in a more quantitative setting and see to what extent we can reproduce the observed behaviour.

\section*{Acknowledgements}
We are indebted to the tireless efforts of the UKIRT and Joint Astronomy Centre staff over the many years that UKIDSS observations were carried out and extend our deepest gratitude. We would also like to thank the Cambridge Astronomical Survey Unit and WFCAM Science Archive scientists for their crucial contributions to the success of the UDS. AM acknowledges the support from a European Research Council consolidator grant (PI: McLure). The United Kingdom Infrared Telescope is operated by the Joint Astronomy Centre on behalf of the Science and Technology Facilities Council of the U.K. This work is also based in part on data collected at Subaru Telescope, which is operated by the National Astronomical Observatory of Japan.
\section*{Appendix}
\subsection*{Dry / moist merger rates for size evolution of passive galaxies}

\begin{figure}
\includegraphics[angle=0, width=240pt]{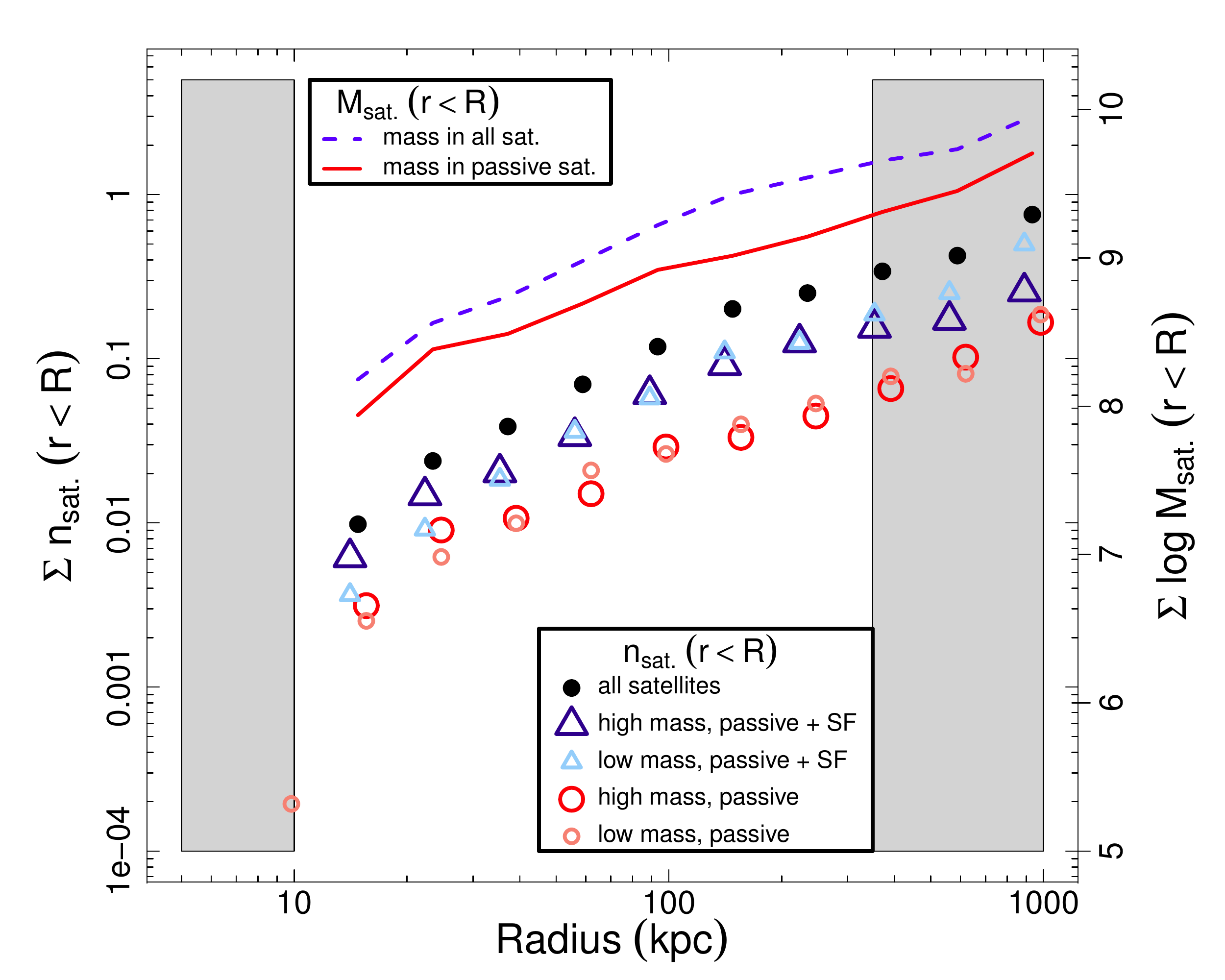}
\includegraphics[angle=0, width=240pt]{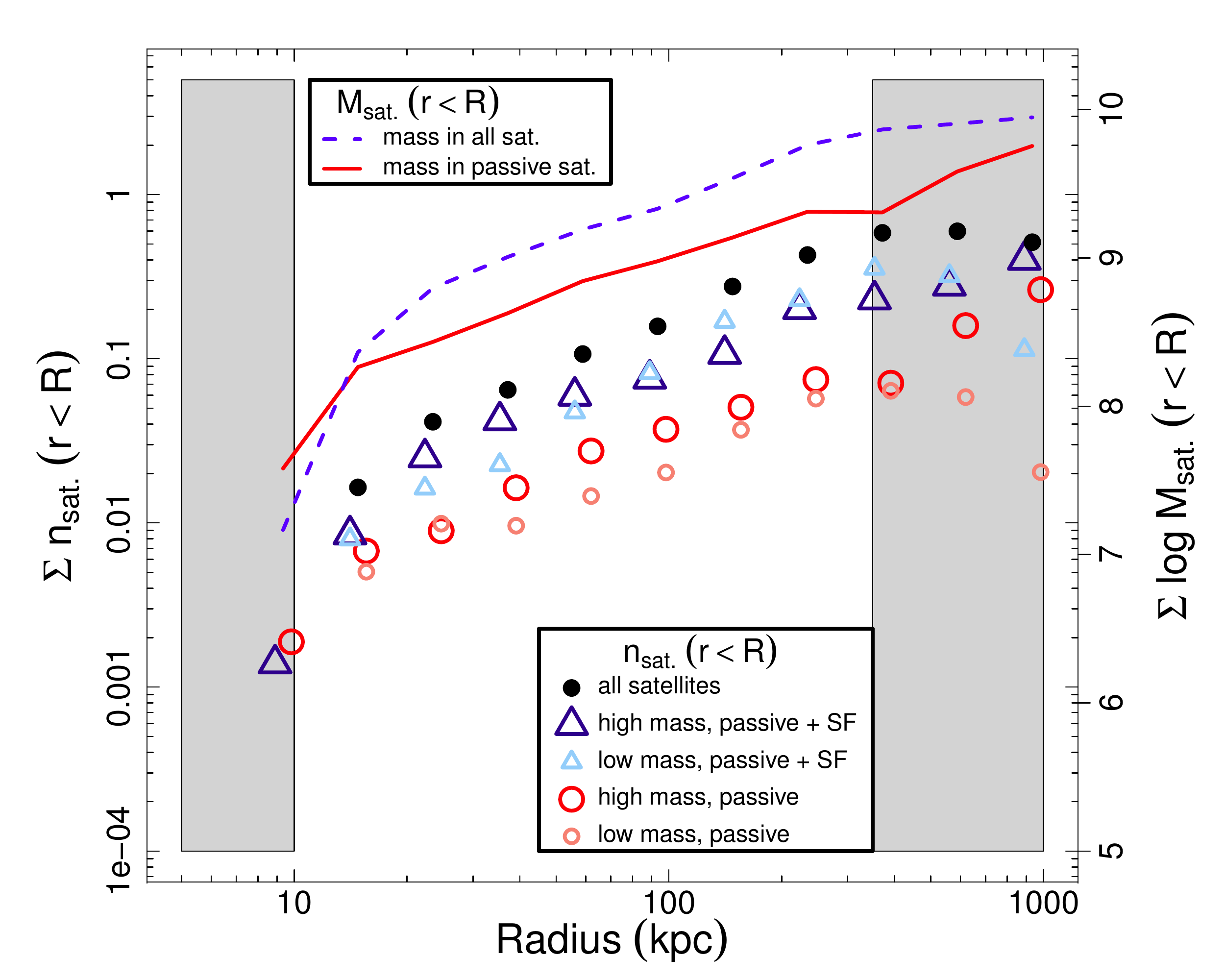}
\caption{Cumulative number and stellar mass of satellites around $\M\sim \M^*$ passive central galaxies as a function of radius for the various satellite samples used in this work. The upper panel is for centrals in the lower redshift interval, $0.4<z<1.3$, while the lower panel shows the measurements for the higher redshift galaxies ($1.3<z<1.9$). Circle (passive) and triangle (mass-selected) symbols correspond to the cumulative numbers of satellites per central galaxy (left-hand scale), while the solid red and dashed blue line are the integrated stellar mass in these satellite samples (right-hand scale), as given in the legend.}
\label{cuml_pass}
\end{figure}

One of the most striking and unexpected discoveries in extragalactic astrophysics over the last decade has been the very compact structure of massive passive galaxies at high redshift \citep[e.g.][]{Daddi05,Zirm07,Trujillo07,Buitrago08,Cimatti08}. The absence of significant numbers of similarly compact passive galaxies at low redshift implies a rapid increase in the effective radius with cosmic epoch. Two mechanisms were proposed to explain this size evolution: a `puffing up' by the stellar orbits resettling following the expulsion of a significant fraction of the galaxy mass by an AGN \citep{Fan08}; and minor dry mergers piling-up stellar light in the outer regions of a galaxy \citep{Naab09}. 

Dry or moist mergers are favoured because there is no clear evidence for the sort of star-formation that might be expected as a result of a gas-rich merger. Furthermore, a dissipational merger could result in an increase in the central density and, perhaps, a smaller resulting size. The community has largely converged on minor dry mergers as the dominant solution, with typical estimates of $\sim 4-5$ such mergers required per galaxy between $0<z<2$, in addition to a single major merger \citep[e.g.][]{Conselice06,Bluck12,McLure13}. A fuller appreciation of the likely descendants of these high-z passive galaxies \citep{Poggianti13}, and the inclusion of the structure of galaxies quenched at lower redshift in the modelling \citep[e.g.][]{Cassata13,Carollo13}, has caused the number of required mergers to be revised downwards. Nevertheless, the rarity of quenched low-mass galaxies at high-z means a suitable galaxy population for minor dry mergers is yet to be confirmed.

In Figure \ref{cuml_pass} we show, for $\M\sim \M^{\ast}$ passive galaxies, the cumulative number of satellite galaxies within radius, $\R$, for the four satellite samples we have used in this work (open symbols as shown in the figure key). In addition we show the combined satellite number by solid black circles, and the summed masses in all (star-forming + passive) and passive satellites, using the mean stellar mass of the samples. The total number of satellites per object is $<1$ in both redshift intervals, which includes satellites with mass ratios that would be considered `major' mergers (a ratio 3:1 is usually considered the major / minor cut off). The combined mass in satellites for high-redshift passive centrals is {\em at least} a factor of three below the mass of the central. This is important, because passive galaxies at $z>1$ are not expected to increase in mass by more than a factor of two or so by $z=0$ \citep[e.g.][]{Cimatti06}.

If the estimates from size evolution studies for the required number of minor mergers are correct, then further satellite accretion is required. Moreover, these accreted satellite would have to be subsequently stripped of much of their gas in order to fulfil the dry / moist merger criterion. In the picture for galactic conformity which we have presented, this would certainly be possible. An environment capable of processing satellites is a natural consequence of the quenching of the central galaxy. The shape of the radial dependence of the satellite passive fraction, if found to be robust, also suggests that by the time a low-mass satellite is close enough to merge with the central, it is likely to have lost much of its gas. 

\subsection*{Example of galaxy weighting}

\begin{figure*}
\centering
\noindent\begin{minipage}{180mm}
\includegraphics[angle=0, width=180mm]{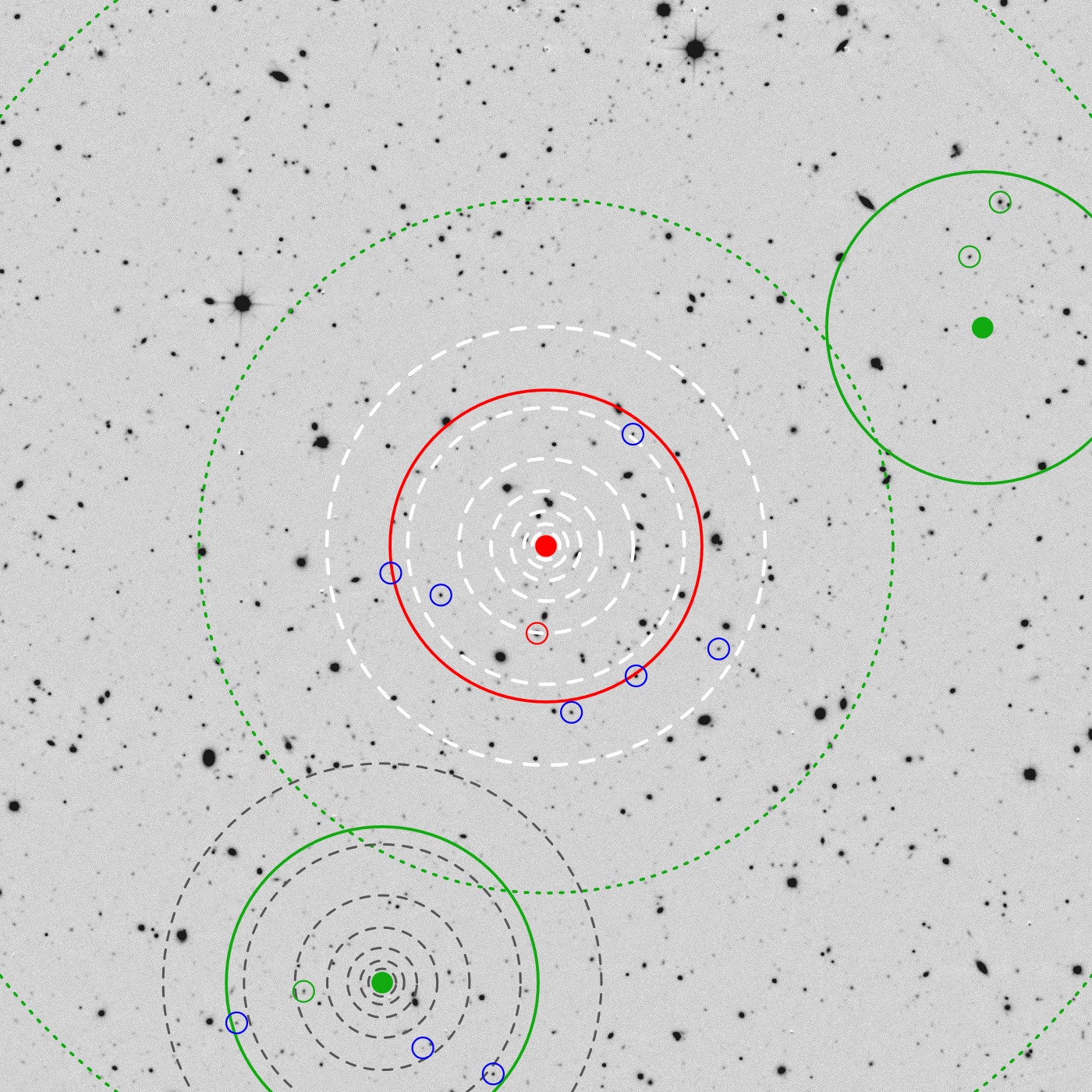}
\caption{A small subregion of the UDS field, demonstrating the weighting and counting procedure used in this work (detailed in the text). A colour version of this Figure is available in the on-line journal.}
\label{weighting}
\end{minipage}
\end{figure*}

In this section we provided an illustrative example of the method and weighting procedure used throughout this work. Figure \ref{weighting} displays a small sub-region ($\sim6\arcmin\times6\arcmin$) of the UDS field, centred on a candidate central galaxy (filled red point). In this particular example the target central sample comprises passive galaxies with stellar masses $10.5 < {\rm log (M_*/M_{\odot})} \leq 11.0$, and this candidate has stellar mass, ${\rm log (M_*/M_{\odot})} = 10.71$.

For each candidate central galaxy we search for other galaxies within a projected radius, ${\rm R}=450~{\rm kpc}$ (large red circle), that have stellar mass, ${\rm log (M_*/M_{\odot})}\geq {\rm log (M_{cen}/M_{\odot}}) - 0.3$, and satisfy the photometric redshift condition, $|z_{{\rm phot,1}}-z_{{\rm phot,2}}| < \sqrt{2}~ \sigma_z(z)(1+z)$. We find one such galaxy, with mass, ${\rm log (M_*/M_{\odot})} = 10.62$ (open red point). Where there are two similar mass galaxies with small projected separation and at consistent $z_{{\rm phot}}$, we have no way of knowing from photometric data alone which galaxy is a genuine central, or which galaxies are satellites of which central. To overcome this limitation of photometric data each central galaxy is given an initial weight, determined by the number, N, of these similar-mass galaxies within the search cylinder: ${\rm weight} = 1/(1+N)$. This weight accounts for the possible double counting of satellite galaxies, without resorting to the unphysical assumption that all possible satellites are satellites of the slightly more massive galaxy. Had this secondary galaxy been of mass, ${\rm log (M_*/M_{\odot})} > 11.01$ (i.e. 0.3 dex greater than the candidate central), then the candidate central galaxy would have been considered a probable satellite and therefore excluded from the central sample.

Even within a relatively narrow stellar mass range, the masses of passive and star-forming galaxies are different, with passive galaxies dominating at higher masses. We must therefore adjust the galaxy weights to account for this difference. Once the target central sample and their weights have been defined, we construct a weighted histogram of their stellar masses in bins of $\Delta {\rm log (M)} = 0.05$. This histogram is compared with one constructed in an identical manner, but for a purely stellar mass-selected central sample. The weights of the target sample are then re-normalised, bin by bin, so that the resulting weighted histogram is identical to that of the mass-selected sample. Take the stellar mass bin that contains our example central galaxy, $10.70 < {\rm log (M_*/M_{\odot})} \leq 10.75$. The sum of weights for the target passive sample in this bin is $82.3$ from $1491$ objects, while for the mass-selected sample it is $175.8$ from $2697$ objects. The weight of our example galaxy is therefore multiplied by $\frac{175.8}{82.3}\frac{1491}{2697}=1.18$, and has a final weight, $w=0.59$.

With the central galaxy sample trimmed of likely satellite contaminants we can proceed to the satellite counting. Counting is performed in concentric annuli of log radial intervals. These annuli are shown by the dashed white circles in Figure \ref{weighting} (the largest two annuli are not shown). All galaxies with $|z_{{\rm phot,1}}-z_{{\rm phot,2}}| < 1.5~ \sigma_z(z)(1+z)$ are considered potential satellites should they lie within one of the annuli. The photometric redshift of the central galaxy is assumed to be correct, as redshift accuracy is dependent on signal-to-noise and the centrals are typically brighter than the possible satellites. If the candidate satellite is indeed physically associated with the central galaxy, then it should be at the same redshift. We therefore scale the stellar mass of each satellite, using the difference in luminosity distance, assuming that it lies at the redshift of the central. Galaxies are included as satellites if, after scaling, their mass falls into one of the stellar mass intervals, $9.7 < {\rm log (M_*/M_{\odot})} \leq 10.1$ or $10.1 < {\rm log (M_*/M_{\odot})} \leq 10.5$. In Figure \ref{weighting} there are six galaxies which satisfy these criteria. The number of satellites in an annulus is multiplied by the weight of the central galaxy and added to the running total.

Most of the galaxies which are counted as satellites will not be physically associated with the central galaxy. To estimate the correct number we must subtract off the background / foreground contaminants. As described in section \ref{background}, \cite{Chen06} found that the most reliable method to use to do so, is to select random locations within a correlation length of the central galaxy. In this work we allow random galaxies to be placed between $1$ and $2~{\rm Mpc}$ from the central. These limits are shown by the green short-dashed lines in Figure \ref{weighting}. Each central galaxy has $20$ random locations chosen, in the following we will only describe one such random object.

Random objects inherit all characteristics (redshift, stellar mass etc.) from their parent central galaxy, and are subjected to the exact same procedure as the central galaxy candidates, including the search for other high-mass galaxies within $\R=450~{\rm kpc}$. In our illustration, a random location in the upper right is first chosen (solid green point and large green circle). However, two galaxies are found within the search criteria (open green points), and in one case the stellar mass is ${\rm log (M_*/M_{\odot})} = 11.10$. This location is therefore rejected, and a new random location is chosen, this time in the lower part of our diagram. For this new location a single galaxy with mass, ${\rm log (M_*/M_{\odot})} = 10.83$, is found, and so the random object is given a weight, $w=0.5$. 

Rather than re-normalising the weights by their overall stellar mass histogram, we demand that each set of $20$ random objects contribute between them the same weight to the random satellite counts, as their parent central galaxy does to the `real' satellite counts. For example, if the sum of the weights of the $20$ random locations for our central object is $12.3$, then each random will have its weight multiplied by $0.59/12.3$, such that their sum is now equal to the weight of the central. Satellites around these random locations are counted in the same way as the central (dark grey dashed circles and open blue points), and added to a running total of random satellite counts. Finally, the random counts are subtracted from the `real' satellite counts to leave a statistical number of satellites as a function of radius from the central galaxy sample.

\bibliographystyle{mn2e.bst}
\bibliography{mn-jour,papers_cited_by_WH}

\label{lastpage}

\end{document}